\documentclass[12pt, a4paper]{article}

\pdfoutput=1
\usepackage{hyperref}

\usepackage{color}
\usepackage{float}
\usepackage{slashed}
\usepackage{bbold}
\usepackage{mathtools}
\usepackage{amsmath}
\usepackage{amssymb}
\usepackage{multirow}
\usepackage{arydshln}
\usepackage[usenames,dvipsnames]{xcolor}
\usepackage{caption}
\usepackage{graphicx}
\usepackage{inputenc}
\usepackage{tikz}
\usetikzlibrary{matrix}
\usetikzlibrary{calc,fit}
\usepackage{cite}

\usepackage{dsfont}

\DeclareFontEncoding{U}{}{}
\DeclareFontFamily{U}{bbold}{}
\DeclareFontShape{U}{bbold}{m}{n}
{  <5> <6> <7> <8> <9> gen * bbold
	<10> <10.95> bbold10
	<12> <14.4> bbold12
	<17.28> <20.74> <24.88> bbold17
}{}
\DeclareSymbolFont{bbold}{U}{bbold}{m}{n}
\DeclareSymbolFontAlphabet{\mathbbold}{bbold}
\newcommand{\nn}{\nonumber}
\newcommand{\ad}{a^\dagger}
\newcommand{\be}{\begin{equation}}
\newcommand{\ee}{\end{equation}}
\newcommand{\bea}{\begin{eqnarray}}
\newcommand{\eea}{\end{eqnarray}}
\newcommand{\eq}[1]{Eq.~(\ref{#1})}

\def\bma#1{\mbox{\boldmath{$#1$}}}

\def\bra#1{\mathinner{\langle{#1}}}
\def\ket#1{\mathinner{{#1}\, \rangle}}

 \def\cE{{\cal E}}

 \def\eps{\epsilon}

 \def\om{\omega}
 \def\Om{\Omega}

\def\bma#1{\mbox{\boldmath{$#1$}}}

\begin{document}

\thispagestyle{empty}
\begin{titlepage}

\today
\vspace{.3in}

\vspace{1cm}
\begin{center}
{\Large\bf\color{black} 
The renormalized Hamiltonian truncation  

\vspace{0.42cm}

method in the large $E_T$ expansion
% The Renormalized Hamiltonian Truncation Method and its Large $E_T$ expansion
% A systematic approach at any order of the Hamiltonian Truncation Method
% Study of the Renormalized Hamiltonian Truncation Method and its Large $E_T$ expansion
% Renormalized Hamiltonian Truncation Method and study of its Large $E_T$ expansion
% Strong coupling, Hamiltonian Truncation Method and the Large $E_T$ expansion
} \\
\medskip\color{black}
\vspace{1cm}{
{\large  J.~Elias-Mir\'o$^{a}$\footnote{jelias@sissa.it}, M.~Montull$^{b,}$\footnote{mmontull@ifae.es}, M.~Riembau$^{b,c,}$\footnote{marc.riembau@desy.de}}
\vspace{0.3cm}
} \\[7mm]
{\it {$^a$\, SISSA and INFN, I--34136 Trieste, Italy}}\\
    {\it {$^b$\, Institut de F\'isica d'Altes Energies (IFAE), Barcelona Institute of Science \\ and Technology (BIST) Campus UAB, E-08193 Bellaterra, Spain}}\\
{\it $^c$ {DESY, Notkestrasse 85, 22607 Hamburg, Germany}}
\end{center}
\medskip

\begin{abstract}
Hamiltonian Truncation Methods are a useful numerical tool to  study strongly coupled QFTs. In this work we present a new method to compute the exact corrections, at any order, in the Hamiltonian Truncation approach presented  by Rychkov et al. in  Refs.~\cite{Rychkov:2014eea,Rychkov:2015vap,Hogervorst:2014rta}.  The method is general but as an example we calculate the exact $g^2$ and some of the $g^3$ contributions for the $\phi^4$ theory in two dimensions. 
The coefficients of the local expansion calculated in Ref.~\cite{Rychkov:2014eea}   are shown to be given by phase space integrals.  In addition we find new approximations to speed up the numerical calculations and implement them to compute the lowest energy levels at strong coupling.  
 A simple diagrammatic representation of the corrections and various tests are  also introduced. 
\end{abstract}
\medskip

\end{titlepage}

%%to format line stance of tableofcontents

\tableofcontents

\section{Introduction and review}

An outstanding problem in theoretical  physics is to solve strongly coupled Quantum Field Theories (QFT). 
When they are not amenable to analytic calculations  one can resort to numerical approaches. 
The two most used numerical approaches are lattice simulations and direct diagonalization of truncated Hamiltonians.  In this paper we further develop the Hamiltonian truncation method recently presented in Ref.~\cite{Rychkov:2014eea,Rychkov:2015vap,Hogervorst:2014rta}, that renormalizes the truncated Hamiltonian $H_T$ to improve the numerical accuracy. 

The Hamiltonian truncation method consists in truncating the Hamiltonian $H$ into a large  finite  matrix $(H_T)_{ij}$ and then diagonalizing it numerically. 
There is a systematic error with this approach that vanishes as the size of the truncated Hamiltonian $H_T$ is increased.  
There are different versions of the Hamiltonian truncation method that mainly differ on the frame of quantization and the  choice of basis in which $H$ is truncated. Two broad categories within the Hamiltonian truncation methods are the  Truncated Conformal Space Approach~\cite{Yurov:1989yu} and Discrete Light Cone Quantization~\cite{Brodsky:1997de}. A less traveled route  consists in using the Fock-Space basis to truncate the Hamiltonian~\cite{Lee:2000ac,Lee:2000xna,Salwen:2002dx,Windoloski,Brooks:1983sb,Rychkov:2014eea,Rychkov:2015vap}. Lately there have been many advances in the Hamiltonian Truncation methods, see for instance \cite{
Hogervorst:2014rta,Katz:2013qua,Katz:2014uoa,Chabysheva:2014rra,Giokas:2011ix,Lencses:2014tba,Coser:2014lla,Lencses:2015bpa}.
\medskip

We review the truncated Hamiltonian approach following the discussion of Ref.~\cite{Hogervorst:2014rta,Rychkov:2014eea}.  The problem we are interested  in  is finding the spectrum of a strongly coupled QFT. Therefore we want to solve  the eigenvalue   equation
 \be
H |\ket{\cE} = \cE | \ket{\cE}  \, , \label{eigen1}
 \ee
 where $H=H_0+V$,   $H_0$ is a solvable Hamiltonian or the free Hamiltonian and  $V$ is the potential.  $H_0$  is diagonalized by the states $H_0 |\ket{E_n} = E_n | \ket{E_n} $.
Suppose we are interested in studying the  lowest energy states of the theory. One way to do it is separating the  Hilbert  space ${\cal H}$  into ${\cal H}= {\cal H}_l \oplus{\cal H}_h $, where  ${\cal H}_l$ is of  finite dimension and  it is spanned by  the states $|\ket{E_n}$ with  $E_n\leq E_{T}$. Then, the Hilbert space ${\cal H}_h$ is an infinite-dimensional Hilbert space containing the rest of the  states $E_n> E_{T}$. 
 The states are projected as  $P_l|\ket{x}\equiv |\ket{x_l} \in{\cal H}_l $ and $(\mathds{I}-P_l)|\ket{x} = P_h |\ket{x}\equiv |\ket{x_h} \in{\cal H}_h$.  
 Then, the eigenvalue problem   can be replaced by 
 \be
 H_{eff}(\cE)|\ket{\cE_l}=\cE |\ket{\cE_l}  \, ,
 \label{eigen2}
 \ee
 where  $H_{eff}\equiv H_T + \Delta H(\cE)$,  the  truncated Hamiltonian is   $H_T=P_l H P_l$ and 
 \be
 \Delta H(\cE)= V_{lh}\frac{1}{\cE-H_{0\, hh}-V_{hh}}V_{hl} \, ,
 \label{deltaHg}
 \ee
 with   $O_{ij}\equiv P_{i}OP_{j}$ for $i,j\in \{h,l\}$. 
 To derive \eq{eigen2}, project \eq{eigen1} into the two equations
 \be
H_{ll} |\ket{\cE_l}+ H_{lh}  |\ket{\cE_h} = {\cal E} | \ket{\cE_l} \, , \quad \quad
 H_{hl} |\ket{\cE_l}+H_{hh}  |\ket{\cE_h} = {\cal E} | \ket{\cE_h}\, ,\label{heigen}
\ee
and then substitute $|\ket{\cE_h}=\left( \cE - H_{hh}\right)^{-1}  H_{hl}|\ket{\cE_l}$  from the second equation in (\ref{heigen}) into the first.
 
 Notice that  \eq{eigen2} is an exact equation and that a complete knowledge of $\Delta H(\cE)$ would render the original eigenvalue problem of \eq{eigen1} solvable by an easy numerical diagonalization. 
 In the limit where $E_T \rightarrow \infty$ the corrections $\Delta H$ to $H_T$ can be neglected, but it is computationally very costly to increase the size of $H_T$ and then diagonalize it.  Therefore it is interesting to calculate $\Delta H$ to improve the numerical accuracy for a given $E_T$.
  A first step to compute  $\Delta H$ is to  perform an expansion of \eq{eigen2} in powers of $V_{hh}(\cE-H_0)^{-1}$,
\be
\Delta H(\cE,\, E_T)=\sum_{n=0}^\infty \Delta H_n(\cE,E_{T}) \ , \ \  \text{where}\quad     \Delta H_n(\cE,\, E_T)=V_{lh}\frac{1}{\cE-E_{hh}}\left(V_{hh}\frac{1}{\cE-E_{hh}} \right)^nV_{hl} \, ,
\nonumber% not cited \label{deltaHseries} 
\ee
where the matrix elements of $\Delta H_n$ are given by
\bea
\Delta H_n(\cE)_{rs}&=&\sum_{\substack{j_1, \cdots \, ,j_n-1 :\,  E_{j_i}>E_T}} V_{rj_1} \frac{1}{{ \cE-E_{j_1} }}V_{j_1 j_2} \frac{1}{ \cE -E_{j_2}} V_{j_2 j_3} \cdots V_{j_{n-2}j_{n-1}}\frac{1}{ \cE-E_{j_{n-1}} }V_{j_{n-1}s} \, ,\quad \quad  \label{deltaHfock}
\eea
in the  $H_0$ eigenbasis and the sums run over all labels $j_1,\, \dots, \,  j_{n-1}$ of states belonging to ${\cal H}_h$ with  $r,s$  denoting  the matrix elements (corresponding to eigenstates  of $H_0$ with $E_s, E_r\leq E_T$ eigenvalues). 
 Naively the truncation of the series in  \eq{deltaHfock} is  justified  for $V_{hh}/H_{0\, hh}<1$ which for large enough $E_T$ and $\cE \ll E_T$ is fulfilled, and allows to go to strong coupling. This is discussed in detail in Sec. \ref{sc2}. The operator $\Delta H$ depends on the exact eigenvalue and in practice the way \eq{eigen2} is solved is by diagonalizing iteratively $H_{eff}(\cE ^*)$ starting with an initial seed $\cE ^*$. It is convenient to take $\cE^*$ close to the exact eigenvalue $\cE$, a simple and effective choice is to take the  eigenvalue obtained from diagonalizing $H_T$.
 
 In Ref.~\cite{Rychkov:2014eea} the $\phi^4$ theory in two dimensions was studied at strong coupling using the  Hamiltonian truncation method just presented in the Fock basis. There, the leading terms of $\Delta H_2$ doing a local expansion were computed and shown to improve the results with respect to the ones found by only diagonalizing $H_T$. The main result of our work is to explain a way to calculate the exact corrections to $\Delta H$ at any order $\Delta H_n$.   As an example we calculate the $\Delta H_2$ correction and some of the $\Delta H_3$ terms for the $\phi^4$ theory in two dimensions and present various approximation schemes for a faster numerical implementation. This can be seen as an extension of the method presented in Ref.~\cite{Rychkov:2014eea} which we believe to be very promising.

\medskip  

The paper is organized as follows. In Sec.~\ref{sec:themethod} we introduce a general formula to compute $\Delta H_n(\cE,\, E_T)$ at any order $n$. Then we apply the method to the $\phi^2$ and $\phi^4$ scalar field theories in $d=2$ space-time dimensions which we first define in Sec.~\ref{sec:phi4}. The method is tested in Sec.~\ref{phi2res} by studying the spectrum of the solvable $\phi^2$ perturbation with the calculation of  $\Delta H_2$ and $\Delta H_3$.  Other numerical tests are also performed in this section. Next, in Sec.~\ref{phi4res} we give the   $\Delta H_2$ correction  for the $\phi^4$ theory, and discuss the $\Delta H_3$ calculation with some examples.  There we also discuss the convergence of the $\Delta H_n$ expansion and compute the lowest energy levels of the theory at strong coupling. In Sec.~\ref{last}, we conclude and outline future directions of the   method that are left open. 
In Appendix~\ref{diags} we introduce a simple diagrammatic representation to compute $\Delta H_n$.
Lengthy derivations and  results are relegated to the Appendices~\ref{app1} and \ref{deltahforphi4app}.
All the numerical calculations for this work have been done  with  \texttt{Mathematica}.
%%%%%%%%%%%%%%%%%%%%%%%%%%%%%%%%%%%%%%%
%%%% DELTA H definition 
%%%%%%%%%%%%%%%%%%%%%%%%%%%%%%%%%%%%%%%
\section{Calculation of $\Delta H$ at any order}
\label{sec:themethod}
In this section we present one of the main results of this paper which is the derivation of the $n$th-order correction $\Delta H_n$ of \eq{deltaHfock} to the Truncated Hamiltonian.
We start by defining the operator
\be
\Delta \widehat H(\cE)=\sum_{n=2}^\infty\Delta\widehat H_n(\cE) \, ,\quad \text{where}\quad  \Delta \widehat H_n(\cE) = \left(V  \frac{1}{\cE-H_0} \right)^{n-1}  V \, 
\label{deltaHnWe}
\ee
which in the $H_0$ eigenbasis is given by
\bea
\Delta \widehat H_n(\cE)_{rs}&=&\sum_{j_1,\, \dots, \,  j_{n-1}=1}^\infty V_{rj_1} \frac{1}{{ \cE-E_{j_1} }}V_{j_1 j_2} \frac{1}{ \cE -E_{j_2}} V_{j_2 j_3} \cdots V_{j_{n-2}j_{n-1}}\frac{1}{ \cE-E_{j_{n-1}} }V_{j_{n-1}s} \, ,  \quad  \quad  \label{deltaHhatfock} 
\eea
where the indices $j_1, \, j_2,\, \dots \, , j_{n-1}$  run over the states of the full Hilbert space ${\cal H}$.  Notice that the only difference between $\Delta H_n$ and $\Delta \widehat H_n$ is that  the later receives contributions from all the eigenstates of $H_0$ while $\Delta H_n$ only from those with  $E_{j}$ energies  $E_j>E_T$. This translates into the fact that each term in $\Delta H_n(\cE)$ has all the poles located at $\cE>E_T$ as seen in  \eq{deltaHfock}.  

From here the derivation of $\Delta H_n$  follows from the observation that  \eq{deltaHhatfock} can be rewritten as the improper Fourier transform of the product of potentials restricted to positive times
\be
\Delta \widehat H_n(\cE)_{rs}=\lim_{\epsilon\rightarrow 0 }\,  (-i)^{n-1}  \int_0^\infty dt_1\cdots dt_{n-1} \ e^{i(\cE-E_r+i\epsilon)(t_1+\cdots t_{n-1})} \, {\cal T}\left\{  V\left( T_1\right) \cdots V\left( T_n\right)  \right\}_{rs} \, , \label{deltaWe}
\ee
where $T_k= \sum_{i=1}^{i=n-k} t_i$, $V(t)= e^{iH_0 t} V e^{-i H_0 t}$ and  ${\cal T}$ denotes the time ordering operation \footnote{This can be seen by introducing the indentity $\mathds{I}=\sum_n |\ket{E_n}\bra{E_n}|$ between each pair of $V$'s  in  \eq{deltaWe}  and integrating over all times $t_1,\dots t_n$. Also notice that the time ordering operation is trivial because the $V$ operators are time ordered  in all the integration domain. The $\lim_{\eps\rightarrow 0}$ is taken at the end of the calculation.}.
Then, our method consists in applying the Wick theorem to \eq{deltaWe} to calculate $\Delta \widehat H_n$  and obtaining $\Delta H_n$ by keeping only the terms of $\Delta \widehat H_n$ corresponding to states with $E_j>E_T$, i.e. by keeping only the terms of $\Delta \widehat H_n$ which have all poles above $E_T$.~\footnote{This procedure can be formalized as follows. The first correction can be written as $
\Delta H_2(\cE) = \int_{\cal C} \frac{dz}{2\pi i } \frac{\Delta \widehat H_2(z)}{\cE-z}
$, where ${\cal C}$ is any path than encircles  only all the poles above $E_T$. For $\Delta H_3(\cE) = \int_{\cal C} \frac{dz}{2\pi i }   \frac{1}{\cE-z} \int_{\cal C} \frac{dz^\prime}{2\pi i } \frac{1}{\cE-z^\prime}  \Delta \widehat H_3(z^\prime,z)$ where we have generalized the operator $\Delta \widehat H_3(z,z^\prime )_{rs}=-\lim_{\epsilon\rightarrow 0 }\,    \int_0^\infty dt_1dt_2 \ e^{i(z-E_r+i\epsilon)t_1}e^{i(z^\prime -E_r+i\epsilon)t_2} \, {\cal T}\left\{  V\left( T_1\right)V\left( T_2\right) V\left( T_3\right)  \right\}_{rs}$. The generalization to the $n$th correction is straightforward.} 
%%%%%%%%%%%%%
   In the following sections we show how to carry this procedure for the cases of the $\phi^2$ perturbation and $\phi^4$ theory.

%%%%%%%%%%%%%%%%%%%%%%%%%%%%%%%%%%%%%%%
%%%% SCALAR THEORY 
%%%%%%%%%%%%%%%%%%%%%%%%%%%%%%%%%%%%%%%

\section{Scalar  theories}
\label{sec:phi4}
We  study scalar theories in two space-time dimensions  defined by the Minkowskian action $S=S_0+S_{I}$ where
\be
 S_0=\frac{1}{2} \int_{-\infty}^\infty  dt  \int_{0}^{L} dx \, :  (\partial \phi)^2   -m^2 \phi^2 :\,  \, ,
 \label{S0}
\ee
\be
S_{I}=- \int_{-\infty}^\infty  dt \,   V(\phi) =-g_\alpha  \int_{-\infty}^\infty  dt   \int_{0}^L dx \, :\phi^\alpha:\, .
\label{SI}
\ee
For simplicity we consider the cases where  $\alpha=2, \, 4$ and $m^2>0$. The symbol $:\, :$ stands for normal ordering which for $S_0$ means that we set the vacuum energy to zero;   while the interaction term is normal ordered with respect to $S_0$, which in perturbation theory is equivalent to renormalize to zero the UV divergences from closed loops with propagators starting and ending on the same vertex.

To study these theories using the Hamiltonian truncation method we begin by defining them on the cylinder $\mathbb{R}\times S_1$ where the circle corresponds to the space direction which we take to have a length $L m \gg 1$, and $\mathbb{R}$ is the time. We impose periodic boundary conditions $\phi(t,x) = \phi(t,x+nL)$ for $n\in{\mathbb Z}$ on $S_1$.  The compact space direction makes the spectrum of the free theory discrete and regularizes the infra-red (IR) divergences. 

In canonical quantization the scalar operators can be expanded in terms of creation and annihilation operators as
\be
\phi(x) =  \sum_{k}\frac{1}{\sqrt{2L\om_k}} \left(a_ke^{ikx}+a_k^\dagger e^{-ikx} \right) \,  , \label{defphi}
\ee
where $\om_k=\sqrt{m^2+k^2}$, $k=\frac{2\pi n}{L}$ with  $n\in \mathbb{Z}$ and the creation and anihilation operators  satisfy the commutation relations
\be
[a_k,\, a_{k^{\prime}}^\dagger] =\delta_{kk^\prime} \, , \quad \quad [a_k,\, a_{k^{\prime}}] =0 \, . 
\ee
The Hamiltonian then reads $H=H_0+V$, where 
\be
H_0 = \sum_k \om_k \,  a_{k}^\dagger a_{k}
\label{H0as}
\ee
and the potentials for a $\phi^2$ and  a $\phi^4$ interaction are given by 
%in \eq{Vphi2} and \eq{Vphi4}.
\be
V = g_2 \sum_{k_1k_2} \frac{L\, \delta_{k_1+k_2,0}}{ \sqrt{2L\om_{k_1}}\sqrt{2L\om_{k_2}}} \left(a_{k_1}a_{k_2}+a_{-k_1}^\dagger a_{k_2} \right) +h.c. \, ,
\label{Vphi2}
\ee
and
\be
V = g \sum_{k_1,k_2,k_3,k_4} \frac{L\, \delta_{\sum_{i=1}^4k_i,0}}{\prod_{i=1}^4\sqrt{2L\om_{k_i}}} \left(a_{k_1}a_{k_2}a_{k_3}a_{k_4}+ 4a_{-k_1}^\dagger a_{k_2}a_{k_3}a_{k_4}+3a_{-k_1}^\dagger a_{-k_2}^\dagger a_{k_3}a_{k_4}\right)+h.c. \, , \quad
\label{Vphi4}
\ee
respectivley, where $g\equiv  g_4$ and    $ \delta_{k_1+k_2,0}$, $\delta_{\sum_{i=1}^4k_i,0}$  stand for Kronecker deltas. 
\medskip

We implement the Hamiltonian truncation using the basis of $H_0$ eigenstates
\be
|\ket{E_i} =    \frac{ a^{\dagger\, n_N}_{k_N}}{\sqrt{n_N!}}\cdots \frac{a^{\dagger\, n_2}_{k_2}}{\sqrt{n_2!}}  \frac{ a^{\dagger\, n_1}_{k_1}}{\sqrt{n_1!}}|\ket{0} \, .
\label{fockstates}
\ee
 which satisfy $
\mathds{I}= \sum_{i} |\ket{E_i}\bra{E_i}|
$, 
where $E_i=\sum_{s=1}^N n_s\sqrt{k_s^2+m^2}$ and $H_0|\ket{0}=0$ . The Hilbert space is divided into ${\cal H}={\cal H}_l\oplus{\cal H}_h$ with ${\cal H}_l$ spanned by the states $|\ket{E_r}$ such that $E_i\leq E_T$ while ${\cal H}_h$ is spanned by the rest of the basis.  Then, the truncated Hamiltonian is
\be
(H_T)_{rs} =\bra{E_r}| H |\ket{E_s} \, , \quad \text{for} \quad E_i\leq E_T \, . 
\ee
In this basis, the operator $\Delta H$  is given by
\be
\Delta H(\cE)_{rs}=\sum_{j,\,j^\prime} V_{rj} \,  \Big(\frac{1}{\cE-H_0-V} \Big)_{jj'} \, V_{j's} \, 
\ee
where the labels $r,\,s$ denote entries with $E_r,E_s\leq E_T$ and the sum over $j,\,j'$ runs over all  states with $E_j, E_{j'} > E_T$. 

The Hamiltonian $H$  can be diagonalized by sectors with given quantum numbers associated with operators that commute with $H$.  These are the total momentum $P$, the spatial parity ${\cal P}: x\rightarrow -x$ and the field parity $\mathds{Z}_2:\, \phi(x)\rightarrow -\phi(x)$, which  act on the $H_0$-eigenstates as $P|\ket{E_i}=\sum_s n_s k_s |\ket{E_i}$, ${\cal P}    \prod_{i=1}^N\frac{ a^{\dagger\, n_i}_{k_i}}{\sqrt{n_i!}} |\ket{0}= \prod_{i=1}^N\frac{ a^{\dagger\, n_i}_{-k_i}}{\sqrt{n_i!}} |\ket{0}$ and $\mathds{Z}_2 |\ket{E_i} = (-1)^{\sum_s n_s}|\ket{E_i}$. We work in the orthonormal basis of eigenstates of $H_0$, $P$, $\cal P$ and $\mathds{Z}_2$ given by 
  \be
  |\ket{\widetilde E_i}=\beta \cdot \left( |\ket{E_i}+{\cal P} \, |\ket{E_i}\right) \, ,
  \ee
  where $\beta=1/2$, $1/\sqrt{2}$ for ${\cal P} \, |\ket{E_i}=|\ket{E_i}$ and ${\cal P} \, |\ket{E_j}\neq |\ket{E_j}$, respectively. 
As done in Ref.~\cite{Rychkov:2014eea}, in the whole paper we focus on the sub-sector with total momentum $P|\ket{\widetilde E_i}=0$, spatial parity ${\cal P}|\ket{\widetilde E_i}=+|\ket{\widetilde E_i}$ and diagonalize separately the  $\mathds{Z}_2 = \pm $ sectors.~\footnote{ For the $V=\int dt :\phi^2:$ theory, the matrix element $\bra{E_i}|V|\ket{E_j}=0$ with  ${\mathcal P}|\ket{E_i}=|\ket{E_i}$ and   ${\mathcal P}|\ket{E_j}\neq|\ket{E_j}$. Therefore, one can diagonalize the ${\mathcal P}|\ket{E_i}=|\ket{E_i}$ and ${\mathcal P}|\ket{E_i}\neq |\ket{E_i}$ sectors separately.}
In this paper we do not investigate the dependence of the spectrum as a function of  the length $L$ of the compact dimension which we leave for future work, and always consider it to be finite.~\footnote{To match the $L\rightarrow\infty$ spectrum one has to take into account the Casimir energy difference between the $L\rightarrow\infty$ and the finite $L$ theory and inspect how various states converge as $L$ is increased. See  Refs.~\cite{Luscher:1985dn,Luscher:1986pf} and Ref.~\cite{Rychkov:2014eea} for a thorough study of the $L$ dependence. } All the numerical calculations are done for  $m=1$ and $L=10$.

%%%%%%%%%%%%%%%%%%%%%%%%%%%%%%%%%%%%%%%%%%%%%%%%%%%%%%%%%%%%%%%%%%%%%%%%%%%%% 
%%%%%%%%%%%%%%%%%%%%%%%%%%%%%%%%%%%%%%%%%%%%%%%%%%%%%%%%%%%%%%%%%%%%%%%%%%%%% 
            %                PHI2              %
%%%%%%%%%%%%%%%%%%%%%%%%%%%%%%%%%%%%%%%%%%%%%%%%%%%%%%%%%%%%%%%%%%%%%%%%%%%%% 
%%%%%%%%%%%%%%%%%%%%%%%%%%%%%%%%%%%%%%%%%%%%%%%%%%%%%%%%%%%%%%%%%%%%%%%%%%%%% 

\section{Case study $\bma\phi^{2}$ perturbation}
\label{phi2res}

In this section we apply the method introduced in Sec.~\ref{sec:themethod} to the scalar theory $H=H_0+V$ with a potential 
\be
V=g_2\int_0^L dt \,  :\phi^2:
\ee
  This is a  simple theory that allows to illustrate various aspects of the calculation of $\Delta \widehat H$ in  \eq{deltaWe} and its relation to $\Delta H$. Also since the theory is solvable we can compare our procedure with the exact results.   The theory is solved by using the eigenstates of $H$, 
 %$H|\ket{\cE_i}=(H_0+V)|\ket{\cE_i}=\cE_i |\ket{\cE_i}$, 
given by 
\be
|\ket{\cE_i} =    \frac{ b^{\dagger\, n_N}_{k_N}}{\sqrt{n_N!}}\cdots \frac{b^{\dagger\, n_2}_{k_2}}{\sqrt{n_2!}}  \frac{ b^{\dagger\, n_1}_{k_1}}{\sqrt{n_1!}}|\ket{\Omega} \, , 
\ee
where  $|\ket{\Om}=|\ket{\cE_0}$ is the vacuum of the theory and $b^\dagger$/$b$ are the creation/annihilation operators so that 
\be
H=\sum_k b^\dagger_k b_k \Om_k +\cE_0 \, ,
\ee
with $\Om_k=\sqrt{\om^2_k+2g_2}$.  Then, one can relate the operators $b^\dagger/ b$ to the $a^\dagger/a$ in $H_0$ (given in \eq{H0as} and \eq{Vphi2}) by the Bogolyubov transformation  
$
b_k=\sinh\alpha_k\, a_{-k}^\dagger+\cosh\alpha_k\, a_{k}
$ 
provided  that 
$ \Om_k\sinh 2 \alpha_k = \om_k^{-1} \, g_2$, $ \Om_k \cosh 2 \alpha_k =\om_k+ g_2/\om_k$.
Then, since  $\bra{0}|H|\ket{0}=0$ we have that~\cite{Rychkov:2014eea}:
\be
 \cE_0(g_2) =\frac{1}{2}\sum_k  \big( \sqrt{\om_k^2 +2g_2} -\om_k -\frac{g_2}{\om_k}\big) =  \frac{L \left(m^2+2g_2\right)}{8\pi}\Big[\log\Big(\frac{m^2}{m^2+2g_2}\Big)+\frac{2g_2}{m^2+2g_2}\Big] \, , \label{veff}
\ee
 where the sum can be done by means of the Abel-Plana formula, which is the exact vacuum energy of the theory.
 \medskip

A brief summary of the rest of this section is the following. In Sec.~\ref{anal} and Sec.~\ref{anal2} we calculate the $2$ and $3$-point corrections to the operator $\Delta H$.  In Sec.~\ref{numtest} we perform a numerical test to check that our expressions for $\Delta H$ are correct. Then, in Sec.~\ref{sc} we discuss the numerical results and the convergence of the expansion $\Delta H(\cE_i)=\sum_n \Delta H_n(\cE_i)$ by comparing with the exact spectrum $\cE_i$.

\subsection{Two-point correction}
\label{anal}

Following the steps explained in   Sec. \ref{sec:themethod} we begin the calculation of the two-point correction by first computing $\Delta \widehat H_2$. From \eq{deltaWe} we have that 
 \be
 \Delta \widehat H_2(\cE)_{rs} = \sum_{j}V_{rj}\frac{1}{\cE-E_j}V_{js}=  \lim_{\eps\rightarrow 0} -i \int_0^\infty dt \, e^{i (\cE-E_r+i\eps)t}  {\cal T}\left\{ V(t) V(0)\right\}_{rs}\, . \label{2pt}
 \ee
 Then, applying  the Wick theorem to \eq{2pt} we find
\be
 \lim_{\eps\rightarrow 0} \,  -ig_2^2 \int_0^\infty dt \,  e^{i (\cE-E_r+i\eps)t} \int_{-L/2}^{L/2} dxdz \,  \sum_{m=0}^{2} s_{2-m} D_F^{2-m}(z,t)     : \phi^{m}(x+z,t)\phi^{m}(x,0): _{rs}  \, ,  \label{wick1}
\ee
where  $s_p=\binom{2}{p}^2p!$  are the symmetry factors and $D_F(z,t)$ is the Feynman propagator with discretized momenta. Henceforth we label the terms $m=0,1,2$ by $\Delta \widehat{H}^{\phi^{2m}}_2$ so that $\Delta \widehat{H}_2 = \Delta \widehat{H}^{\mathds{1}}_2+\Delta \widehat{H}^{\phi^{2}}_2+\Delta \widehat{H}^{\phi^{4}}_2$  and similarly for $\Delta H_2$; the labels only inform about the  total number of fields in each term which do not need to be local. 
 Due to the time integration domain,  it is convenient to use \emph{half} Feynman propagator
 \be
D_L(z,t) \equiv D_F(z,t)\theta(t) =  \frac{1}{2L}  \sum_{n=-\infty}^{n={\infty}}\frac{1}{\om_k}  e^{-i\om_k t} e^{i\frac{2\pi n z}{L} }\theta(t)  \, ,  \label{prop1}
 \ee
 the  momentum of the propagator is discretised due to the finite extent of the space. 
Next, we proceed to calculate the  operators in  \eq{wick1}, 
 starting with the detailed calculation of the coefficient of  the identity operator $\Delta\widehat H_2^{\mathds{1}}$:
\be
 \Delta \widehat H_2^{\mathds{1}}(\cE)_{rs} = \lim_{\eps\rightarrow 0} \,  - is_2 g_2^2 \int_0^\infty dt  \int_{-L/2}^{L/2} dz  \ e^{i({\cal E}-E_r+i\epsilon)t_1} D_L^2(t,z)\, \mathds{1}_{rs} \, ,
 \ee
where $\mathds{1}_{rs}\equiv\delta_{rs}\int_{-L/2}^{L/2}dz$ has dimensions of $[E]^{-1}$. Then, upon inserting the propagator of \eq{prop1} and performing the space-time integrals we find
 \be
%%%%%%%%%%%%%%%%%%%%%%%%%%%%%%%%%%
 \Delta \widehat H_2^{\mathds{1}}(\cE)_{rs}  =   \frac{s_2g_2^2}{4L} \sum_{k}\frac{1}{\om_k^2}\frac{1}{ \cE-E_r-2\om_k} \label{2ptId}\, \mathds{1}_{rs}\, .
\ee
The  operator in \eq{2ptId}  has poles from all  possible intermediate states and, as  explained in Sec. \ref{sec:themethod},  the operator $\Delta H_2^{\mathds{1}}(\cE)$  is found by keeping only those terms  with poles  located at  $E_r+2\om_k>E_T$, therefore
 \bea
\Delta H_2^{\mathds{1}}(\cE)_{rs}&=& \frac{s_2g_2^2}{L} \sum_{k:\, E_r+2\om_k>E_T}\frac{1}{4\, \om_k^2}\frac{1}{ \cE-E_r-2\om_k}\, \mathds{1}_{rs} \ \, . \label{ID3}
\eea
The calculations of $\Delta H_2^{\phi^2}$ is similar to the one for \eq{ID3}, we start by computing
\be
 \Delta \widehat H_2^{\phi^2}(\cE)_{rs} = \lim_{\eps\rightarrow 0} \,  - is_1 g_2^2 \int_0^\infty dt  \int_{-L/2}^{L/2} dx dz  \ e^{i({\cal E}-E_r+i\epsilon)t_1} D_L(z,t)\, : \phi (x+z,t )  \phi (x,0) :_{rs} \, ,
 \ee
where  we expand $:\phi (x+z,t )  \phi (x,0):$ in modes, as in \eq{defphi}, and do the simple space-time integrals.
For the full expressions of $\Delta \widehat H_2^{\phi^2}$ see  Appendix~\ref{app1}. Then, keeping only the terms with poles at $\cE>E_T$ we get 
\be
\Delta H_2^{\phi^2}(\cE)_{rs} = s_1 g_2^2 \sum_{q:\, 2\om_q+E_r>E_T} \frac{1}{4\, \om_q^2}\frac{1}{\cE-E_r-2\om_q}  (a_{q}^\dagger a_{q})_{rs}  \, .  \label{nonl1} 
\ee
The operator $\Delta H_2^{\phi^4}$ is obtained in a similar way,
\be
\Delta  H_2^{\phi^4}(\cE)_{rs} =s_0g_2^2\sum_{q_1,q_2:\,    2 \om_{q_2}+E_r>E_T}\frac{  1}{4\, \om_{q_2}\om_{q_1}}\frac{1}{\cE-E_r-2\om_{q_2}}    \left( a_{q_1}^\dagger a_{-{q_1}}^\dagger a_{q_2}a_{-{q_2}}\right)_{rs}    \, . \label{nonl2}
\ee
In  Appendix~\ref{diags} we give a simple way to derive these expressions from diagrams, and for the full expressions of $\Delta \widehat H_2^{\phi^2}$ and $\Delta \widehat H_2^{\phi^4}$  see  Appendix~\ref{app1}.  
Notice that the values of $q_{1}$, $q_2$ and $q$ appearing in the sums of \eq{nonl1} and \eq{nonl2} can take only the momenta of the states $|\ket{E_s}\in{\cal H}_l$ on which $a$ and $a^{\dagger}$ act, and therefore are bounded. On the other hand, the values of the $k$'s in \eq{ID3} go all the way to infinity. Also, even though the operators in \eq{nonl1} and \eq{nonl2} may seem not hermitian due to the $E_r$ appearing in the expressions, one can see that the operator $(\Delta H_2^{\phi^2})_{rs}$ is diagonal and therefore $E_r=E_s$, while $\Delta H_2^{\phi^4}$  is not diagonal, but one can check that $E_r+2\om_{q_2}=E_s+2\om_{q_1}$, making it hermitian as well.
\medskip

We end this section by noticing that the  operator of  \eq{ID3} can be rewritten as
\be
(\Delta H_2^{\mathds{1}})_{rs}=\int_{E_T}^\infty   \frac{dE}{\cE-E}\frac{s_2g_2^2 }{L} \sum_{k=-\infty}^\infty \frac{ \delta(E-E_r-2\om_k)}{(2\om_k)^2}\,  \mathds{1}_{rs} = s_2g_2^2  \int_{E_T}^\infty \frac{dE}{2\pi} \frac{ \Phi_2( E-E_r )}{\cE-E} \, \mathds{1}_{rs}\, , \label{2ptID}
\ee
where $\Phi_2$ is  the two-particle phase space with discretized momenta,
\be
\Phi_2(E-E_r)= \sum_{k_1,k_2}  \frac{L\, \delta_{ k_1+k_2,\, 0}}{(2L\om_{k_1})\, (2L\om_{k_2})} \,  2\pi\, \delta(E-E_r- \om_{k_1}- \om_{k_2})\, ,
\ee
where  from \eq{2ptID} one has that $E-E_r> 2m$.~\footnote{The lower limit in \eq{2ptID}  should be taken slightly above $E_T$ to reproduce  the lower limit $q:\, 2\om_q +E_r> E_T$ in the sum of \eq{2ptId}. }
\eq{2ptID} can be evaluated by means of the Abel-Plana formula, which for $LE_T \gg 1$ is well approximated by its continuum limit~\footnote{The difference between the continuum limit and discrete result ranges from ${\cal O}( g^2 L^{-1}E_T^{-3})$ to ${\cal O}( g^2 L^{-1}E_T^{-1}m^{-2})$ depending on the matrix entry. }. The continuum two-body phase space is given by
\be
\Phi_{2}(E) =\int_{-\infty}^\infty \frac{d^2p_1}{(2\pi)^2\, 2\om_{p_1}}\frac{d^2p_2}{(2\pi)^2\, 2\om_{p_2}} \, (2\pi)^2 \, \delta^{(2)}(P^\mu-p_1-p_2) =\frac{1}{E\sqrt{E^2-4m^2}} \, ,\label{phase2}
\ee
where $P^\mu=(E,0)$ and $E>2m$. Therefore (for $LE_T \gg 1$) we find
\be
\Delta H_2^{\mathds{1}}(\cE)_{rs} 
\simeq s_2 g_2^2 \int_{E_T}^\infty\frac{dE}{2\pi}\frac{1}{\cE-E} \frac{1}{E-E_r}\frac{1}{\sqrt{(E-E_r)^2-4m^2}}\, \theta(E-E_r-2m)  \,   \mathds{1}_{rs} \, .  \label{we}
\ee
This result is useful for numerical implementation since  \eq{we} can be integrated in terms of logarithmic functions. Finally, we notice that upon expanding the function $s_2 /(2\pi)\Phi_2(E)$ around $m/E=0$ we find agreement  with Ref.~\cite{Rychkov:2014eea} that computed it by  other means (there called $\mu_{220}(E)=1/(\pi E^2)$).

\subsection{Three-point correction}
\label{anal2}

The calculation of the three-point correction $\Delta H_3$ also starts  from the expression in   \eq{deltaWe}
 \be
 \Delta \widehat H_3(\cE)_{rs} = -\lim_{\eps\rightarrow 0} \int_0^\infty dt_1dt_2 \, e^{i (\cE-E_r+i\eps)(t_1+t_2)}  {\cal T}\left\{ V(T_1)V(T_2) V(T_3)\right\}_{rs}\, , \label{3pt} 
 \ee
 where $T_k=\sum_{n=1}^{3-k}t_n$.
 Next we apply the Wick theorem and find that  the  time ordered product   $ {\cal T}\left\{ V(T_1)V(T_2) V(T_3)\right\}$  is given by
 \be
g_2^3 \int_{-L/2}^{L/2}dx_1dx_2dz \sum_{m,n,v=0}^{2} s^{mnv}_2 D_F^{m}(x_1,t_1)D_F^{n}(x_2,t_2)D_F^{v}(x_1+x_2,t_1+t_2)    : \phi^{2-n-m}_{X_1,T_1}\phi^{2-n-v}_{X_2,T_2}\phi^{2-v-m}_{X_3,T_3}: \label{nonl3}
 \ee
 where we have introduced the notation $X_k=z+\sum_{n=1}^{3-k}x_n$ and  $\phi_{x,t}=\phi(x,t)$; while  the symmetry factor is given by
\be
s^{mnv}_p=\frac{p!^3}{(p-m-n)!(p-m-v)!(p-n-v)! m!n!v!} \label{symfact}\, .
\ee
We use  the same notation as in the previous section $\Delta \widehat H_3= \Delta \widehat H^{\mathds{1}}_3+\Delta \widehat H^{\phi^2}_3+\Delta \widehat H^{\phi^4}_3+\Delta \widehat H^{\phi^6}_3$, and similarly for $\Delta H_3$. 
Then, upon performing the space-time integrals in  \eq{nonl3} and only keeping the terms with all the  poles above $E_T$ we find $\Delta H_3$. Then, for the term $\Delta  H_3^{\mathds{1}}$ we get
\be
\Delta  H_3^{\mathds{1}}(\cE)_{rs} =  s_2^{111} g_2^3\frac{1}{L}\sum_{k: E_{rs}+2\om_k>E_T}\frac{1}{(2\om_k)^3}  \frac{1}{\left(\cE-E_{r}-2\om_k\right)^2}  \,  \mathds{1}_{rs} \, . \label{3ptID}
\ee
The expressions for $\Delta H_3^{\phi^2}$, $\Delta H_3^{\phi^4}$ and $\Delta H_3^{\phi^6}$ are lengthy but straightforward to obtain and are relegated to Appendix~\ref{app1}. 
\medskip

As done in the previous section, \eq{3ptID} can be written as 
\be
\Delta  H_3^{\mathds{1}}(\cE)_{rs} =s_2^{111} g_2^3\int_{E_T}^\infty\frac{dE}{(\cE-E)^2}\frac{1}{L}\sum_{k}\frac{1}{(2 \om_k)^3} \delta(E-E_r-2\om_k)  \, \mathds{1}_{rs}  \, ,  \label{3ptID2}
\ee
which for $L^{-1}E_T\gg 1$ is well approximated by its continuum limit
\be
\Delta  H_3^{\mathds{1}}(\cE)_{rs} \simeq s_2^{111} \, \frac{ g_2^3}{2 \pi} \int_{E_T}^\infty\frac{dE}{(\cE-E)^2} \frac{1}{(E-E_r)^2}\frac{1}{\sqrt{(E-E_r)^2-4m^2}}   \, \mathds{1}_{rs} \, ,  \label{3ptID4}
\ee
and can be integrated in terms of logarithmic functions. This is useful for a fast numerical implementation.

\subsection{A numerical test}
\label{numtest}
\begin{figure}[t]
\begin{center}
  \includegraphics[scale=0.35]{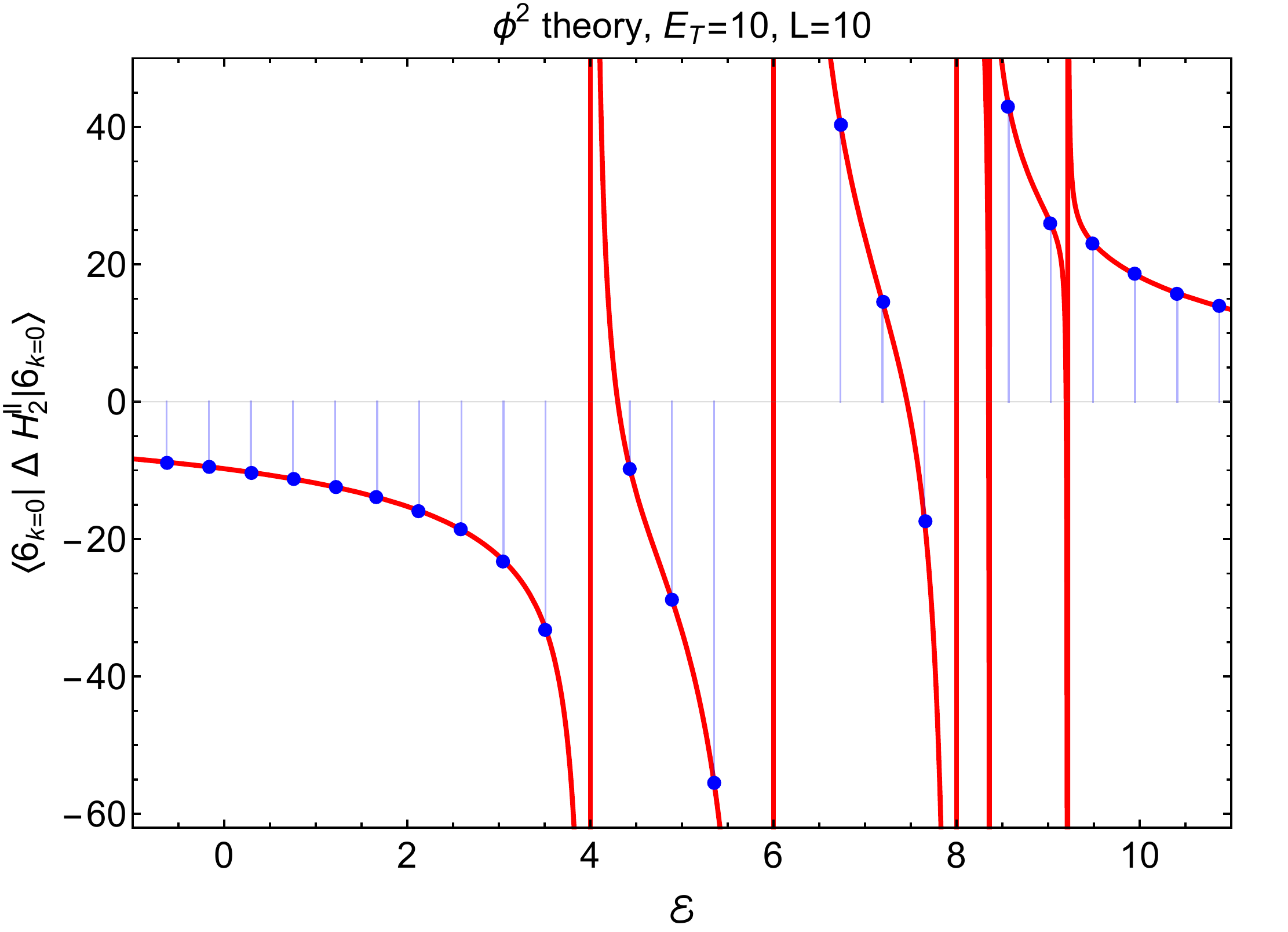} \hspace{0.2cm}
  \includegraphics[scale=0.35]{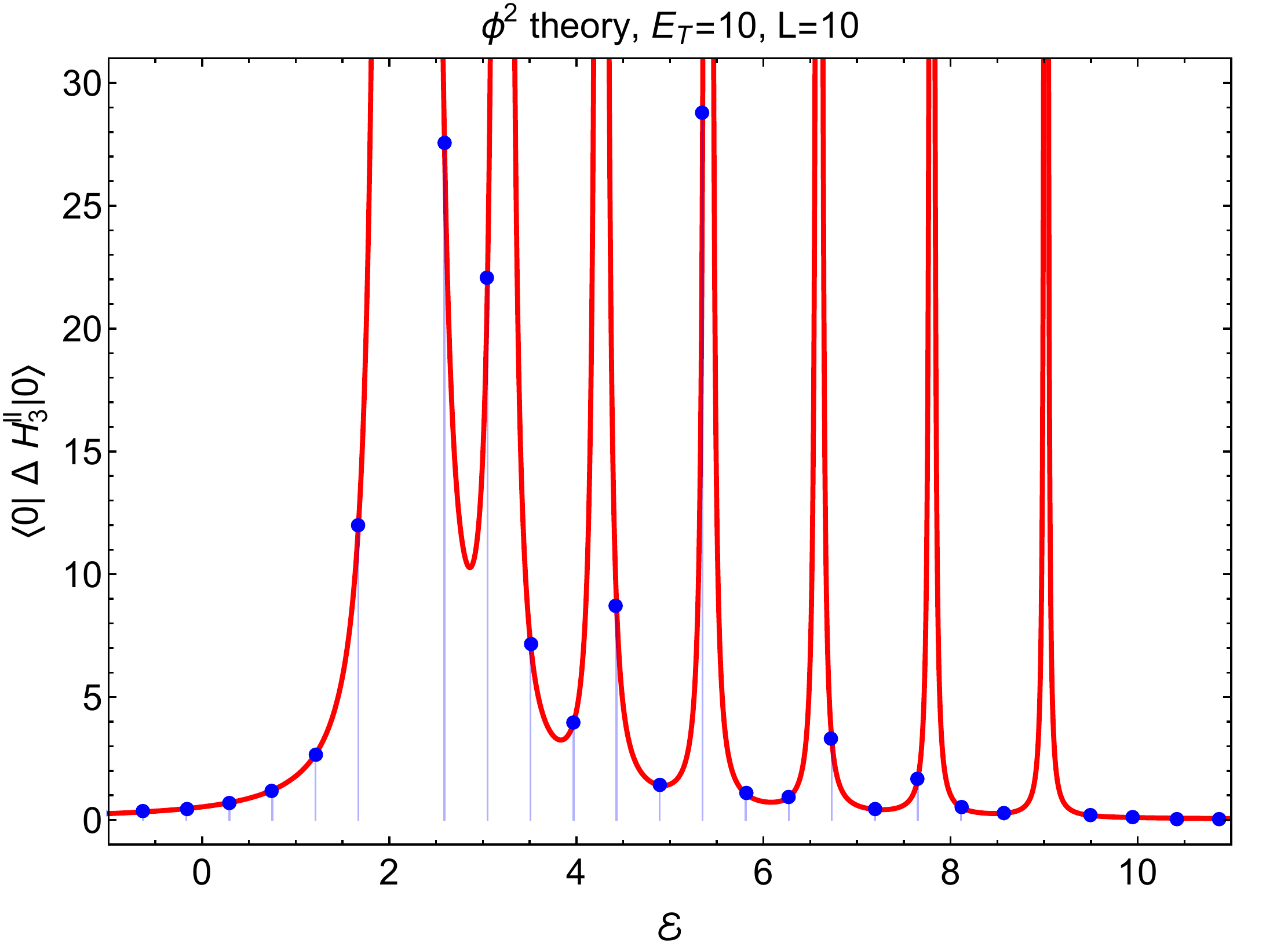}
\end{center}
\vspace{-6mm}
\caption{Comparison of both sides of Eqs.~(\ref{ch1})  and (\ref{ch22}).}
\label{fig:checkpoles}
\end{figure}
We perform a numerical check to test our prescription to select the poles of $\Delta \widehat H_n(\cE)$ to get $\Delta H_n$, i.e.  that we can select the desired intermediate states of $H_0$ by looking at the poles of the terms of $\Delta \widehat{H}_n$. The check consists in computing $\Delta \widehat{H}_2$ as explained, and then selecting only the terms with all poles at $\cE \le E_T$. We refer to the expression as $\Delta H^{ll}_n$ to differentiate it with $\Delta H_n$ that only receives corrections from terms with poles at $\cE >E_T$. $\Delta H^{ll}_2$  is then compared with the matrix elements of $VP_l ( \cE-H_0)^{-1}P_l V$, finding an exact agreement.  The same is done for $\Delta \widehat H_3(\cE)$ by comparing it against  $VP_l ( \cE-H_0)^{-1}V ( \cE-H_0)^{-1} P_l V$.
This check has been done for all the matrices used in the present work, both for $\phi^2$ and $\phi^4$. For brevity we only show the check for two matrix entries of the $\phi^2$ theory. These are
\bea
\bra{6_{k=0}}| \, VP_l\frac{1}{\cE-H_0}P_l V |\ket{6_{k=0}} &=&  \sum_{k:\, 2\om_k+6m<E_T}\frac{g_2^2}{2\, \om_k^2}\frac{1}{ \cE-6m-2\om_k}\nonumber\\[.2cm]
&+& \frac{3\, g_2^2 }{2m^2} \left( \frac{5}{\cE-4m} + \frac{24}{\cE-6m}+\frac{9}{\cE-8m}\right),\label{ch1}\\[.2cm]
\bra{0}|\,  V P_l \frac{1}{{\cal E}-H_0}V\frac{1}{{\cal E}-H_0}P_lV\, |\ket{0}&=& g_2^3 \sum_{k:\, 2\om_k<E_T}\frac{1}{\om_k^3}\frac{1}{(\cE-2\om_k)^2} \, .\label{ch22}
\eea
In Fig.~\ref{fig:checkpoles} we compare both sides of equations Eq.~(\ref{ch1}) and (\ref{ch22}). The  red  curves correspond to the right hand side of Eqs.~(\ref{ch1})-(\ref{ch22}), which are our analytical results, and the blue dots are given by the product of the matrices in the left hand side of the equations.
In the left plot, done for $\bra{6_{k=0}}|\Delta H^{ll}_2|\ket{6_{k=0}}$,  the first pole arises at  the four-particle threshold and subsequent poles appear for higher excited states. Instead, the first pole in the right plot, done for $\bra{0}|\Delta H^{ll}_3 |\ket{0}$, occurs at $\cE=2m$. Notice that in both figures there are no poles for $\cE>E_T$.
% The plots of Fig.~\ref{fig:checkpoles}  also provide a check of the codes generating the Fock space basis and the matrices created for the potential. 
%%%%%%%%%%%%%%%%%%%%%%%%%%%%%%%%%%%%%%%%%%%%%%%%%%%%%

  \begin{figure}[t]
\centering
  \includegraphics[width=0.48\textwidth]{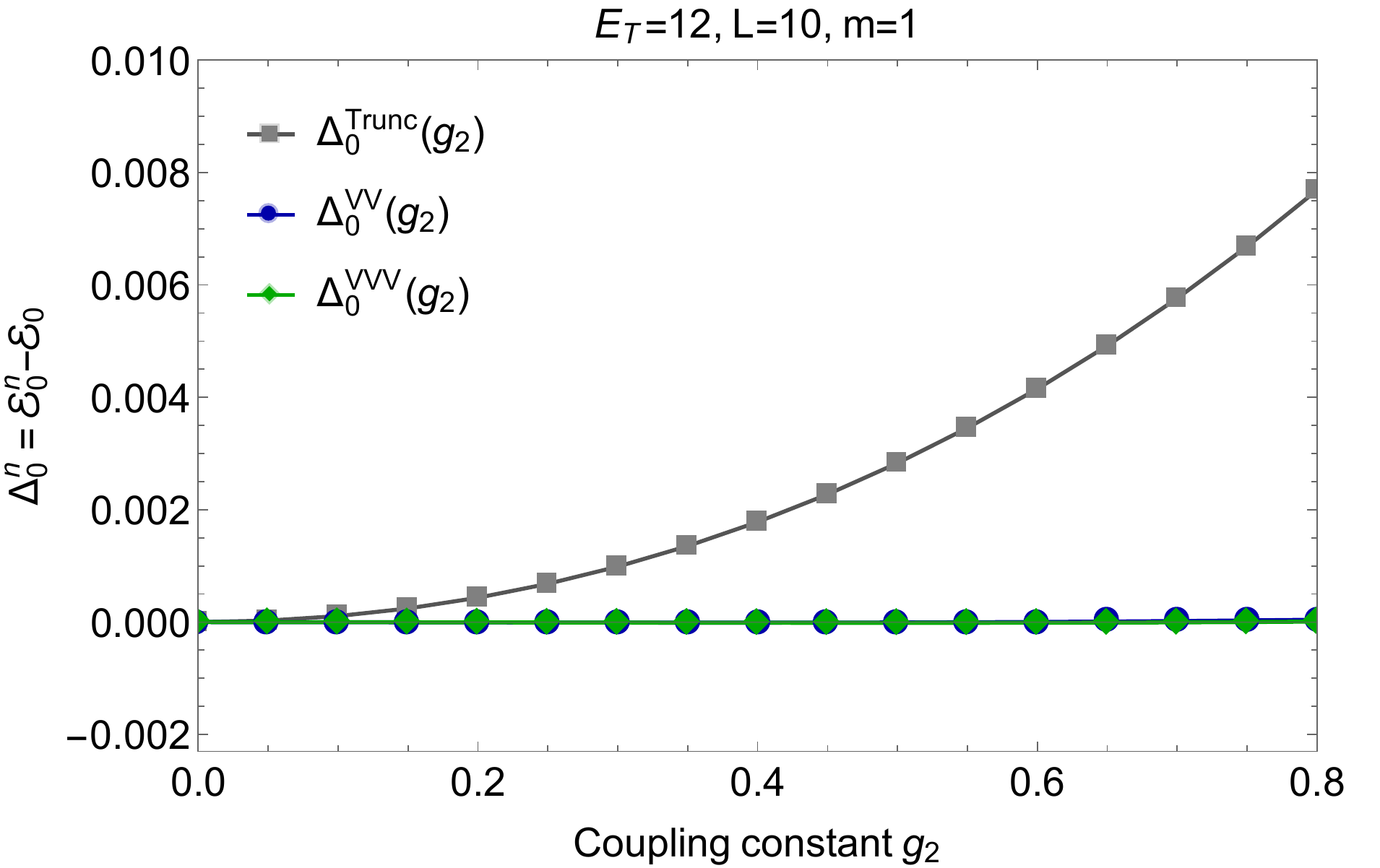} \quad  \includegraphics[width=0.4848\textwidth]{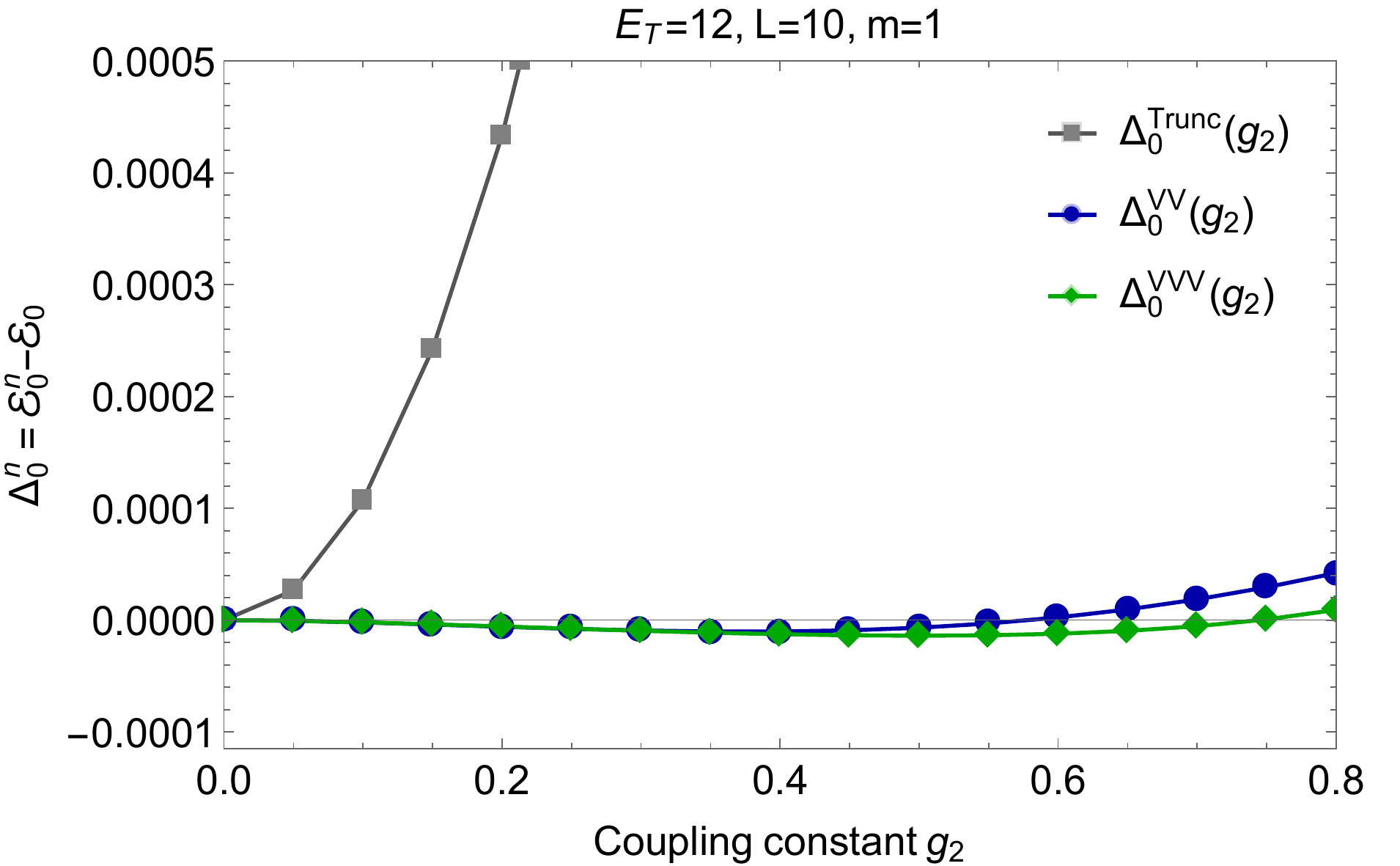} 
\caption{\em  {\bf \em Left:} comparison  of the exact vacuum energy with the numerical result as a function of the coupling constant $g_2$ (for $V=g_2\int dx \phi^2$).   {\bf \em Right:} left plot  with  the $y$-axis zoomed in a factor $\times 20$. \label{vacuumPHI2}}
\end{figure}

\subsection{Spectrum and convergence}
\label{sc}

We perform a numerical study of the  convergence of the energy levels as a function of the truncation energy $E_T$ and their convergence as higher order corrections $\Delta H_n$ are calculated for a fixed $E_T$.
We use the formulas in Eqs.~(\ref{ID3})-(\ref{nonl2}), (\ref{3ptID}) and (\ref{app31})-(\ref{app33})  to numerically compute $\Delta H_2$  and $\Delta H_3$.~\footnote{
The sums over $k$ in Eqs.~(\ref{ID3})-(\ref{nonl2}), (\ref{3ptID}) and (\ref{app31})-(\ref{app33})  have been done with a cutoff  $k= 250$. We have checked  that increasing the cutoff has little impact on the results and find agreement with analytic formulas like \eq{2ptID}.}

We begin by comparing the vacuum eigenstate $\cE_0^i$ obtained by numerically diagonalizing $H_T+ \sum_{n=2}^N\Delta H_n$  (for $N=2$ and 3) with  the exact vacuum energy $\cE_0$. In Fig.~\ref{vacuumPHI2} we show a plot  of  
$\Delta _0^i = \cE_0^i-\cE_0$ as a function of the coupling constant $g_2$.  The plot is done for a truncation energy of $E_T=12$ and $L=10$  (recall that we work in $m=1$ units). For an easier comparison with previous work, these plots have been done with the same choice of parameters and normalizations as in Fig.~2 of Ref.~\cite{Rychkov:2014eea}.
The gray curve in Fig.~\ref{vacuumPHI2} is obtained  by numerically diagonalizing $H_T$, whose lowest eigenvalue is $\cE_0^\text{T}$. The blue curve  is obtained by  diagonalizing the renormalized hamiltonian $H_T+ \Delta H_2(\cE_0^\text{T})$, whose lowest eigenvalue is $\cE_0^{VV}$. Lastly,  the green curve is obtained by diagonalizing  $H_T+ \Delta H_2(\cE_0^{VV})+ \Delta H_3(\cE_0^{VV})$ (we find little difference in evaluating the latter operator in $\cE_0^\text{Trunc}$ instead of $\cE_0^{VV}$).
 The right plot of Fig.~\ref{vacuumPHI2} is a zoomed in version of the left plot in order to resolve the difference between the $\Delta_0^{VV}$ and $\Delta_0^{VVV}$ curves. 
 
The right plot shows that  overall $\Delta_0^{VVV}$ performs better than  $\Delta_0^{VV}$,  this indicates that the  truncation of the  series expansion $\Delta H= \sum_{n=2}^\infty \Delta H_n$ at $n=3$ is perturbative in the studied range. The effect is more pronounced for the highest couplings $g_2\simeq[0.6,0.8]$. As a benchmark value $\cE_0(g_2=0.8)=-0.351864$, see \eq{veff}. Therefore the  relative error at $g_2=0.8$ is $2\%$, $0.01\%$ and $0.002\%$ for the Truncated, the $VV$ and the $VVV$ corrections, respectively.
\begin{figure}[h!!]
\centering
  \includegraphics[width=0.47\textwidth]{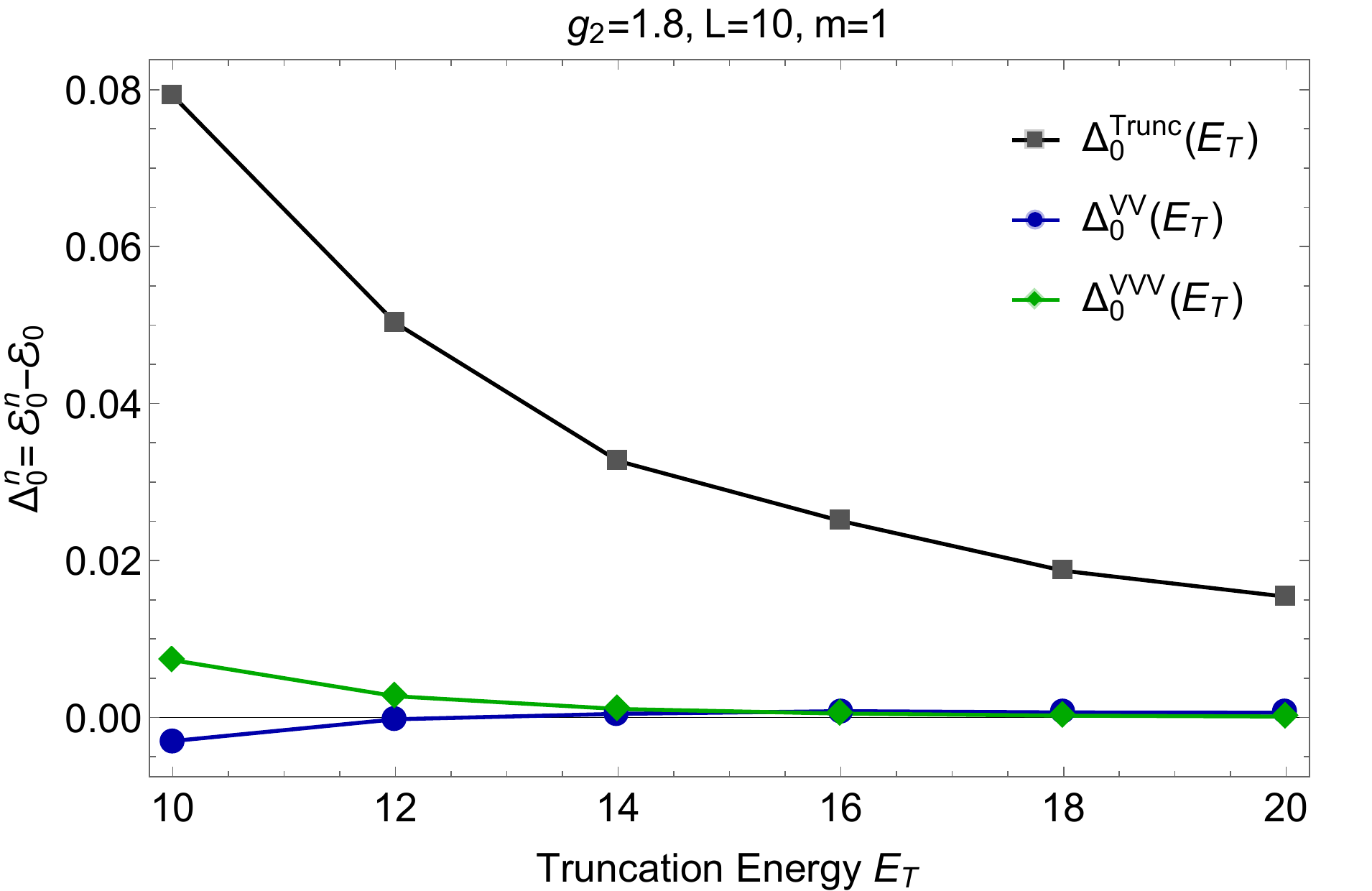} \quad  \includegraphics[width=0.4868\textwidth]{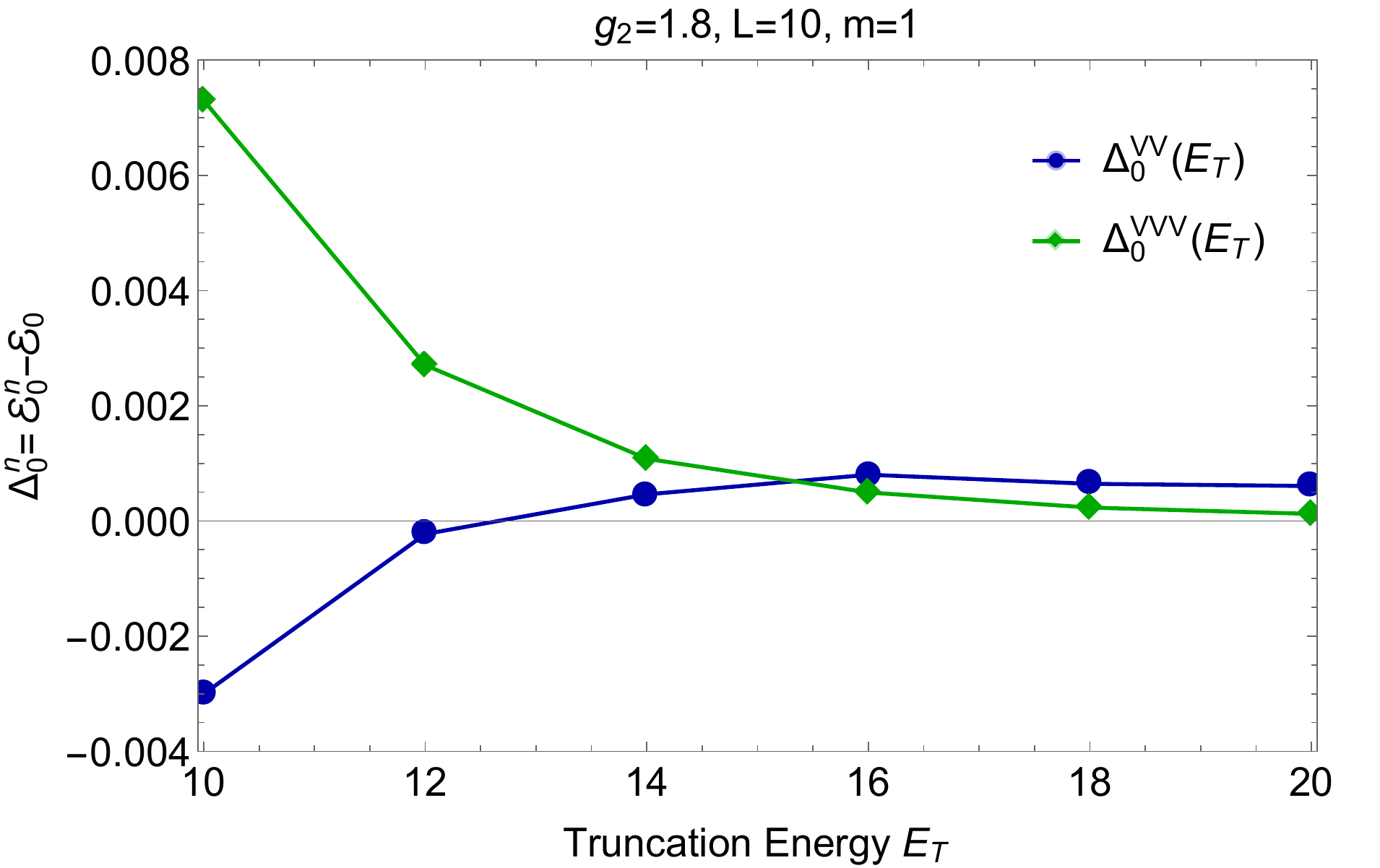} 
\caption{\em  {\bf \em Left:} comparison  of the exact vacuum energy with the numerical result as a function of the truncation energy $E_T$.   {\bf \em Right:} left plot zoomed in. }  \label{exactConvergence}
\end{figure}   

Next, we check the convergence of the energy levels as a function of the truncation energy $E_T$. 
In Fig.~\ref{exactConvergence}, in the left plot  
we show   $\Delta _0^i = \cE_0^i-\cE_0$ as a function of the truncation energy $E_T$, for $i=$Trunc, $VV$ and $VVV$.
Both the $\Delta _0^{VV}$ and $\Delta _0^{VVV}$ curves give better results than  $\Delta _0^\text{Trunc}$ for the whole range. 
Also,  the curves  $\Delta _0^{VV}$ and $\Delta _0^{VVV}$  have a better convergence behavior and, when converged, they are closer to  zero than $\Delta _0^\text{Trunc}$.
The right plot is a zoomed in version to resolve the difference between $\Delta _0^{VV}$ and $\Delta _0^{VVV}$.
The plot shows that for $E_T\lesssim 15$ the curve $\Delta _0^{VV}$ gives better results than $\Delta _0^{VVV}$ while for  larger $E_T$  the behavior is reversed. This indicates that for $E_T\lesssim 15$ (and $g_2=1.8$) the truncation of the series  $\Delta H= \sum_{n=2}^\infty \Delta H_n$  is not a good approximation, and adding more terms will not improve the accuracy. However, as $E_T$ is increased it pays off to introduce higher order corrections to get a better result. This is because $\Delta _0^{VVV}$ has a faster converge rate than $\Delta _0^{VV}$ to the real eigenvalue. 
The value is $\cE_0(g_2=1.8)=-1.360719$, see \eq{veff}. Therefore the  relative error at $E_T=20$ is $1\%$, $0.04\%$ and $0.009\%$ for the Truncated, the $VV$ and the $VVV$ corrections, respectively.

   \begin{figure}[h]
\centering
  \includegraphics[width=0.489\textwidth]{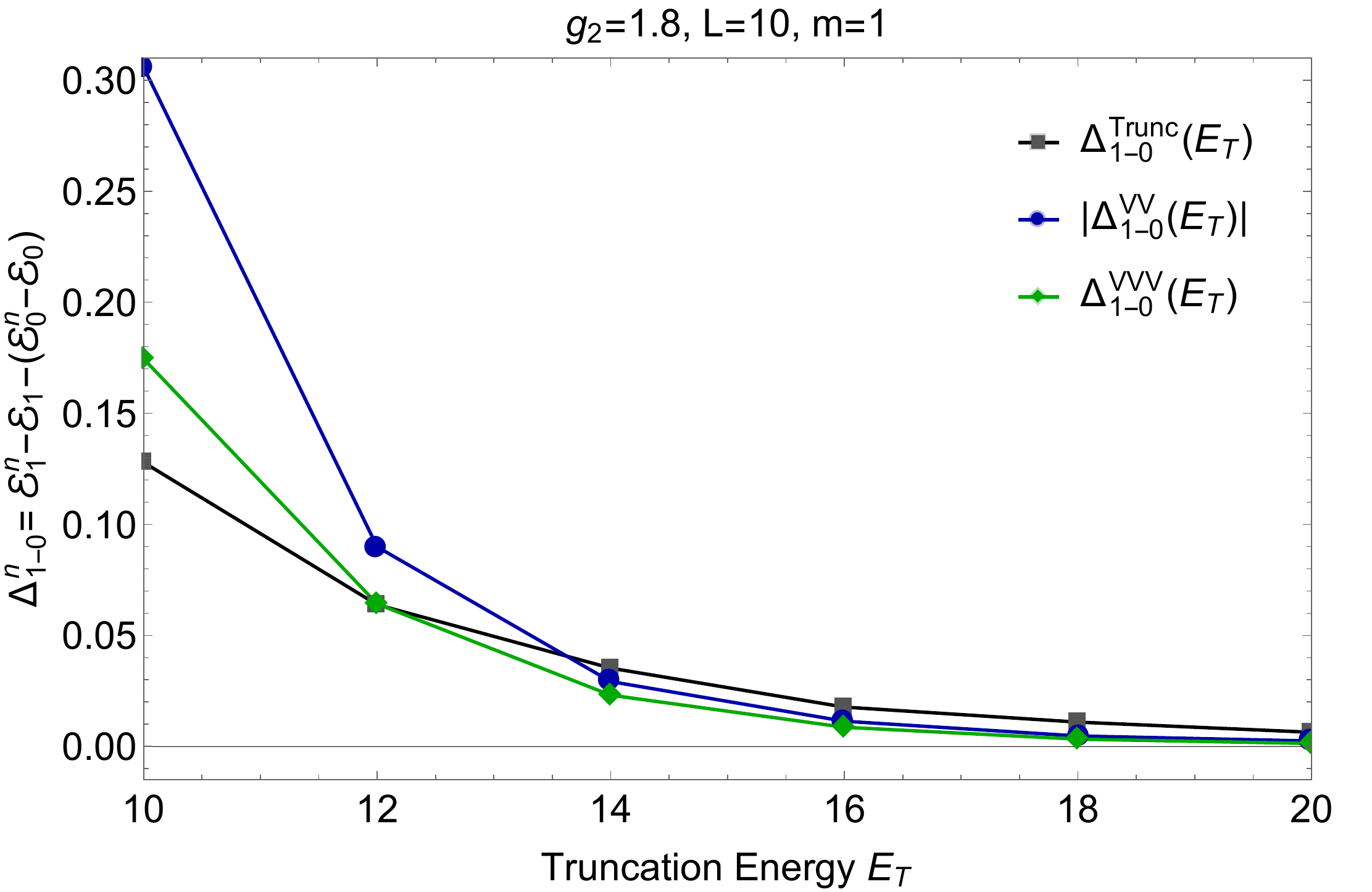} \quad  \includegraphics[width=0.476\textwidth]{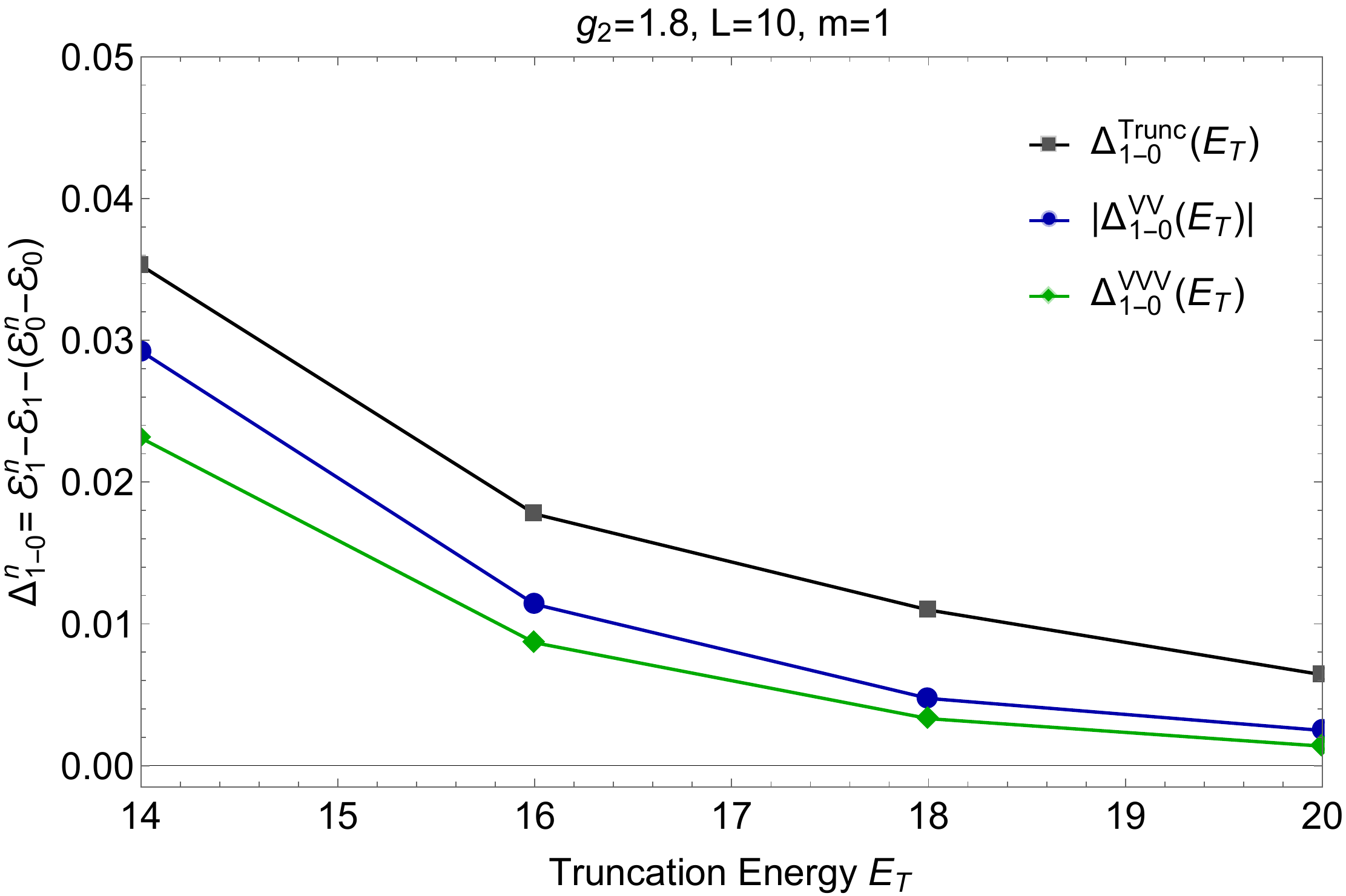} 
\caption{\em  {\bf \em Left:} comparison  of the exact energy difference $\cE_1-\cE_0$  with respect the numerical result as a function of the truncation energy $E_T$.  {\bf \em Right:} left plot zoomed in. On both plots we have taken the absolute value of the curve corresponding to the $VV$ corrections, in blue.}  \label{exactConvergence2}   
\end{figure}  

In Fig.~\ref{exactConvergence2} we repeat the plots of Fig.~\ref{exactConvergence}   for the first $\mathds{Z}_2$-even excited state but taking the absolute value of the $\Delta _1^{VV}$ curve for clarity.  
The plots show a similar convergence rate for  the three $\Delta _1^{i}$ curves. However, there is a similar pattern compared to Fig.~\ref{exactConvergence}: for $E_T\lesssim 15$ introducing higher order corrections of the series $\Delta H= \sum_{n=2}^\infty \Delta H_n$ gives worse results, while for larger values of $E_T$ adding  higher $\Delta H_n$ corrections improves them.
The value is $\cE_1(g_2=1.8)=0.784042$, hence the  relative error at $E_T=20$ is $0.8\%$, $0.3\%$ and $0.17\%$ for the Truncated, the $VV$ and the $VVV$ corrections, respectively.

\section{The  $ \bma\phi^4$ theory}
\label{phi4res}

Next we apply the method presented in previous sections to the $\phi^4$ theory. We  start by deriving the exact expressions for $\Delta H_2$ in detail,  then we perform various useful approximations for a faster numerical implementation and discuss general aspects of the method. We also discuss the pertubativity of the $\Delta H_n$ expansion and compute the spectrum of the theory at different couplings while studying its behaviour in $E_T$ and $g$ using the results of $\Delta H_2$. We end the section with some comments on future work and a discussion of the calculation of $\Delta H_3$.
 %Finally we discuss the numerical  implementation of the method  to extract the spectra at strong coupling.

\subsection{Two-point correction}
\label{phi42point}
Again, we follow Sec.~\ref{sec:themethod} to derive $\Delta H$ by first computing $\Delta \widehat{H}$. From \eq{deltaWe} we have
 \be
 \Delta \widehat H_2(\cE)_{rs} = \sum_{j}V_{ rj}\frac{1}{\cE-E_j}V_{js}=  \lim_{\eps\rightarrow 0} -i \int_0^\infty dt \, e^{i (\cE-E_{r}+i\eps)t}  {\cal T}\left\{ V(t) V(0)\right\}_{rs}\, . \label{2pt4phi4A}
 \ee
It is convenient to re-write the two-point correction in the following equivalent form
  \be
 \Delta \widehat H_2(\cE)_{rs} = \sum_{j}V_{ rj}\frac{1}{\cE-E_j}V_{js}= \lim_{\eps\rightarrow 0} -i \int_0^\infty dt \, e^{i (\cE-E_{rs}+i\eps)t}  {\cal T}\left\{ V(t/2) V(-t/2)\right\}_{rs}\, , \label{2pt4phi4B}
 \ee
 where $E_{rs}=(E_r+E_s)/2$.
 Applying  the Wick theorem we find
\be
-ig^2 \int_0^\infty dt \,  e^{i (\cE-E_{rs}+i\eps)t} \int_{-L/2}^{L/2} dxdz \,  \sum_{m=0}^{4} s_{4-m} D_F^{4-m}(z,t)     : \phi^{m}(x+z,t/2)\phi^{m}(x,-t/2): _{rs}  \, ,  \label{wick1phi2}
\ee
where  $s_p=\binom{4}{p}^2p!$  are the symmetry factors. By integrating \eq{wick1phi2} and keeping only the contributions from high energy intermediate states $E_j> E_T$ we obtain the exact expression for $\Delta H_2$. We use the shorthand notation $\Delta H_2= \Delta  H^{\mathds{1}}_2+\Delta  H^{\phi^2}_2+\Delta  H^{\phi^4}_2+\Delta  H^{\phi^6}_2+\Delta  H^{\phi^8}_2$ for $m=0,1,2,3,4$, and similarly for $\Delta \widehat H_2$. For $\Delta  H^{\mathds{1}}_2$, $\Delta  H^{\phi^2}_2$ we obtain:
\bea
\Delta  H_2^{\mathds{1}}(\cE,E_T) &=&\frac{s_4g^2}{2^4L^2}\sum_{k_1k_2k_3k_4} \frac{1}{\omega_{k_1}\omega_{k_2}\om_{k_3}\om_{k_4}} F_{0}(k_1,k_2,k_3,k_4,{\cal E},E_T)  \, , \label{nlid1}  \\[.2cm]
\Delta  H_2^{\phi^2} (\cE,E_T)&=&\frac{s_{3}g^2}{2^4L^2} \sum_{k_1,k_2,k_3 }\sum_{q_1,q_2}\,  \frac{1}{\omega_{k_1}\omega_{k_2}\om_{k_3}}\frac{1}{\sqrt{\om_{q_1}\om_{q_2}}} F_{2}(k_1,k_2,k_3,q_1,q_2,{\cal E},E_T)   \label{nlphi2} \, , 
\eea
where $F_0(k_1,k_2,k_3,k_4,{\cal E},E_T)$ is given by
\be
F_{0\, rs}= \delta_{\Sigma_{i=1}^4k_i,0} \, \frac{\theta( \om_{k_1}+\om_{k_2}+\om_{k_3}+\om_{k_4} +E_{rs}-E_T)}{\cE-\om_{k_1}-\om_{k_2}-\om_{k_3}-\om_{k_4}- E_{rs} }  \, \mathds{1}_{rs} \, ,
\ee
and the operator $F_2(k_1,k_2,k_3,q_1,q_2,{\cal E},E_T)$ is given by 
\bea
F_{2\,rs}&=&\delta_{k_1+k_2+k_3, q_1}  \, \delta_{q_1,-q_2}  \frac{\theta(E_{rs}+\om_{k_1}+\om_{k_2}+\om_{k_3}-E_T)}{\cE-E_{rs}-\om_{k_1}-\om_{k_2}-\om_{k_3}} \,  ( a_{q_1} a_{q_2})_{rs} \nn \\[.1cm]
%%%%%%%%%%%%%%%%%%%%%%%%%%%%%%%%%%%%%%%%%%%%%%%%%%
&+&  \delta_{k_1+k_2+k_3, q_1} \, \delta_{q_1,-q_2}  \frac{\theta(E_{rs}+\om_{k_1}+\om_{k_2}+\om_{k_3}-E_T)}{\cE-E_{rs}-\om_{k_1}-\om_{k_2}-\om_{k_3}}\, (a^\dagger_{q_1}\ad_{q_2} )_{rs}\nn \\[.1cm]
%%%%%%%%%%%%%%%%%%%%%%%%%%%%%%%%%%%%%%%%%%%%%%%%%%
&+&  \delta_{k_1+k_2+k_3, q_2}  \, \delta_{q_1,q_2}    \frac{\theta(E_{rs}+\om_{k_1}+\om_{k_2}+\om_{k_3}+\om_q-E_T)}{\cE-E_{rs}-\om_{k_1}-\om_{k_2}-\om_{k_3}-\om_q}\, ( \ad_{q_1} a_{q_2})_{rs}\nn \\[.1cm]
%%%%%%%%%%%%%%%%%%%%%%%%%%%%%%%%%%%%%%%%%%%%%%%%%%
&+&   \delta_{k_1+k_2+k_3, q_2}   \, \delta_{q_1,q_2}   \frac{\theta(E_{rs}+\om_{k_1}+\om_{k_2}+\om_{k_3}-\om_q-E_T)}{\cE-E_{rs}-\om_{k_1}-\om_{k_2}-\om_{k_3}+\om_q} \, (\ad_{q_1} a_{q_2})_{rs} \,.\quad \quad \, 
\label{f2rs}
\eea
In Eqs.~(\ref{nlid1})-(\ref{nlphi2}), all $q_i$'s are bounded from above ($q_i \leq q_{max}$) because they correspond to the momenta of creation/annihilation operators that act on the light states (i.e. states in ${\cal H}_l$). Instead the $k_i=2\pi n_i/L$ run over all possible values $n_i \in \mathbb{Z}$. Similar expressions for $\Delta  H^{\phi^4}_2,\, \Delta  H^{\phi^6}_2, \, \Delta  H^{\phi^8}_2$ are given in Appendix~\ref{deltahforphi4app}. As mentioned before, a simple way to derive these expressions from diagrams is given in Appendix \ref{diags}.
 We have performed the same kind of numerical checks done in Sec.~\ref{numtest} for all the operators $\Delta \widehat{H}_2$ in the $\phi^4$ theory. 
  %%%%%%%%%%%%%%%%%
  %%%%%%%%%%%%%%%%%
  %%%%%%%%%%%%%%%%% 
\subsubsection*{Approximations}
The exact expressions for $\Delta H_2$ are computationally demanding. Here we present different approximations that speed up the calculations and simplify their analytic structure. These basically consist in approximating the contribution from the highest energy states to $\Delta H$ in terms of a local expansion (as normally done in Effective Field Theory calculations), while keeping the contributions from lower energy states in their original non-local form. This is achieved by defining an energy $E_L$ and then by separating  $\Delta H_2$ into two parts, $\Delta H_{2 \, +}$ where we only sum over intermediate states with $E_j \geq E_L$  and $\Delta H_{2 \, -}$ where we sum over those with $E_T < E_j < E_L$. 
\bea
\Delta  H_{2\, +}(\cE,E_L)_{rs}&=& \Delta H_{2}(\cE,E_L)_{rs} ,, \label{DeltaHplusT}\\[.2cm]
\Delta  H_{2\, -}(\cE,E_T,E_L)_{rs}&=&\Delta H_{2}(\cE,E_T)_{rs}-\Delta H_{2}(\cE,E_L)_{rs} \, .
\label{DeltaHminusT}
\eea
We choose $E_L \gg E_T$ so that $\Delta H_{2 \, +}$ is well approximated by local operators~\footnote{In the cases where we are only interested in having a good approximation for the lower energy entries $r,s$ of the matrix, then $E_L$ can be taken to be similar to $E_T$. }. As an example we show how to implement this procedure for the contribution of $\Delta H_2^{\phi^2}$  given in \eq{nlphi2} and \eq{f2rs}. We start by examining the term  $\Delta  H_{2\, +}^{\phi^2}(\cE,E_L)=\Delta H_{2}^{\phi^2}(\cE,E_L)$, which  is obtained by replacing $E_T$ by $E_L$ in  \eq{f2rs}. In this case $\sum_i\om_{k_i} \gtrsim E_L \gg E_T \gtrsim \om_q, E_{rs}$, and then it can be well approximated by
\be
\Delta  H_{2\,+}^{\phi^2}\simeq  c_2  \,  V_2
\label{1stlocalsunrise}
\ee
with 
\be
c_2(\cE,E_L) = \frac{s_{3}g^2}{(2L)^3}   \sum_{k_1,k_2,k_3}  \frac{L\, \delta_{k_1+k_2+k_3,0}}{  \om_{k_1} \om_{k_2} \om_{k_3} }\, \frac{\theta(\om_{k_1}+\om_{k_2}+\om_{k_3}-E_L) }{\cE-\om_{k_1}- \om_{k_2} - \om_{k_3}  }  \label{ch2}\, ,
\ee
and  $V_2=\int_0^L dx\, \phi^2 (x)$ which has dimensions of $[E]^{-1}$. 
The approximation in  \eq{1stlocalsunrise} receives corrections of at most ${\cal O}(E_T/E_L)$. 
The expansion  of  $\Delta H_{2+}^{\phi^2}$ in terms of local operators   can be   obtained by  expanding the  term  $\Delta \widehat H^{\phi^2}_2$   in \eq{wick1phi2} around 
$t,z=0$
  \be
 \Delta \widehat H_2^{\phi^2}(\cE)_{rs} =-ig^2  s_{2} \int_0^\infty dt \,  e^{i (\cE-E_{rs}+i\eps)t} \int_{-L/2}^{L/2} dz  D_F^{2}(z,t)   \int_{-L/2}^{L/2} dx   \left[ : \phi^{2}(x,0): _{rs} +{\cal O}(t^2,z^2) \, \right] , \label{llocal}
\ee
and, after integrating, keeping only the contributions from those states  that produce poles at $\cE>E_L$, when $E_{rs}$ is neglected. On the other hand $\Delta  H_{2\, -}^{\phi^2}(\cE,E_T,E_L)=\Delta H_{2}^{\phi^2}(\cE,E_T)-\Delta H_{2}^{\phi^2}(\cE,E_L)$ is given by the same expressions as in \eq{nlphi2} and  \eq{f2rs} but now the sums to perform are much smaller since the momenta of the intermediate states are restricted between $E_T$ and $E_L$.
%The approximation done for $\Delta H_+^{\phi^2}$ in terms of a local operator  is standard in Effective Field Theory: whenever  there is an energy gap between the low and the high energy physics, one can  focus on low energy processes and take into account  the effects of the   high energy modes  in terms of  local operators. 

The same exercise done for $\Delta H_{2+}^{\phi^2}$ can be done for   $\Delta  H^{\mathds{1}}_{2+}$ and  $\Delta  H^{\phi^{4}}_{2+}$ and one has that in the limit $E_L\gg E_T$
\be
\Delta  H_{2\,+}^{\mathds{1}} \simeq  c_{0}  \,  \mathds{1} \, \, , \quad   \quad     \Delta  H_{2\,+}^{\phi^2} \simeq  c_{2}  \,  V_2  \,  \,  ,\quad \quad    \Delta  H_{2\,+}^{\phi^4}   \simeq  c_{4}  \,  V_4 \, ,  \label{llocal2}
\ee
where   $V_\alpha =\int_0^L dx\, \phi^\alpha (x)$ and has dimensions of $[E]^{-1}$, 
\bea
c_0(\cE,E_L) &=& \frac{s_{4}g^2}{(2L)^4}   \sum_{k_1,k_2,k_3,k_4}  \frac{L\, \delta_{k_1+k_2+k_3+k_4,0}}{  \om_{k_1} \om_{k_2} \om_{k_3} \om_{k_4} }\, \frac{\theta(\om_{k_1}+\om_{k_2}+\om_{k_3}+\om_{k_4}-E_L) }{\cE-\om_{k_1}- \om_{k_2} - \om_{k_3}  - \om_{k_4}  }\label{COL} \, , \\[.2cm]
c_4(\cE,E_L) &=& \frac{s_{2}g^2}{(2L)^2}   \sum_{k_1,k_2}  \frac{L\, \delta_{k_1+k_2,0}}{  \om_{k_1} \om_{k_2}  }\, \frac{\theta(\om_{k_1}+\om_{k_2} -E_L) }{\cE-\om_{k_1}- \om_{k_2}  }  \label{se457}\, ,
\eea
and $c_2$ is given in \eq{ch2}. 
On the other hand the operators $\Delta  H_{2}^{\phi^6}$ and $\Delta  H_{2}^{\phi^8}$ are  of the tree-level and disconnected type because they involve one and zero propagators respectively, see \eq{wick1phi2}. 
 Therefore  the operators  $\Delta  H_{2+}^{\phi^6}$ and $\Delta  H_{2+}^{\phi^8}$  are not well approximated by a  local expansion, and we do not approximate them. For $E_L$ sufficiently big though,  $\Delta  H_{2+}^{\phi^6} = \Delta  H_{2+}^{\phi^8} =0$ and all the contribution to $\Delta H_2^{\phi^6}$, $\Delta H_2^{\phi^8}$ comes from  $\Delta  H_{2-}^{\phi^6}$, $\Delta H_{2-}^{\phi^8}$, as can be explicitly seen from Eqs.~(\ref{F6})-(\ref{F8}).  Notice that these operators only contribute to the entries of $\Delta H_{rs}$ with high values for $E_r$, $E_s$. 
Again, the coefficients of the local operators in \eq{llocal2} can be   obtained by  expanding  $\Delta \widehat{H}_2$ in \eq{wick1phi2} around $t,z=0$
  \be
 \Delta \widehat H_2(\cE)_{rs} =-ig^2 \int_0^\infty dt \,  e^{i (\cE-E_{rs}+i\eps)t} \int_{-L/2}^{L/2} dxdz \,  \sum_{m=0}^{4} s_{4-m} D_F^{4-m}(z,t)  : \phi^{2m}(x,0): _{rs} +{\cal O}(t,z)^2 \, , \label{llocal}
\ee
and, after integrating, keeping only the contributions from those states  that produce poles at $\cE>E_L$, when $E_{rs}$ is neglected.
The evaluation of the coefficients in \eq{llocal2} can  still be hard to evaluate numerically. In the next section we explain an alternative and simpler derivation of the coefficients $c_{2m}$  and further approximations to evaluate them.

\subsection{Local expansion and the phase-space functions}
\label{localphase}
 From the first term in the local expansion of \eq{llocal} the coefficients of the local operators are given by:
\be
\hat c_{2n}(\cE)=-i g^2 s_{4-n} \int_0^\infty dt e^{i(\cE+i\eps)t}\int_{-\infty}^\infty \, dx\,  D_F^{4-n}(x,t)\, , \label{rj}
   \ee
   where $s_{4-n}$  is the symmetry factor and, as explained above,     the common $E_{rs}$-shift on the eigenvalue $\cE$ is neglected.~\footnote{The derivation of the coefficients $\hat c_{2n}(\cE)$ in \eq{rj}    applies to any $\phi^\alpha$ theory.}
    Next, applying  the Kramers-Kronig dispersion relation to $c_n(\cE)$ in  \eq{rj}
   \be
\hat c_{2n}(\cE)=- \int_{-\infty}^\infty \frac{dE}{\pi}\frac{1}{\cE-E+i\eps} \text{Im} \, \hat  c_{2n}(\cE) \label{psps}\, .  
\ee
Next, we compute $\text{Im } \hat c_{2n}$. First we do the space integral which, up to $g^2 s_{4-n}$, yields 
\be
  \text{Im}-i  \sum_{k's} \frac{L\, \delta_{\sum_i k_i,\, 0}}{\prod_{i}2L\om_{k_i}} \int_{0}^\infty dt  e^{i(E-\sum_i\om_{k_i}+i\eps)t} = -\frac{1}{2}   \sum_{k's}  \frac{L\, \delta_{\sum_i k_i,\, 0}}{\prod_{i}2L\om_{k_i}} \,  2\pi\delta\Big(E-\sum_i\om_{k_i}\Big) \, , \label{rj2}
\ee
where we have used   $D_F(t,x)\theta(t)=D(t,x)\theta(t) $ with $D(t,x)=\sum_k (2L\om_k)^{-1}e^{ikx-i\om_kt}$.
Therefore we find~\footnote{\eq{ps} can also be derived from the  optical theorem, with careful treatment of the symmetry factors.},
\be
\hat c_{2n}(\cE)= \frac{g^2 s_{4-n}}{2\pi} \int_{-\infty}^\infty  \frac{dE}{\cE-E+i\eps} \Phi_{{4-n}}(E)\label{ps}
\ee
where $\Phi_{m}(E)$ is the $m$-particle phase space
\be
\Phi_m(E)= \sum_{k_1,k_2,\dots, k_m}  \frac{L\, \delta_{\sum_{i=1}^m k_i,\, 0}}{\prod_{i=1}^m 2L\om_{k_i}} \,  2\pi\delta\Big(E-\sum_{i=1}^m \om_{k_i}\Big)\, . 
\ee
Finally,   the coefficients in \eq{llocal2} are obtained by including only the contributions from  poles located at $\cE\geq E_L$
    \bea
   %%%%%%%%%%%%%%%%%%%%%%%%%%%%%%%%%%%
c_0(\cE) &=&   s_4\,   g^2   \ \int_{E_L}^\infty   \frac{  dE}{2\pi}\,  \frac{1}{\cE-E}  \, \Phi_4(E)   \label{phifirst}  \,   \, ,\\[.2cm]
       %%%%%%%%%%%%%%%%%%%%%%%%%%%%%%%%%%%
 c_2(\cE) &=&     s_3  \, g^2   \ \int_{E_L}^\infty   \frac{  dE}{2\pi}\,  \frac{1}{\cE-E}  \, \Phi_3(E)      \ ,  \\[.2cm]
    %%%%%%%%%%%%%%%%%%%%%%%%%%%%%%%%%%%
 c_4(\cE)&=&       s_2\,   g^2   \ \int_{E_L}^\infty   \frac{  dE}{2\pi}\,  \frac{1}{\cE-E}  \, \Phi_2(E)\label{philast}\, .   
   \eea
It would be interesting to see if in general, higher $\Delta H_{n+}$ corrections can also be written in terms of phase space functions. In the rest of the section we explain useful approximations   to evaluate  Eqs.~(\ref{phifirst})-(\ref{philast}).
   
   \subsubsection*{Continuum  and  high energy limit of the phase space}
   
We start by approximating the phase space by its continuum limit.~\footnote{This is a good approximation for $Lm\gg 1$ and we have checked it explicitly in our numerical study.}  
   \noindent   Recall that in the continuum limit the  relativistic phase-space for $n$-particles is given by
    \be
    \Phi_n(E) = \int \prod_{i=1}^n \frac{dk^1_i}{(2\pi)\, 2\om_{k_i} } \, (2\pi)^2 \delta^{(2)}(P^\mu+\sum_{i=1}^n k_i^\mu) \, , \label{relphase}
    \ee  
  where  $P^\mu=(E,0)$ and $k_i^\mu=(\om_{k_i},k_i)$. 
  Then, for  the 2-body phase space one has
  \be
\Phi_{2}(E) =\frac{1}{E\sqrt{E^2-4m^2}} \label{2body1}\, .
\ee
Next, solving for the Dirac delta's in \eq{relphase},  the 3-body phase-space is given by
   \be
\Phi_3(E)   =  \frac{1}{2\pi }\int_{4m^2}^{(E-m)^2}  \frac{d s_{23}}{\sqrt{s_{23}\left(s_{23}-\left[E+m\right]^2\right)\left(s_{23}-\left[E-m\right]^2\right)\left(s_{23}-4m^2\right)}}\, ,\label{inte1}
\ee
 with  $E\ge3m$.  This integral can be solved by standard Elliptic integral transformations and we obtain,
\be
   \Phi_3(E)=\frac{  g^2}{\pi}    \frac{1}{(E-m)} \frac{1}{\sqrt{(E+m)^2-4m^2}} K\left(\alpha  \right)\label{ellipticphi3} \, ,
   \ee
where $ \alpha = 1- \frac{16 Em^3}{(E-m)^3(E+3m)}$ and  $K(\alpha)=\int_0^{\pi/2}\frac{d\varphi}{\sqrt{1-\alpha\sin^2(\varphi)}} $  is an elliptic integral.   

\medskip 
  
In general though, finding the exact phase space functions  $\Phi_n(E)$ is difficult but can be simplified in the  limit $E\gg m$. In our case, this limit is justified because the phase space functions are evaluated for $E\geq E_L \gg m$.   Notice that to take the high energy limit of $\Phi_n(E)$  one can not expand the integrand of  \eq{relphase}    because, after solving for the Dirac delta's constraints,  it is  of ${\cal O}(1)$ at the integral limits, see for instance the elliptic integral in \eq{inte1}. Instead, we use the following relation for the phase space 
\be
I_n(\tau) \equiv \int_{-\infty}^{\infty} dx \, D_E^n(x,\tau) =\frac{1}{2 \pi} \int_0^\infty dE \, e^{-E \tau} \Phi_n(E)\label{phaserel}
\ee
where $D_E(x,\tau)$ is the euclidean propagator and $\Phi_n(E)$ is only non vanishing for $E\ge n \, m$. The Euclidean propagator in $d=2$ is given by the special Bessel function of second kind $K_0(m\rho)$ with  $\rho=\sqrt{x^2+\tau^2}$ and $I_n(\tau) \equiv \int_{-\infty}^\infty dx \, K^n_0(m\rho)(2\pi)^{-n} $.
  At this point we can use a clever trick done in Ref.~\cite{Rychkov:2014eea} to find the leading terms of the inverse Laplace transform of $I_n(\tau)$ in the limit $E\rightarrow\infty$. Since the phase space $\Phi_n(E)$ is the inverse Laplace transform of $I_n(\tau)$,  the leading parts of $\Phi_n(E)$ as $E\rightarrow\infty$ come from the non-analytic parts of $I_n(\tau)$ as $\tau\rightarrow 0$. To find the non-analytics parts of $I_n(\tau)$ first one notices that
  %and then can compute $\Phi_n(E)$ using the derivation done in \ref{Rychkov:2014eea} where they find the inverse Laplace transform of $I_m(\tau) \equiv \int_{-\infty}^\infty dx \, K^m_0(m\rho)(2\pi)^{-m} $.  where  
  \be
  K_0(m\rho)=\left\{\begin{array}{l l }
 - \log\left( \frac{e^\gamma m \rho}{2}\right) \left[1+{\cal O}(m^2\rho^2 )\right] & ,\quad \rho \ll 1/m \\[.2cm]
 \sqrt{\frac{\pi}{2m\rho}}  \, e^{-m\rho} \left[1+{\cal O}(m^{-1}\rho^{-1} )\right]  &, \quad \rho\gg 1/m \, 
  \end{array} \right.\label{besselexpansion}
  \ee
  where $\gamma$ is the Euler constant. 
Then, the contributions to $I_n(\tau) = \int_{-\infty}^\infty dx \, K^n_0(m\rho)(2\pi)^{-n}$ when $\tau \rightarrow 0$ are dominated by the region where $\rho\ll 1/m$ and the integrand  can be approximated by $K_0(m\rho)\approx - \log\left( \frac{e^\gamma m \rho}{2}\right) $.~\footnote{This method is like the method of regions which is used to get the  leading terms of  multi-loop Feynman diagrams in certain kinematical limits or mass hierarchies.    } This approximation introduces spurious IR divergences in the region of integration $\rho \gg 1/m$ where the approximation of the integrand is not valid.~
 These divergences can be regulated with a cutoff  $\Lambda$ or, equivalently, one can take derivatives with respect to the external coordinate $\tau$ to regulate the integral $I_n(\tau)$.~\footnote{This is similar to the fact that the UV divergences of multi-loop Feynman diagrams are polynomial in the external momenta because taking enough derivatives with respect to the external momenta the integrals  are UV finite.}  Hence, approximating  $K_0(m\rho)\approx - \log\left( \frac{e^\gamma m \rho}{2}\right) $ and integrating over $x$ one can find the non-analytic terms of $\partial_\tau I_n(\tau)$ as $\tau \rightarrow 0$. For instance, for $n=4$  
  \bea
 \partial_\tau I_4(\tau) =\frac{1}{4\pi^3}\log(m\tau e^{\gamma})\left[\log(m\tau)\log(m\tau e^{2\gamma})+\gamma^2+\frac{\pi^2}{4} \right]+const.+ {\cal O}(\tau)\, ,   \label{ftr3} 
  \eea
  where the constant does not depend on $\tau$.  Lastly from \eq{phaserel}, $\partial_\tau I_n(\tau)$ is related to the phase space $\Phi_n(E)$ by the Laplace transform, 
    \be
   \int_0^\infty  dE  [-E\Phi_n(E)] \, e^{-\tau E}  =2\pi\,   \partial_\tau I_n(\tau)
  \ee
so that for $n=4$ one has
    \bea
\Phi_4(E) &=&\frac{3}{2\pi^2}\frac{1}{E^2}\left[ \log^2\left(E/m\right)-\pi^2/12\right] + {\cal O}\left(m^2/E^4\right)    \, .
\label{phi4expanded}
  \eea
Therefore using \eq{phi4expanded} and expanding Eqs.~(\ref{2body1}), (\ref{ellipticphi3}) at large $E$,
 \bea
   %%%%%%%%%%%%%%%%%%%%%%%%%%%%%%%%%%%
c_0(\cE) &\simeq&   s_4\,   g^2   \ \int_{E_L}^\infty \frac{  dE}{2\pi}\,  \frac{1}{\cE-E}  \, \frac{3}{2\pi^2}\frac{1}{E^2}\left[ \log^2\left(E/m\right)-\pi^2/12\right]    \,   \, \label{c0app} ,\\[.2cm]
       %%%%%%%%%%%%%%%%%%%%%%%%%%%%%%%%%%%
 c_2(\cE) &\simeq&     s_3  \, g^2   \ \int_{E_L}^\infty   \frac{  dE}{2\pi}\,  \frac{1}{\cE-E}  \,  \frac{3}{2 \pi} \frac{1}{E^2} \log(E/m)     \ ,  \\[.2cm]
    %%%%%%%%%%%%%%%%%%%%%%%%%%%%%%%%%%%
 c_4(\cE)&\simeq&       s_2\,   g^2   \ \int_{E_L}^\infty   \frac{  dE}{2\pi}\,  \frac{1}{\cE-E}  \, \frac{1}{E^2}  \, \label{c4app} ,
   \eea
where the error made  in the approximations is of the order ${\cal O}\left(m^2/E_L^2\right)$. We end this section by noticing that the leading terms of the phase space functions $\Phi_2(E)$ and $\Phi_3(E)$ in the large $E$ expansion agree with the corresponding result of  Ref.~\cite{Rychkov:2014eea} (there called $\mu_{444}(E)=s_2 \Phi_2(E)/(2\pi)$, $\mu_{442}(E)=s_3 \Phi_3(E)/(2\pi) $). The local approximation in Eqs.~(\ref{c0app})-(\ref{c4app}) can be refined by taking into account the $E_{rs}$ shift, see Ref.~\cite{Rychkov:2014eea}. 

% For the calculation of $\Delta H_{-}$ an approximation that could be made to increase the speed of the calculations would be to compute only a small number of diagonal entries with evenly spaced $E_{r,s}$ energies, and assigning these coefficients $c_{rs}$ to all the entries with energies $E_{r,s}$ in between the range of two given computed entries. The error in this case would be ${\cal O} \big(\frac{\Delta E_{r,s}}{E_{r,s}} \big)$ where $\Delta E_{r,s}$ is the difference between the exact and approximated value assigned to the entry $r,s$. We didn't use this approximation. (Improve)

 \subsection{Spectrum and convergence}
\label{sc2}  

Before starting with the numerical results we first discuss the series $\Delta H=\sum_{n=2}\Delta H_n$ in more detail.  The truncation of the $\Delta H$ series  in powers of $(V_{hh}/H_{0\,  hh})^n$     is only justified for    $V_{hh}/H_{0\, hh}<1$. 
Notice that  even for weak coupling $g\ll 1$ the series does not seem to converge. 
Let us consider a particular matrix entry
  \be
 \bra{E_r}|\Delta H_n| \ket{E_s}  = \sum_{j_1,\dots, j_{n-1}}  V_{r j_1} \cdots  V_{j_{i-1} j_i} \frac{ 1}{\cE-E_{j_i}}\cdots       \frac{ 1}{\cE-E_{j_{n-1}}}V_{j_{n-1}s}  \, , \label{exem}
 \ee
where all the terms in the sums have a definite sign depending on whether $n$ is even or odd.  For instance, consider a contribution to \eq{exem} from   states of   high occupation number but low momentum like
 \be
V_{j_{i-1} j_i} \frac{ 1}{\cE-E_{j_i}} \rightarrow  \frac{\bra{N_{k}N_{-k}}|V| \ket{N_{k}N_{-k}} }{ \cE-2N \om_k} = \frac{6g}{4L \om_k^2}\frac{ (2N(N-1)+4N^2)}{\cE-2N\om_k}\label{sdds} \, , 
  \ee
 where $| \ket{N_{k}N_{-k}}$ is a Fock state with $N$ particles of momentum $k$ and $-k$ that satisfy    $2N\om_k>E_T$. The term of \eq{sdds}  gives a non-perturbative contribution even for small $g$ for high enough $N$ and becomes worse for smaller momentum $|k|$.~
Thus the series $(\Delta H)_{rs}=\sum(\Delta H_n)_{rs}$ seems to be non-convergent but we will assume that (when the expansion parameter is small) the first terms of the series are a good approximation to $(\Delta H)_{rs}$. 
Notice that the appearance of the  non-perturbative contributions (like in \eq{sdds}) can be worse for those matrix entries  $(\Delta H)_{rs}$ with energies $E_{r,s}$ closer to $E_T$ because the intermediate  states  in  $V_{jj^\prime}$  can have lower momentum and high occupation number  for a given  $\Delta H_n$.

 For the first terms of the expansion  $(V_{hh}/E_{h})^n$, a naive estimate of the dimensionless expansion parameter is
 $\alpha_{rs} \sim  g/E_T\times 1/(L   \mu^2 _{rs} )$ where the $g$ and $L$  can be read off from the potential; the $E_T^{-1}$ arises because the sums in \eq{exem} are dominated by the first terms, starting at $1/E_T$ (for $\cE<<E_T$); and by direct inspection of the potential  $m/N\lesssim \mu_{rs} \lesssim E_T$ where $N$ is a possibly large occupation number, depending on the matrix entry. 
 %{\roig The estimate above is valid for $d=2$ space-time dimensions; in general for $d$ space-time dimensions an estimate can be obtained by the same reasoning finding $\alpha_{rs}\sim g/E_T\times \frac{1}{L^{d-1} \mu_{rs}^2}$.}
 
 It can happen that entries with energies $E_{r}$, $E_s$ close to $E_T$ do not have a perturbative $(\Delta H)_{rs}=\sum (\Delta H_n)_{rs}$ expansion and even including the first terms of the series is a worse approximation than setting $(\Delta H)_{rs} \rightarrow 0$; these entries can induce big errors on the computed eigenvalues.
 Since the eigenvalues we are interested in computing are mostly affected by the lower $E_{r,s}$-energy matrix entries   we will neglect the  renormalization of the higher $E_{r,s}$ energy entries where the series $(\Delta H)_{rs}=\sum (\Delta H_n)_{rs}$ is not perturbative. One way to  select those entries would be to keep only those that satisfy   $\alpha_{rs}\sim(\Delta H_3)_{rs}/ (\Delta H_2)_{rs}<1$. 
 However, this can be computationally expensive and instead we  take a more pragmatic approach
 and only renormalize those matrix entries $(H_T)_{rs}$ with either $E_{r}$ or $E_s$   below some  conservative cutoff $E_W$, below which  the series is perturbative.
  
Up until this point the discussion has been done for $g\ll 1$. However, for those matrix entries where $\alpha_{rs}$ is a perturbative expansion parameter one can increase $g$ to strong coupling~\footnote{In the $\phi^4$ theory the strong coupling can be estimated to be  $g\gtrsim 1$, see Eqs.~(\ref{spert1}) and (\ref{spert2}).} by increasing $E_T$ at the same time. 
 Increasing $E_T$ means enlarging the size of $H_T$ and  $\Delta H$, and it can happen that the new matrix entries do not have a perturbative $(\Delta H)_{rs}=\sum (\Delta H_n)_{rs}$ expansion. As explained above, in those cases  we set $(\Delta H)_{rs}$ to zero.~\footnote{For the $\phi^2$ perturbation studied in Sec.~\ref{phi2res} we find that  the error in the computed eigenvalues can be decreased by increasing $E_T$ even without introducing $E_W$. For the $\phi^4$ we find that $E_W$ must be introduced.}

\subsubsection*{Numerical results}
 In the rest of the section we perform a numerical study of the spectrum of the $\phi^4$ theory. First we summarize the concrete implementation of the method. We find the spectrum of $H$ by diagonalizing $H_{eff}=H_T+\Delta H_2(\cE^T)$ where  $\cE^T$ is the eigenvalue of $H_T$.~\footnote{The dimension of the Hilbert space $\mathcal{H}_{ll}$ for $E_T=10$, $12$, $14$, $16$ and $18$ is  117(108),  309(305), 827(816), 2160(2084) and 5376(5238) for the  $\mathds{Z}_2$-even(odd) sectors, respectively.} As explained in Sec.~\ref{phi42point}, to calculate $\Delta H_2$ we separate it in $\Delta H_{2+}$ and  $\Delta H_{2-}$ defined in Eqs.~(\ref{DeltaHplusT})-(\ref{DeltaHminusT}) and take $E_L = 3 E_T$.~\footnote{ The choice $E_L = 3 E_T$ is done so that   the local expansion is a good approximation for intermediate states with $E_j \ge E_L$. Also, for this $E_L$ one has that  $\Delta H_{2+}^{\phi^6}= \Delta H_{2+}^{\phi^8}=0$.}. We found little differences when iterating the diagonalization with $\cE$. We also find that increasing $E_L$ does not have a significant effect on the result. For $\Delta H_{2-}(\cE, E_T, E_L)$ we use the expressions in Eqs.~(\ref{F1})-(\ref{F8}) and for $\Delta H_{2+}(\cE, E_L)$ we use the ones in Eqs.~(\ref{c0app})-(\ref{c4app}). We do a conservative estimate of the expansion parameter $\alpha_{rs}$ and set to zero $(\Delta H_2)_{rs}$ for all those entries that are not perturbative.  
        
\begin{figure}[t]
\centering
  \includegraphics[width=0.484\textwidth]{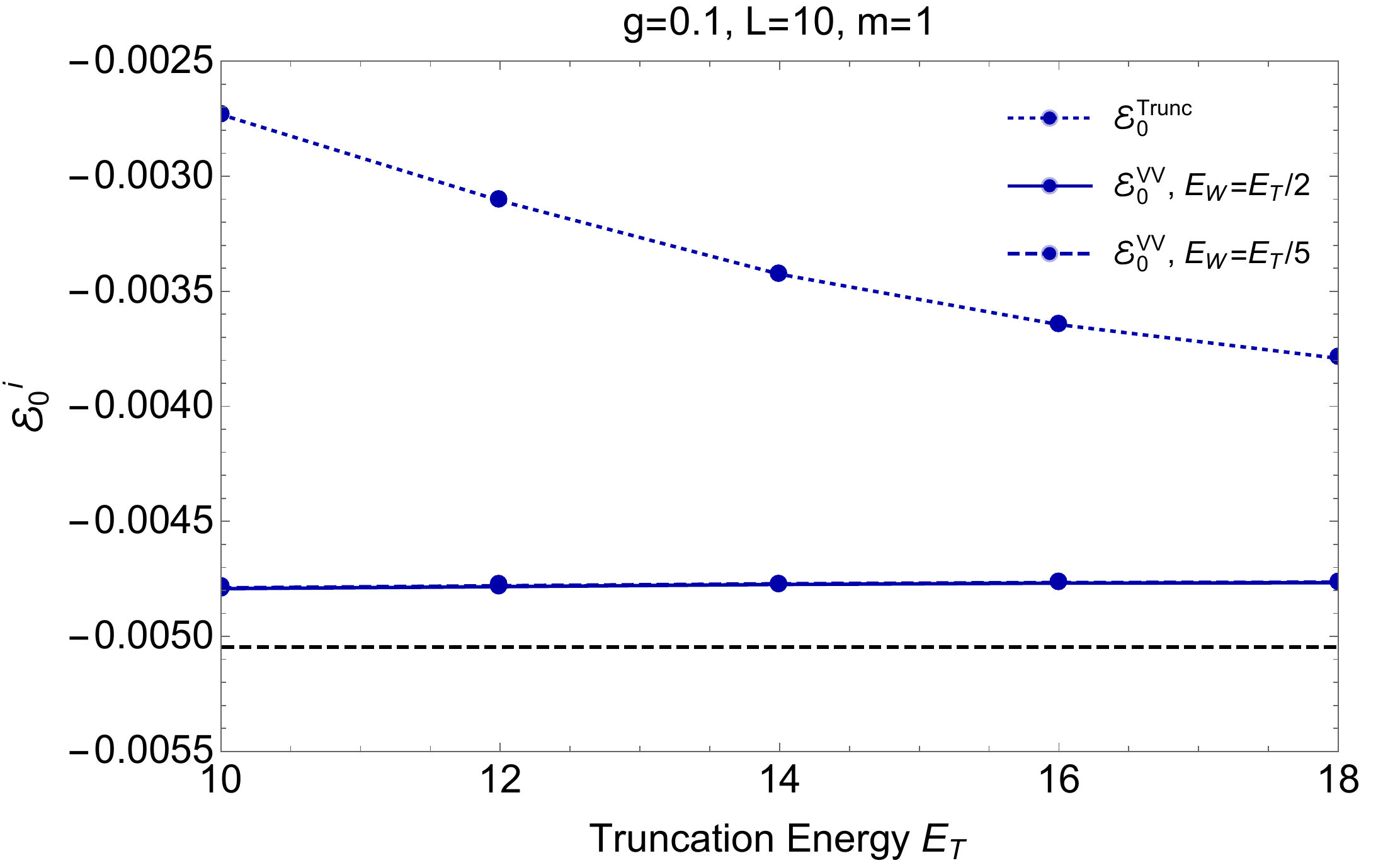} \quad  \includegraphics[width=0.481\textwidth]{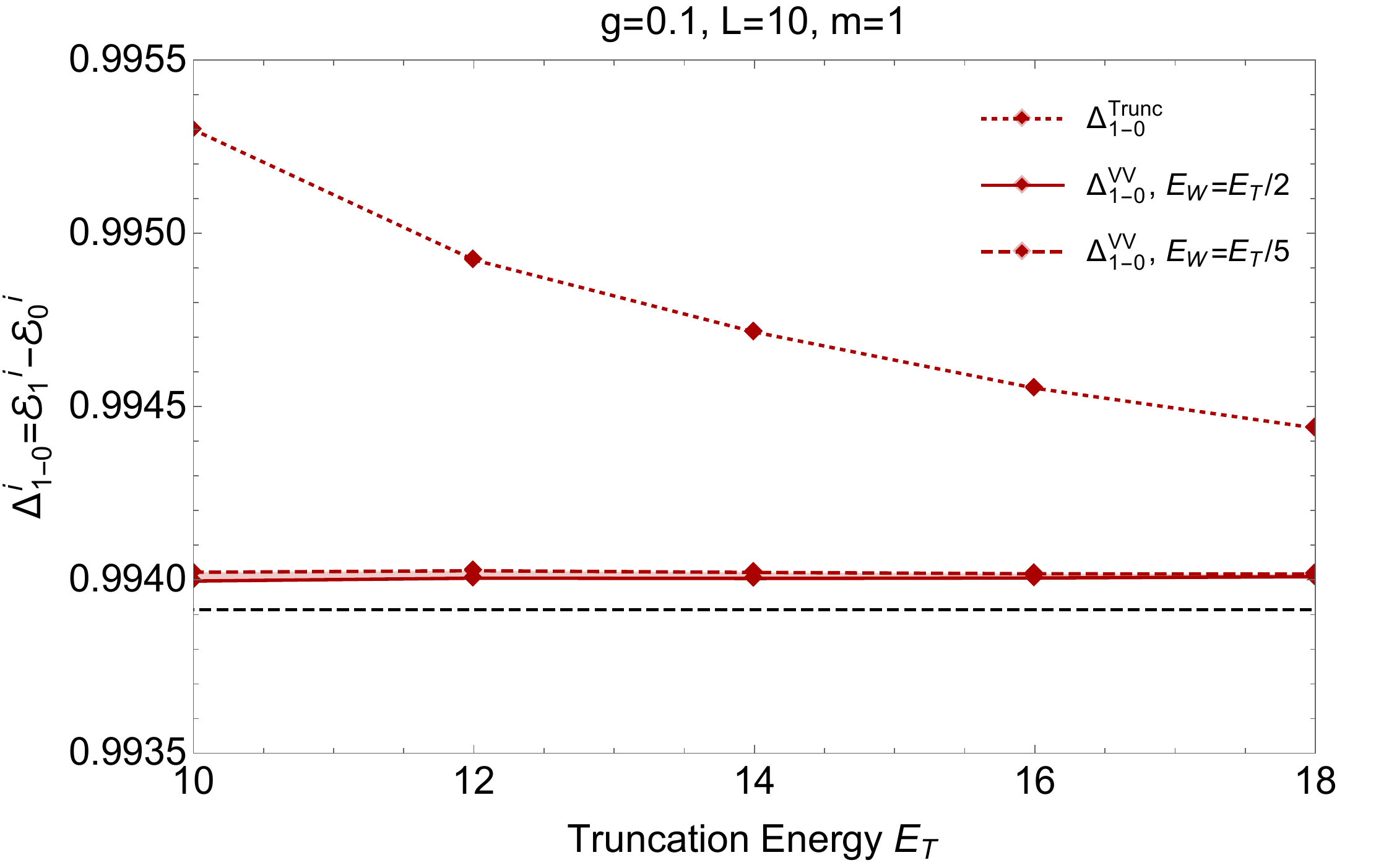} 
\caption{\em  {\bf \em Left:} The vacuum energy $\cE_0^i$   as a function of the truncation energy $E_T$ for a coupling of $g=0.1$. {\bf \em Right:} Energy difference between the  first  $\mathds{Z}_2$-odd excited state  and the vacuum energy $\cE_0^i$   as a function of the truncation energy for $g=0.1$. In both plots, the dotted curves are computed with the truncated Hamiltonian while  the solid and dashed curves are computed with the renormalized hamiltonian at order $VV$. Dashed and dotted lines correspond to  the cutoffs $E_W=E_T/2$ and  $E_W=E_T/5$. We have overlaid two dashed black lines corresponding to the calculation in perturbation theory, see. Eqs.~(\ref{spert1}) and  (\ref{spert2}).}  \label{exactConvergence323}   
\end{figure}   
  
First we study the lowest eigenvalues of  $H$ at weak coupling, where we can compare with standard perturbation theory. 
The perturbative   corrections to the vacuum and the  mass are given by \cite{Rychkov:2014eea}:
   \bea
   \Lambda/m^2 &=&-\frac{21\xi(3)}{16\pi^3}\, \bar g^2 + 0.04164(85) \, \bar g^3+\dots   \, ,\label{spert1}\\[.2cm]
   m_{ph}^2&=&m^2\big[1-\frac{3}{ 2 } \, \bar g^2 + 2.86460(20) \,   \bar g^3+\cdots \big] \label{spert2}\, ,
   \eea
   where $\bar g \equiv g/m$ and  $m_{ph}$ is the physical mass. In Fig.~\ref{exactConvergence323} we show the result for the vacuum energy and $m_{ph}$.
 \begin{figure}[p!]
\centering
  \includegraphics[width=0.484\textwidth]{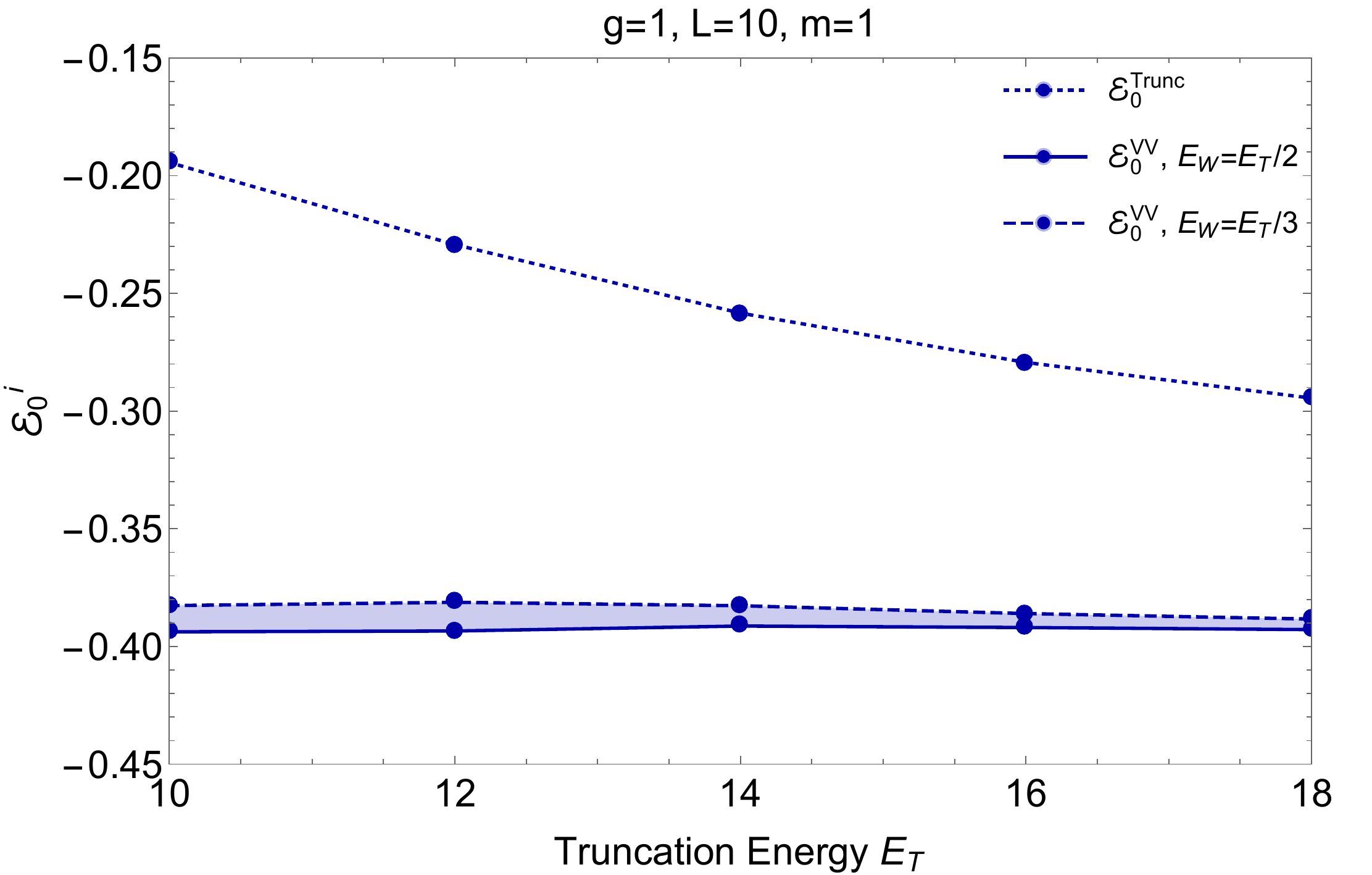} \quad  \includegraphics[width=0.481\textwidth]{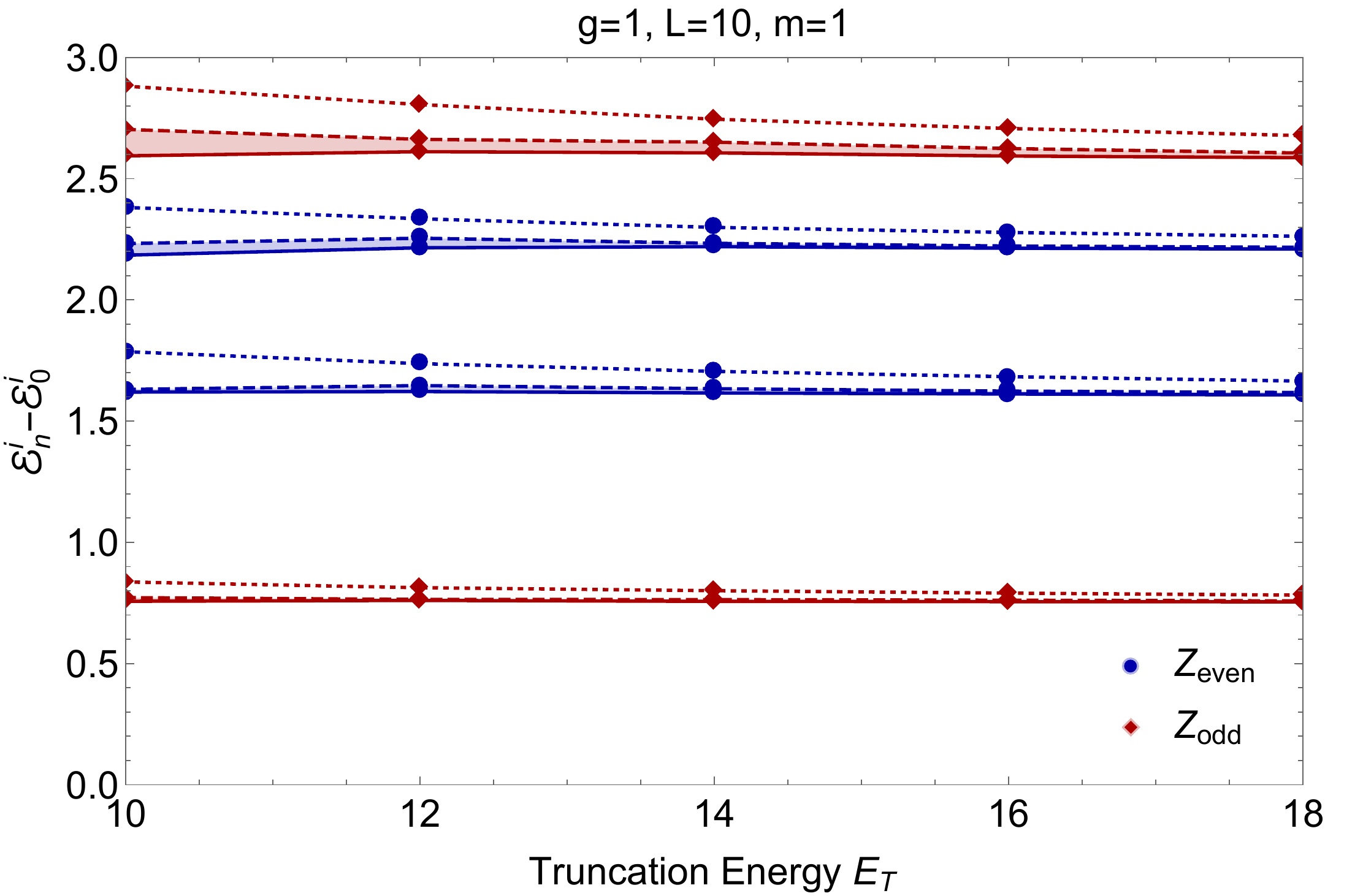} 
  
  \hspace{.2cm}
  
  \includegraphics[width=0.484\textwidth]{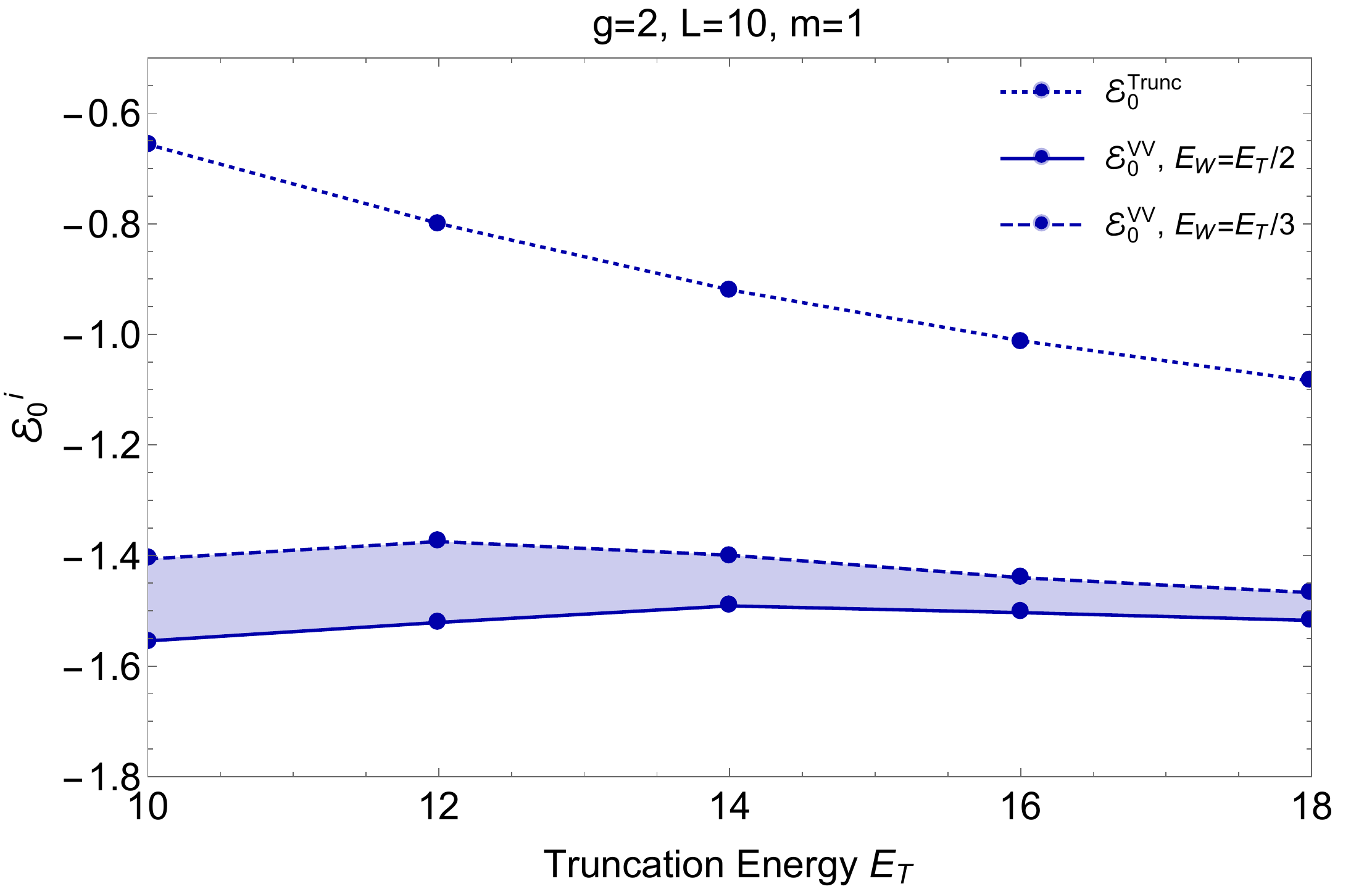} \quad  \includegraphics[width=0.481\textwidth]{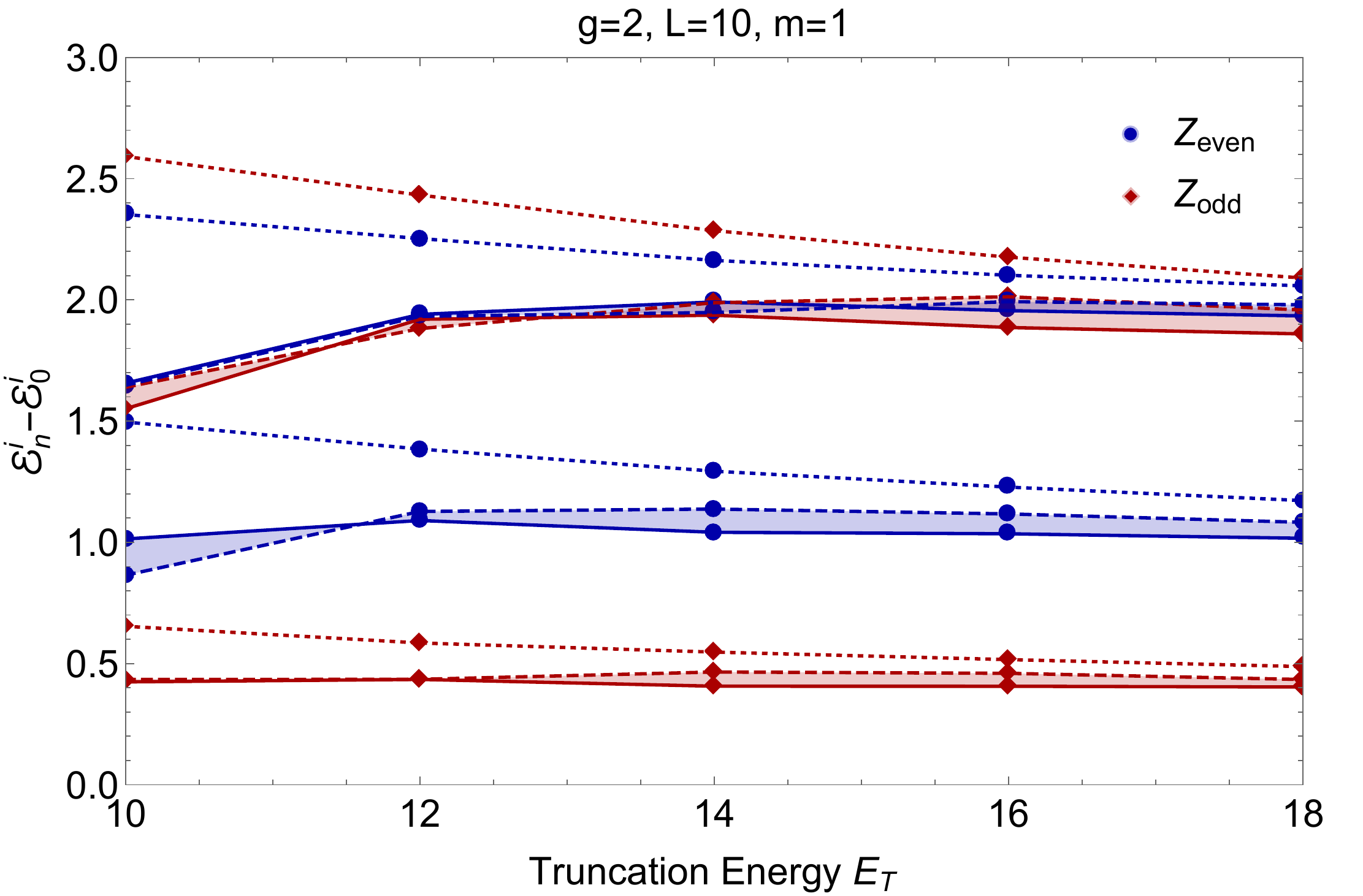} 
  \hspace{.2cm}
  
  \includegraphics[width=0.484\textwidth]{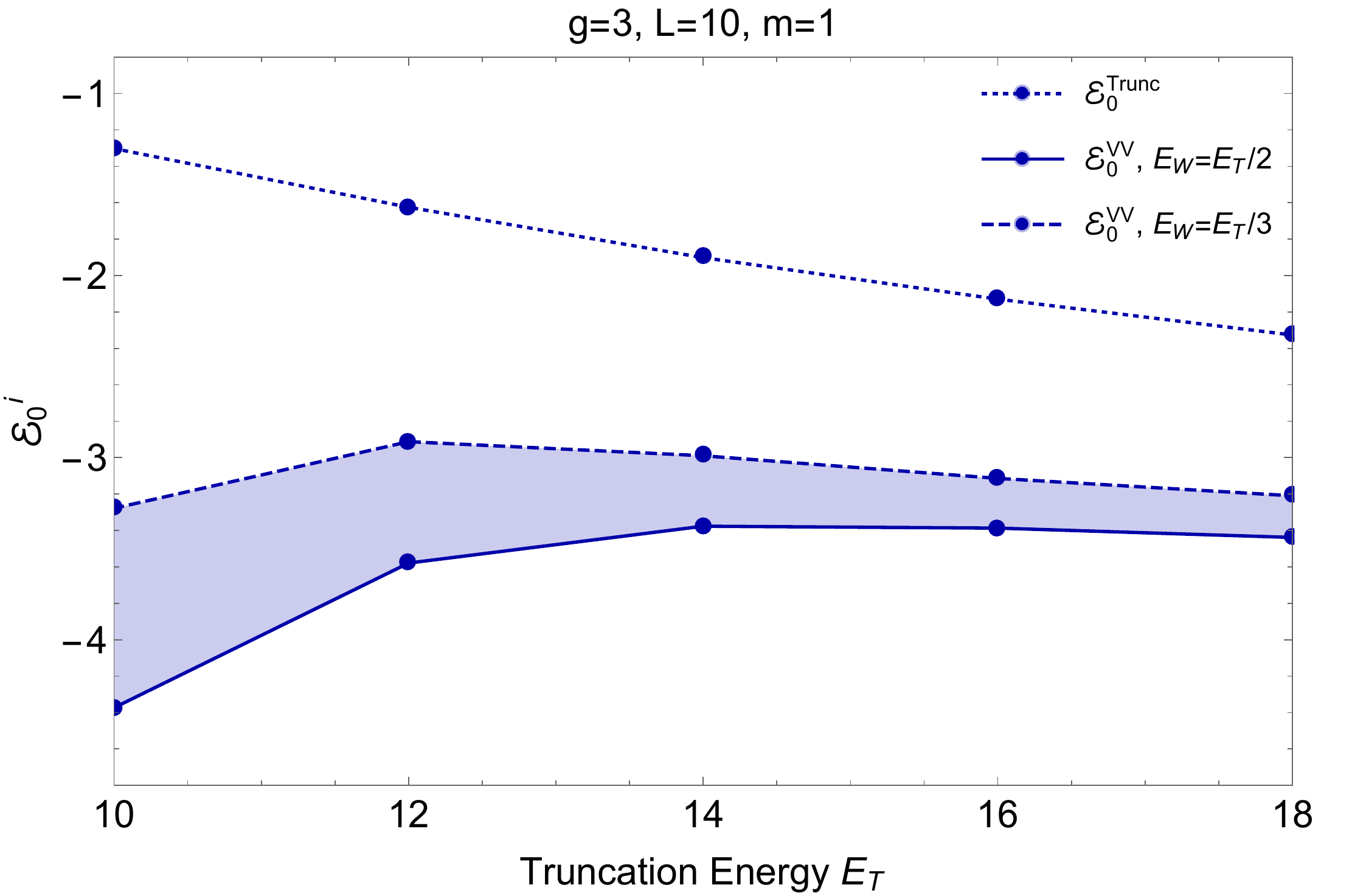} \quad  \includegraphics[width=0.481\textwidth]{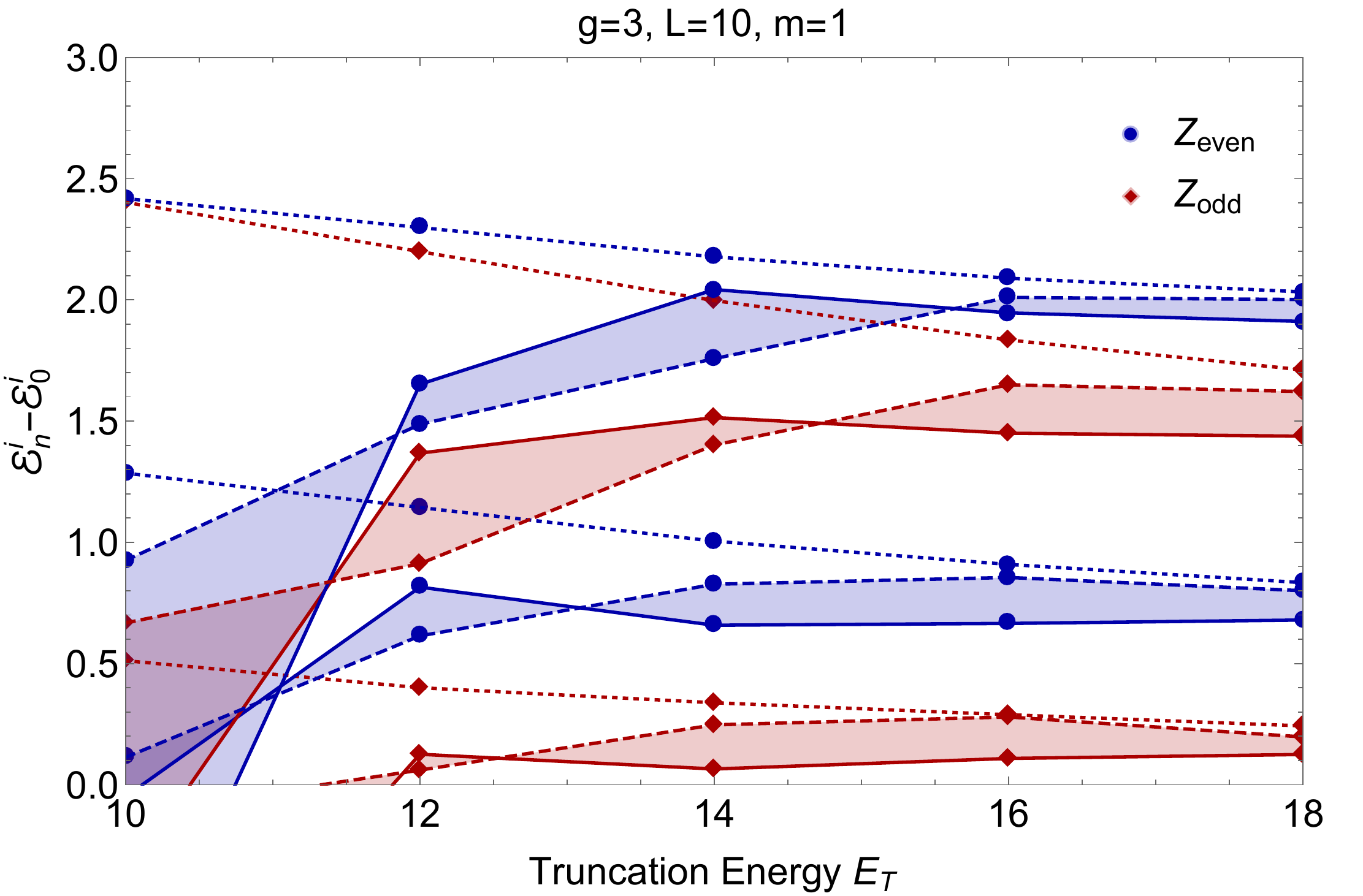} 
\caption{\em   {\bf \em Left:} The vacuum energy $\cE_0^i$  as a function of $E_T$ for  $g=1$, $2$ and $3$ in descending order.  {\bf \em Right:} Energy difference between the  first excited states and the vacuum energy  as a function of the coupling $E_T$ for $g=1$, $2$ and $3$.  In all the plots of the figure the blue curves correspond to  the $\mathds{Z}_2$-even  sector while the red ones to the $\mathds{Z}_2$-odd.  The  dotted curves are computed with the truncated Hamiltonian, while the solid and dashed lines are computed adding $\Delta H_2$ with cutoffs $E_W=E_T/2$ and  $E_W=E_T/3$. }
\label{exactConvergence32}
\end{figure}   
As explained before,  only  those entries with $E_{r,s}$ energies below  a cutoff $E_W$ are renormalized. 
We do the plot for different values of $E_W=E_T/2,E_T/5$ and we find that the vacuum energy and the physical mass do not depend much on this cutoff.   For the left plot the difference between  $E_W=E_T/2$ and $E_W=E_T/5$ is inappreciable.~\footnote{In fact, for this case we have checked that  setting $E_W=E_T$ gives a  result on top of the lines of $E_W=E_T/2$. This is  because at weak coupling there is not much overlap between the lowest lying eigenstates of $H$ and the high $H_0$ eigenstates. }
We find that the spectrum is much flatter as a function of $E_T$ for renormalized eigenvalues than the ones computed with $H_T$. Since the exact spectrum is independent of the truncation energy $E_T$, a flatter curve in $E_T$ indicates a closer value to exact energy levels.
However, it could still happen  that adding $\Delta H_3$ corrections shifted the spectrum by a small amount, as it happens for the $\phi^2$ perturbation seen in Figs.~\ref{exactConvergence} and \ref{exactConvergence2} for the range $16\lesssim E_T\leq 20$.
In the plots we have superimposed constant dashed black lines that are obtained from  the perturbative calculations in \eq{spert1} and \eq{spert2}. We find that the eigenvalues computed with $\Delta H_2$ are much closer to the perturbative calculation than the ones done with $H_T$. The difference between the perturbative result and the one from $\cE^{VV}$ is of ${\cal O}(10^{-4})$ and can be attributed to higher order  corrections in the perturbative expansion. Another source of uncertainty comes from higher order  $\Delta H_n$ corrections not included.

\begin{figure}[t]
\centering
  \includegraphics[width=0.484\textwidth]{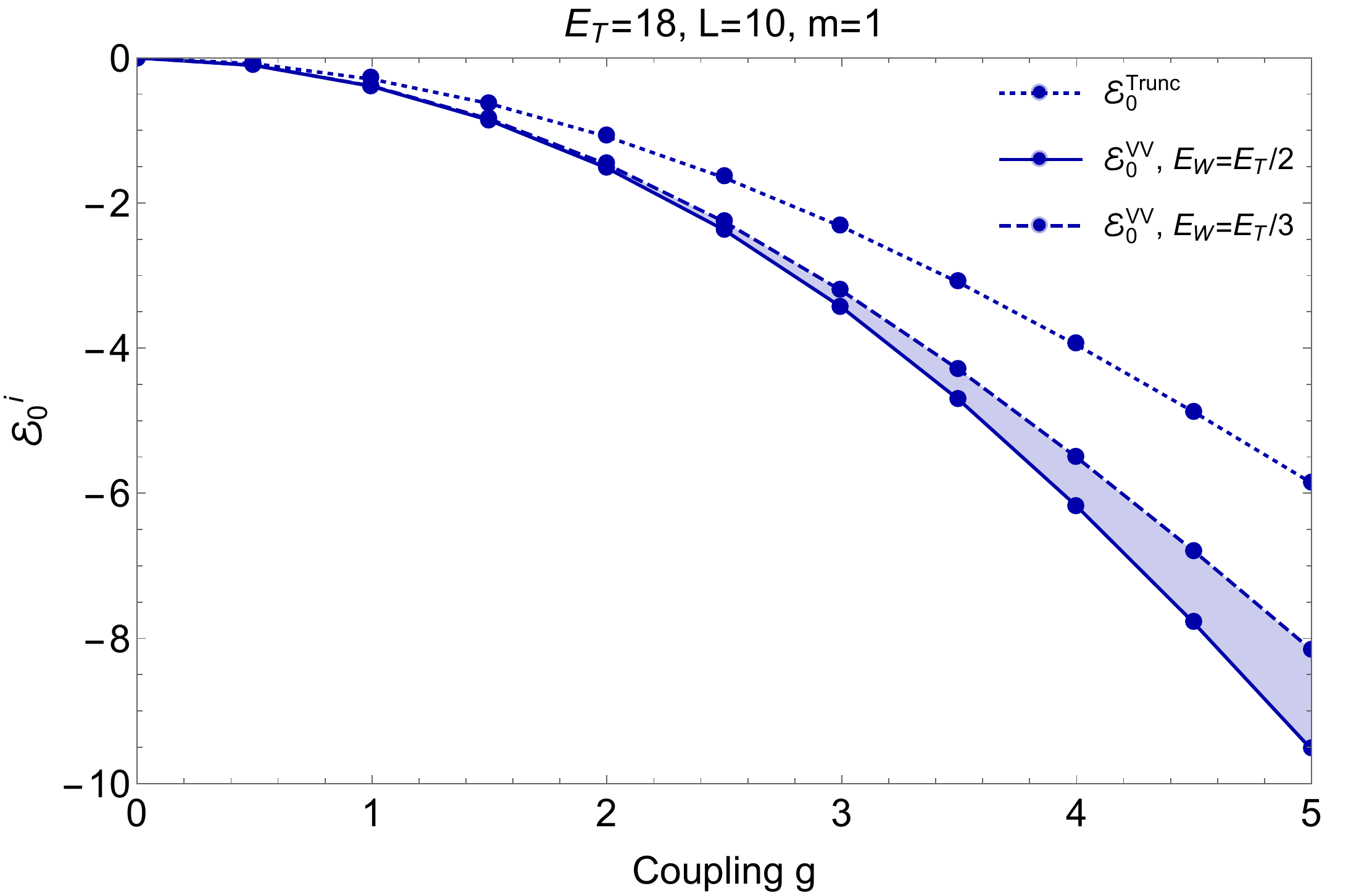} \quad  \includegraphics[width=0.481\textwidth]{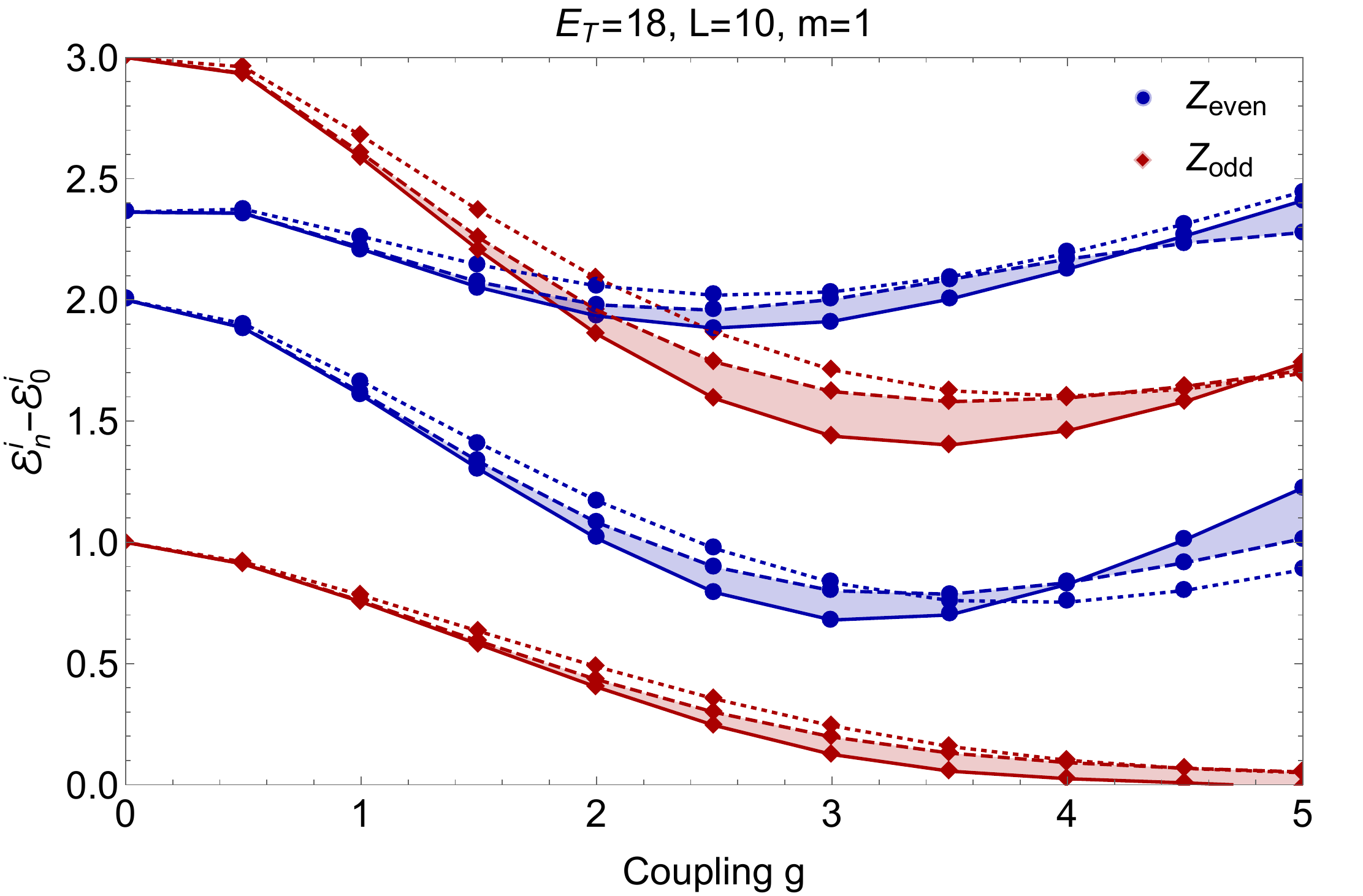} 
\caption{\em   {\bf \em Left:} The vacuum energy $\cE_0$  as a function of the coupling $g$.   {\bf \em Right:} Energy difference between the  first excited states and the vacuum energy  as a function of the coupling constant $g$.  In all the plots of the figure the blue curves correspond to  the $\mathds{Z}_2$-even  sector while the red ones to the $\mathds{Z}_2$-odd.  The  dotted curves are computed with the truncated Hamiltonian for a truncation energy $E_T=18$, while the solid and dashed lines are computed adding $\Delta H_2$ with cutoffs $E_W=E_T/2$ and  $E_W=E_T/3$. }\label{pcoupling}
\end{figure}   

In Fig.~\ref{exactConvergence32} we show  plots with different energy levels as a function of the truncation energy $E_T$ for $g=1$, $2$, $3$. To compare with previous work, these plots have been done with the same choice of parameters and normalizations as in Figs.~(9)-(10) of Ref.~\cite{Rychkov:2014eea}.
In all the plots the  dotted lines are computed using the truncated Hamiltonian while  the solid and dashed lines are computed using $\Delta H_2$ with $E_W=E_T/2$ and $E_W=E_T/3$, respectively. The diamonds and the circles correspond to states in the $\mathds{Z}_2$-even and $\mathds{Z}_2$-odd sectors of the theory. 
We find that in all the plots, for high enough values of $E_T$, the solid lines for the $\Delta H_2$ are   flatter than the truncated ones.
The difference between the dotted and dashed lines is bigger for the plot for $g=3$ than the one for $g=1$. This can be understood because one expects more overlap from higher $H_0$ excited states with the vacuum for higher coupling. The difference between the solid and dashed lines becomes smaller as $E_T$ is increased. This can be understood because as $E_T$ is increased  bigger parts of $(H_T)_{rs}$  are being renormalized, and eventually the difference between using $E_W=E_T/2$ and $E_T/3$ becomes negligible. 
An intrinsic error of our calculation of the eigenvalues is the difference between the values obtained for different choices of $E_W$. This error could be reduced with a more careful  estimate of the expansion parameter $\alpha_{r,s}$, which would be very interesting for the future development of the method. In fact, it seems that for  $E_T\lesssim 12(14)$ for $g=2(3)$ the cutoff $E_W$ is too high (and might include non-perturbative corrections like the one in  \eq{sdds}) as  the eigenvalues deviate a lot from the computation done with $H_T$. 
 Another small source of uncertainty in our calculation comes from not having included higher order $\Delta H_n$ corrections; in the next section we explain the calculation of $\Delta H_3$.

In Fig.~\ref{pcoupling} we show two plots of the vacuum and first excited states as a function of the coupling constant $g$  for $E_T=18$ (cf. Fig.~4 of Ref.~\cite{Rychkov:2014eea}).
There is an intrinsic uncertainty in our procedure in the choice of $E_W$, and as we discussed above it could be lowered by increasing the size of the truncation $E_T$ or ideally by refining the determination of $E_W$. 
Notice that the renormalization of the truncated Hamiltonian matters as the solid lines have a significant difference with respect to the truncated  (as seen in Fig.~\ref{exactConvergence32} the solid lines show a better convergence as a function of $E_T$).
For $g\gtrsim 3.5$ the first $\mathds{Z}_2$-odd excited state seems to become degenerate with the vacuum which is a  signal of the spontaneous   breaking of the $\mathds{Z}_2$ symmetry.
This plot can be used to determine the critical coupling, see Ref.~\cite{Rychkov:2014eea}.

%%%%%%%%%%%%%%%%%%%%%%%%%%%%%%%%%%%%%%
%%%%%%%%%%%%%%%%%%%%%%%%%%%%%%%%%%%%%%
%%%%%%%%%%%%%%%%%%%%%%%%%%%%%%%%%%%%%%
\subsection{Three point correction and further comments}
\label{3pointphi4}
As explained in the previous section we have performed the numerical study of the $\phi^4$ theory without taking into account the three point correction $\Delta H_3$. This would be an interesting point for the future and therefore we give a small preview of the type of expressions one obtains when computing the three point correction. As done throughout the paper, to get the expression for $\Delta H_3$ we start by first computing
 \be
 \Delta \widehat H_3(\cE)_{rs} = -\lim_{\eps\rightarrow 0} \int_0^\infty dt_1dt_2 \, e^{i (\cE-E_r+i\eps)(t_1+t_2)}  {\cal T}\left\{ V(T_1)V(T_2) V(T_3)\right\}_{rs}\, , \label{3pt} 
 \ee
 where $T_k=\sum_{n=1}^{3-k}t_n$. Then we find $\Delta H_3$ by keeping only those terms that have all  poles at $\cE>E_T$. Then, we see that the three point correction can be split into
\be
\Delta H_3= \Delta H_3^{\mathds{1}}+\Delta H_3^{\phi^2}+\Delta H_3^{\phi^4}+\Delta H_3^{\phi^6}+\Delta H_3^{\phi^8}+\Delta H_3^{\phi^{10}}+\Delta H_3^{\phi^{12}}  \, ,
\ee
where the subindices denote the number of fields in each term. The correction $\Delta H_3^{\mathds{1}}$ is given by
\be
\Delta H_3^{\mathds{1}}(\cE)=   \frac{s_{222}\,g^3}{(2L)^6}\sum_{k_i,p_i,l_i} \frac{L^2 \delta_{p_1+p_2+k_1+k_2,0}^{l_1+l_2+k_1+k_2,0}}{\om_{k_1}\om_{k_2}\om_{p_1}\om_{p_2}\om_{l_1}\om_{l_2}}    \frac{\theta(\Sigma_{i=1}^2[\om_{p_i}+\om_{k_i}]-E_T)}{\cE-\Sigma_{i=1}^2[\om_{p_i}+\om_{k_i}]} \frac{\theta(\Sigma_{i=1}^2[\om_{l_i}+\om_{k_i}]-E_T)}{\cE-\Sigma_{i=1}^2[\om_{l_i}+\om_{k_i}]}  \, .
 \ee
 where the symmetry factor is defined in \eq{symfact}. The rest of the terms $\Delta H_3^{\phi^2},\, \cdots,\, \Delta H_3^{\phi^{12}}$ can be computed in a similar fashion as explained in previous sections, but we do not present them here since we did not include them in the numerical analysis.
  
Another interesting thing to study in the future is the local expansion of $\Delta H_3$ and higher orders in $\Delta H_n$. Here we present some of the terms for the $\Delta H_3$ case. As done for $\Delta H_2$, when the local expansion applies the calculation is simplified. 
We use the diagrammatic representation explained  in Appendix~\ref{diags} for the expressions at ${\cal O}(t^0,z^0)$ of the local renormalization. 
As an example the leading local coefficients that renormalize the operators $V_2$, $V_4$ and $V_6$ are
\be
\Delta H_{3+}^{\phi^2}\simeq \Big(  \begin{minipage}[h]{0.13\linewidth}
        \vspace{0pt}
        \includegraphics[width=\linewidth]{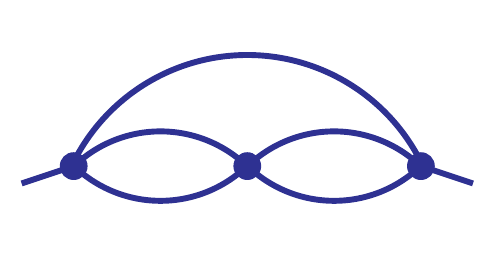} 
   \end{minipage} + \begin{minipage}[h]{0.12\linewidth}
        \vspace{0pt}
        \includegraphics[width=\linewidth]{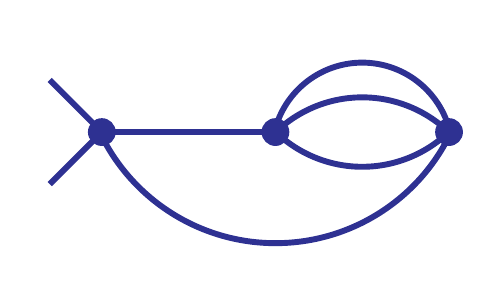} 
   \end{minipage}  +   \begin{minipage}[h]{0.12\linewidth}
        \vspace{0pt}
        \includegraphics[width=\linewidth]{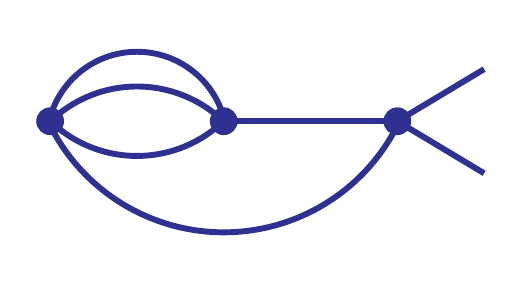} 
   \end{minipage}  + \begin{minipage}[h]{0.12\linewidth}
        \vspace{0pt}
        \includegraphics[width=\linewidth]{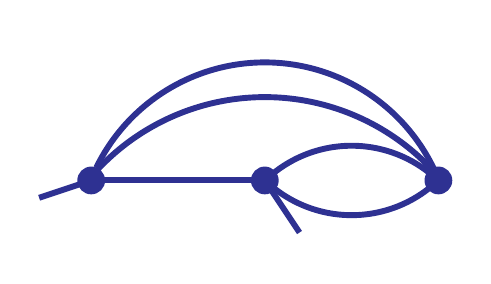} 
   \end{minipage}+\cdots  \Big)\,  V_2
\ee
where for example, 
\be
\begin{minipage}[h]{0.12\linewidth}
        \vspace{0pt}
        \includegraphics[width=\linewidth]{./figs/mass5.pdf} 
   \end{minipage}  =\frac{  s_{131}\, g^3}{(2L)^5}\sum_{k,l,p_i}   \frac{L^2\delta_{p_1+p_2+p_3+k,0}^{l+k,0} }{\om_k\om_{p_1}\om_{p_2}\om_{p_3}}  \frac{\theta(\om_l+\om_k-E_L)}{\cE-\om_l-\om_k}    \frac{\theta(\om_k+\Sigma_{i=1}^3\om_{p_i}-E_L)}{\cE-\om_k-\Sigma_{i=1}^3p_i}   \, .
   \ee
  For the renormalization of the quartic we get
\be
\Delta H_{3+}^{\phi^4}\simeq \Big(  \begin{minipage}[h]{0.135\linewidth}
        \vspace{0pt}
        \includegraphics[width=\linewidth]{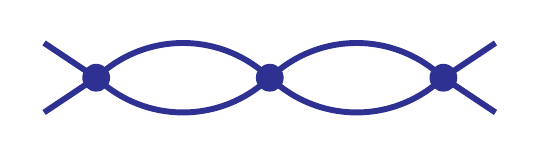} 
   \end{minipage} + \begin{minipage}[h]{0.13\linewidth}
        \vspace{0pt}
        \includegraphics[width=\linewidth]{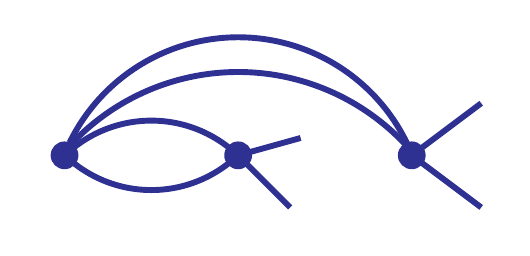} 
   \end{minipage}  +   \begin{minipage}[h]{0.13\linewidth}
        \vspace{0pt}
        \includegraphics[width=\linewidth]{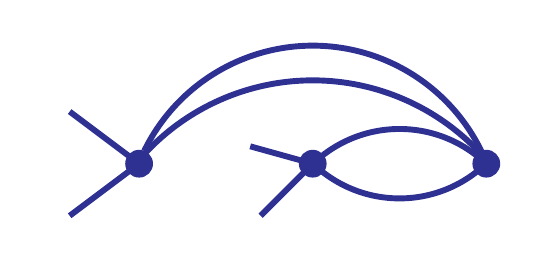} 
   \end{minipage}  +  \cdots  \Big)\,  V_4
   \label{phi4zooVVV}
\ee
where  for example,
   \bea
  \begin{minipage}[h]{0.14\linewidth}
        \vspace{0pt}
        \includegraphics[width=\linewidth]{./figs/lambda1.pdf} 
   \end{minipage} &=&\frac{ s_{220}\, g^3}{(2L)^4}\sum_{l_1l_2p_1p_2}   \frac{L^2\delta_{l_1+l_2,0}^{p_1+p_2,0}}{ \om_{l_1}\om_{l_2}\om_{p_1}\om_{p_2}}  \frac{\theta(\om_{l_1}+\om_{l_2}-E_L)}{\cE-\om_{l_1}-\om_{l_2}}    \frac{\theta(\om_{p_1}+\om_{p_2}-E_L)}{\cE-\om_{p_1}-\om_{p_2}}  \label{gusano}\, . 
    \eea
For  $V_6$
\be
\Delta H_{3+}^{\phi^6}\simeq  ( \begin{minipage}[h]{0.11\linewidth}
        \vspace{0pt}
        \includegraphics[width=\linewidth]{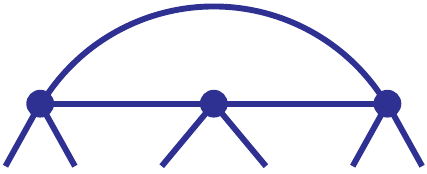} 
   \end{minipage} + \cdots )\, V_6
\ee
where
   \be
    \begin{minipage}[h]{0.11\linewidth}
        \vspace{0pt}
        \includegraphics[width=\linewidth]{./figs/100peus} 
   \end{minipage}  = \frac{  s_{111}\, g^3}{(2L)^3}    \sum_{k}   \frac{L^2}{m\,\om_k^2} \left( \frac{1}{\cE-2\om_k}\right)^2  \theta(2\om_k-E_L)\, . 
   \ee
  
  \medskip
   
    As final remark, notice that the expression in \eq{gusano} is the square of the coefficient of $V_4$ (in $\Delta H_{2+}^{\phi^4}$) up to a numerical factor (see \eq{se457})
    \be
\Big(\begin{minipage}[h]{0.085\linewidth}
        \vspace{0pt}
        \includegraphics[width=\linewidth]{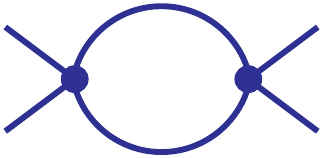} 
   \end{minipage} \Big)^2 \, = 6g \begin{minipage}[h]{0.14\linewidth}
        \vspace{0pt}
        \includegraphics[width=\linewidth]{./figs/lambda1.pdf} 
   \end{minipage}  \, .
    \label{sqrtgusano}
      \ee 
It would be very interesting to investigate whether certain classes of diagrams  in the $\Delta H_+=\sum_n \Delta H_{n+}$ expansion can be resumed. This would reduce the error in the computed spectrum and its dependence  on the arbitrary truncation energy $E_T$. For instance, it could be that the resummation comes only from the leading pieces of the different diagrams.~\footnote{This is the case in standard perturbation theory. For example the Renormalization Group Equations in $d=4$ resum the leading  logs coming from different diagrams.}

\subsection{Summary of the method and comparison with Ref.~\cite{Rychkov:2014eea}}

In this section we summarize our approach to the renormalized Hamiltonian truncation method and briefly comment on the main differences with Ref.~\cite{Rychkov:2014eea}. 
%\medskip

The aim of the renormalized Hamiltonian truncation method is to find   the lowest eigenvalues $\cE$ of $H$. This is done by diagonalizing $H_{eff} \equiv H_T + \Delta H$, where $H_T$ is the  truncated Hamiltonian and $\Delta H$ encodes the contributions from the $H_0$ eigenstates with $E> E_T$. Computing $\Delta H$ is difficult but the problem is   simplified if one expands $\Delta H$ in powers of $V_{hh}/H_{hh}$. One expects that the first terms of the series $\Delta H = \sum_n \Delta H_n$ are a good approximation to $\Delta H$ if the expansion parameter is small. These terms can  be computed as explained in Sec.~\ref{sec:themethod}, by first finding $\Delta \widehat{H}_n$ and keeping only the contributions from the states with $E> E_T$. Then, we notice  that for some entries with $E_r, \, E_s$ close to $E_T$,  the series  $(\Delta H)_{rs} = \sum_n (\Delta H_n)_{rs}$ is not perturbative (for the chosen parameters $g$, $E_T$). We deal with this problem by setting to zero all those entries with $E_r$ or $E_s>E_W$ where $E_W$ is chosen appropriately, see Sec.~\ref{sc2}. 

In order to speed up the numerics and gain analytic insight, we perform several approximations to the exact expression of $\Delta H_2$. First we introduce a scale $E_L$ so that $\Delta H_2=\Delta H_{2 -}+\Delta H_{2+}$ where $\Delta H_{2+}$ only receives contributions of the states  with $E\geq E_L$ while $\Delta H_{2 -}$  only receives contributions of states with $E_T< E < E_L$.
The scale $E_L$ is chosen such that $\Delta H_{2+}$  can be well approximated by the first terms of  a local expansion.  
In our case, we only keep the leading terms  $\Delta H_{2+}=  \sum_{n=0}^{n=2}  c_{2n} \int dx\,  \phi^{2n}(x,t)$ and  we find that the coefficients $c_i$  can be written in terms of phase space functions. 
Lastly, the coefficients $c_i$ are approximated by taking the continuum limit and then expanding them in powers of $m/E_L$. On the other hand $\Delta H_{2-}$ is kept exact because its numerical implementation is less costly and it does not admit an approximation by truncating a local expansion.  The whole procedure has been described in Sec.~\ref{phi4res}  and used to do the plots of Sec.~\ref{sc2}.

\subsubsection*{Comparison with Ref.~\cite{Rychkov:2014eea}}

Refs.~\cite{Rychkov:2014eea,Hogervorst:2014rta} introduced a renormalized Hamiltonian truncation method by diagonalizing $H_{eff}= H_T + \Delta H$ and expanding $\Delta H$ in a series. As explained, we have used this as our starting point. 
In Ref.~\cite{Rychkov:2014eea} though an approximation to $\Delta H_2$ is calculated using a different approach  than in this paper. To get $\Delta H_2$,  Ref.~\cite{Rychkov:2014eea} starts by defining   the following operator  $M(E)$ 
\be
M(E)_{rs} \, dE \equiv \sum_{E_j \le E \le E_j + dE} V_{rj} V_{js}  \quad\quad \text{such that } \quad\quad  \Delta H_2 = \int_{E_T}^{\infty} \,dE \, \frac{M(E)}{\cE-E} \, ,
\label{defME}
\ee
and then noticing that $M(E)$ is related to the matrix element
\be
C(\tau)_{rs} \equiv \bra{r}|V(\tau/2)\, V(-\tau/2) |\ket{s} =  \int_0^\infty dE \, e^{-[E-(E_r+E_s)/2]\tau} \, M(E)_{rs}
\label{ctaueq}
\ee
by a Laplace transform.  In Ref.~\cite{Rychkov:2014eea}, the $E \rightarrow \infty$ behavior of $M(E)$ is  found by doing the inverse Laplace transform of the non-analytic parts of $C(\tau)$ in the limit $\tau \rightarrow 0$.
This is done in the continuum limit, which is a good approximation. The obtained result for $M(E)$ in this limit is taken to compute $\Delta H_2$. Ref.~\cite{Rychkov:2014eea}  differentiates two renormalization procedures, one where the   term ($E_r+E_s)/2$ in \eq{ctaueq}  is approximated to zero (called local), and one where it is taken into account (called sub-leading). In the later case $M(E)$ is given by $M(E-E_{rs})$, and therefore for entries with $E_{rs} \sim E_T$ taking the limit $E-E_{rs} \gg m$ is not justified when $E \sim E_T$. The way in which this problem is dealt with  is by neglecting all the contributions of $M(E-E_{rs})$ for $E \leq E_{rs} + 5 m$; in other words, a $\theta(E-E_{rs}-5 m)$ is multiplied to the integrand in \eq{defME}.~\footnote{They find that $M(E)$ starts to be well approximated by the first terms in the $m/E$ expansion when $E \ge 5m$.}

With this, we can already find the main differences between the two approaches. In our case we calculate the exact expression of $\Delta H_2$ which, if needed, can be approximated. Instead,  Ref.~\cite{Rychkov:2014eea} finds the contributions of $\Delta H_2$ that are leading in the limit where $E \rightarrow \infty$ (which neglects  the  tree and disconnected contributions). From our approach we can recover the local result of Ref.~\cite{Rychkov:2014eea} if  we set $E_L = E_W=E_T$, neglect the tree and disconnected contributions,  take the continuum limit, perform a local expansion to $\Delta H_{2+}$, and make an expansion in $m/E\ll 1$.
The choice $E_L = E_T$  implies $\Delta H_2 = \Delta H_{2+}$ and  $\Delta H_{2-}=0$, while $E_W=E_T$ means that no entries $(\Delta H_2)_{rs}$ are set to zero.  In a similar way we can recover the sub-leading result taking into account the $E_{rs}$ terms, while introducing by hand a $\theta(E-E_{rs}-5 m)$ in the integrals of the coefficients.

Even though the two approaches are quite different, our method and their sub-leading renormalization can still give similar results due to the following. For large enough $E_T$, the low entries of $(\Delta H_2)_{rs}$  only receive contributions from loop-generated operators  \footnote{This can be easily seen from the exact calculations or using the diagrams in Appendix~\ref{diags}.}, and can be well approximated by a local (up to the $E_{rs}$ dependence) expansion  even if $E_L=E_T$. On the other hand, for high energy entries of $(\Delta H_2)_{rs}$ the tree and disconnected operators are non-zero, and none of the operators can be approximated by a truncated local expansion if $E_L=E_T$. However, in many cases these high energy entries become non perturbative and we set them to zero when $E_r$ or $E_s> E_W$. Therefore we find that if $E_W$ is used, it can be a good approximation for large enough $E_T$ to neglect the tree and disconnected terms all together and set $E_L = E_T$ while performing a local expansion. 
With this we connect with Ref.~\cite{Rychkov:2014eea} where the scale $E_W$ is not used to get rid of the non-perturbative contributions. Instead the tree and disconnected terms are  neglected,  all the entries of $(\Delta H_2)_{rs}$ are approximated by the loop-generated local (up to $E_{rs}$) operators only and the $\theta(E-E_{rs}-5 m)$ is introduced in \eq{defME}. As explained, neglecting the tree and disconnected terms is justified, while the introduction of $\theta(E-E_{rs}-5 m)$ and truncating the local expansion in practice largely reduce the values of the high energy entries with respect to the exact result. All of these effectively act as our scale $E_W$. Therefore we see that in many cases our approach and the one in Ref.~\cite{Rychkov:2014eea} can give similar results.

Even though the numerical results are similar, our approach introduces  new tools and insights that we think improve the renormalized Hamiltonian truncation method   and can help to develop it further.
    
\section{Conclusion and outlook}
\label{last}

In this paper we have developed further the Hamiltonian truncation method. 
In particular we have explained a  way to compute the corrections to the truncated Hamiltonian at any order in the large $E_T$ expansion of $\Delta H=\sum_n \Delta H_n$. 
We have applied these ideas to scalar field theory in two dimensions and studied the spectrum of the theory as a function of the truncation energy and the coupling constant.

There are various open directions that are very interesting and deserve further investigation. 
Firstly, it would be a great improvement to the method to find a more precise estimate of the expansion parameter of the series. This  estimate should be easy to implement numerically and  lead to a precise definition of the cutoff $E_W$. In this work we have been pragmatic in this respect, and investigated the behaviour of the spectrum as this cutoff is modified. It might be that only removing the contribution of certain type of matrix elements (like the ones corresponding to high occupation number and zero momentum) the series is greatly improved. 

We have not pushed the numerical aspects of the method very far and all the computations have been done with  \texttt{Mathematica}. With more efficient programming languages it would be interesting to further study and check that as the truncation energy $E_T$ is increased the uncertainty in the precise choice of $E_W$ is reduced.

Another point that should be addressed is the dependence of the spectrum on $L$ as higher $\Delta H_n$ corrections are added; also it could be relevant to inspect if there are diagrams that dominate  for large $Lm\gg 1$. 

Another very interesting path to develop further is to apply renormalization group techniques to resum the fixed order calculations of $\Delta H$. Since the exact eigenvalues do not depend on the truncation energy $E_T$, it may be possible resum the calculation of $\Delta H_n$. Our analytic expressions for the $\Delta H_n$ corrections permit a precise study of the possible resummation of the leading corrections at each order in the perturbation theory of the large $E_T$ expansion.  
One could start by studying the resummation of the leading local corrections, and for that the phase space formulation that we have introduced is useful as there are simple recursion relations for the differential phase space. 

Another fascinating avenue to pursue is the applicability of the method to other theories with higher spin fields and  to increase the number of dimensions. 
In this regard, we  notice that  the derivation of Eqs.~(\ref{phifirst})-(\ref{philast}) seems to be formally valid in any space-time dimension $d$.  
Recall that the $c_i$'s are the coefficients of the local operators added to $H_T$ to take into account the effect of the highest energetic  $H_0$ eigenstates not included in the light Hilbert space ${\cal H}_l$. As $d$ is increased beyond $d=2$ the   UV divergencies appear due to the increasingly rapid growth of the phase space functions $\Phi_i(E)$. One can then regulate the $c_i$ coefficients with a cutoff $\Lambda$. For instance, consider the coefficient $c_4$ of the $\phi^4$ operator
\be 
c_4^\Lambda(\cE)= s_2\,   g_0^2   \ \int_{E_L}^\Lambda   \frac{  dE}{2\pi}\,  \frac{1}{\cE-E}  \, \Phi_2(E) \, ,
 \ee  
in $d=4$.  Then, requiring that the energy levels are independent of the regulator one finds the following  $\beta$-function
 \be
 \beta(g)=  -   \Lambda \frac{\partial c_4^\Lambda}{\partial\Lambda} +{\cal O}(g^3) =   \frac{s_2 g^2}{2\pi}   \Phi_2(\Lambda) +{\cal O}(g^3,\cE) \, , 
 \ee
 where  the $\cE$ corrections can be neglected in the limit of  large $\Lambda \gg \cE$.
Redefining $g\equiv\lambda/4!$ one recovers the known result for the $\lambda\phi^4$ theory
$ \beta(\lambda)=\frac{3}{16\pi^2} \lambda^2 +{\cal O}(\lambda^3)\, ,
$
where we have neglected the mass corrections that for $\Lambda\gg m$ decouple as $\Phi_2(\Lambda)=1/(8\pi)+{\cal O}(m^2/\Lambda^2)$.
A possible way to make contact between the calculation in the renormalized Hamiltonian method and the standard calculation of the beta function is by noticing that the coefficient of the divergent part of the  amplitude  is proportional to the coefficient of its finite imaginary part which in turn (by the optical theorem) is proportional to the two-particle phase space.  It would be very interesting to further study  RG flows from the perspective of the renormalized Hamiltonian truncation method approach.

We think that  the Hamiltonian truncation method is a very promising approach to study strong dynamics, and that there are still open important questions to be addressed.

\section*{Acknowledgments}
 We thank Jos\'e R. Espinosa, Slava Rychkov and Marco Serone for very useful comments on the draft.  We have also benefited from discussions with Christophe Grojean, Antonio Pineda, Andrea Romanino and Giovanni Villadoro. JE-M thanks ICTP for the hospitality during various stages of this work. MR is supported by la Caixa, Severo Ochoa grant. This work has been partly supported by  Spanish Ministry MEC under grants FPA2014-55613-P and FPA2011-25948, by the Generalitat de Catalunya grant 2014-SGR-1450, by the Severo Ochoa excellence program of MINECO (grant SO-2012-0234), and by the European Commission through the Marie Curie Career Integration Grant 631962.
%%%%%%%%%%%%%%%%%%%%%%%%%%%%%%%%%%%%%%%%%%%%%%%%%%%%%%%%%%%%%%%%%%%%%%%%%%%%%%%%%%%%%%%%%%%%%%%%%%%%%%%%%%%%%%%%%%%%%%%%%%%%%%%%%%%
\appendix
   \numberwithin{equation}{section}
    \section{Diagramatic representation}
\label{diags}
There is a simple and powerful diagrammatic representation that permits to easily find the expression for $\Delta H_n$. 
This can be used to either  compute the full operator  $\Delta H_n$ or the leading ${\cal O}(t^0,z^0)$ coefficients in the local expansion of $\Delta H_{n+}$ defined in Sec.~\ref{phi2res}. This representation is valid for any $\phi^\alpha$ theory, but here we give examples only for the $\phi^4$ case for concreteness. 

\subsubsection*{Local coefficients}
Imagine that we want to find the local coefficients ${\cal O}(t^0,z^0)$ for  $\Delta H_{3+}^{\phi^2}$. To find them one puts 3 vertices ordered horizontally~\footnote{The vertices are ordered in a line because the $V(T_s)$'s in  \eq{deltaWe} are time-ordered in the whole integration domain. This is in contrast with the standard Feynman diagrams in the calculation of an $n$-point function, where each space-time integral is over the whole real domain.} and draws all possible diagrams that have only 2 external lines, four lines meeting at each vertex and don't have any lines starting and ending at the same vertex. Next, we assign a momentum for each internal line and draw a vertical line between every pair of vertices. One such diagrams is
\be
  \begin{minipage}[h]{0.18\linewidth}
        \vspace{0pt}
        \includegraphics[width=\linewidth]{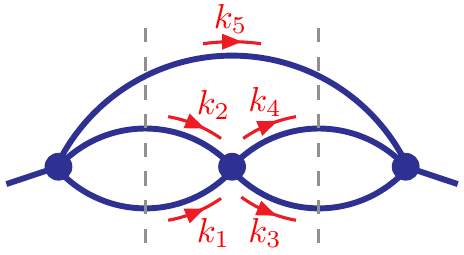} 
   \end{minipage} \, . \label{exdi}
\ee
The expression corresponding to a given diagram with $n$ vertices  and $N$ propagators is given by
\be
 s g^n \sum_{k's}\frac{1}{\prod_{i=1}^N (2L\om_{k_i})}\prod_{p=1}^{n-1}L\, \delta_p \frac{\theta\big(\sum_{k_j \in \{s_p\}}\om_{k_j}-E_L\big)}{\cE-\sum_{k_j \in \{s_p\}}\om_{k_j}} 
\, , \label{digform}\ee 
where $k_j=2 \pi n_j/L$ with $n_j \in \mathbb{Z}$. Each of the  $n-1$ sets of momenta $\{{s_p}\}$ consist in the momenta of the  internal lines that are cut by each vertical line. In (\ref{exdi}) these would be $s_1=\{k_1,k_2,k_5\}$ and $s_2=\{k_3,k_4,k_5\}$. The symbol $\delta_p$ stands for a Kronecker delta that imposes that the total momentum crossing a cut is zero; $s$ is a symmetry factor that counts all the ways that the lines of the vertices can be connected to form the diagram. Applying this recipe to the diagram in (\ref{exdi}) one has
{\small \be
\begin{minipage}[h]{0.16\linewidth}
        \vspace{0pt}
        \includegraphics[width=\linewidth]{./figs/mass1Ex.pdf} 
   \end{minipage} = s^{221}_4\,  g^3\sum_{k's}   \frac{L^2\delta_{k_1+k_2+k_5,0}^{k_3+k_4+k_5,0} }{ \prod_{i=1}^5 (2L\om_{k_i}) }  \frac{\theta(\om_{k_1}+\om_{k_2}+\om_{k_5}-E_L)}{\cE-\om_{k_1}-\om_{k_2}-\om_{k_5}}    \frac{\theta(\om_{k_3}+\om_{k_4}+\om_{k_5}-E_L)}{\cE-\om_{k_3}-\om_{k_4}-\om_{k_5}} 
   \label{A3}
       \, . \vspace{0.13cm}
 \ee
 where the symmetry factor $s_p^{mnv}$ is given in \eq{symfact}. Another example of a contribution to $\Delta H_{3+}^{\phi^2}$ would be
\be
 \begin{minipage}[h]{0.167\linewidth}
        \vspace{0pt}
        \includegraphics[width=\linewidth]{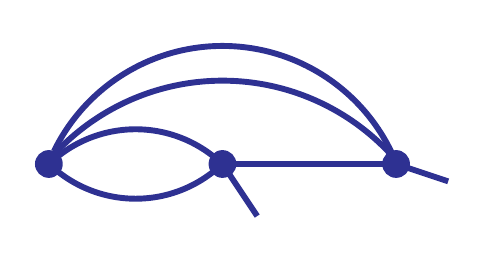} 
   \end{minipage}    =s^{212}_4\,  g^3\sum_{k's}   \frac{L^2\delta_{k_3+k_4+k_5,0}^{k_1+k_2+k_3+k_4,0} }{ \prod_{i=1}^5 (2L\om_{k_i}) }  \frac{\theta(\Sigma_{s=1}^5\om_{k_s}-E_L)}{\cE-\om_{k_1}-\om_{k_2}-\om_{k_3}-\om_{k_4}}    \frac{\theta(\Sigma_{s=1}^4\om_{k_s}-E_L)}{\cE-\om_{k_3}-\om_{k_4}-\om_{k_5}}\,    \label{A4}.
\ee}
Notice that the ordering of the vertices matters since the diagrams of (\ref{A3}) and (\ref{A4}) have the same topology but give different results.

With this prescription one easily recovers Eqs.~(\ref{ch2}), (\ref{llocal2}), and (\ref{se457}) corresponding to the $\Delta H_{2+}$ coefficients in the $\phi^4$ theory
\be
c_0=  \begin{minipage}[h]{0.094\linewidth}
        \vspace{0pt}
        \includegraphics[width=\linewidth]{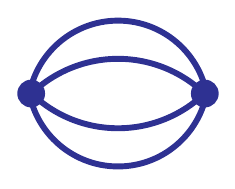} 
   \end{minipage}  \, , \quad \quad \quad c_2=\begin{minipage}[h]{0.133\linewidth}
        \vspace{0pt}
        \includegraphics[width=\linewidth]{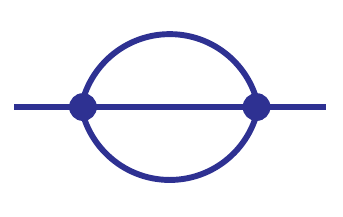} 
   \end{minipage} \, ,  \quad \quad\quad c_4 =\begin{minipage}[h]{0.12\linewidth}
        \vspace{0pt}
        \includegraphics[width=\linewidth]{./figs/Phi42Pt1loop.pdf} 
   \end{minipage} \, . \quad 
\ee 

Notice that to include the contributions $E_{rs}$ mentioned at the end of Sec.~\ref{localphase} the same diagrammatic representation applies but one must then substitute $\cE \rightarrow \cE-E_{rs}$ in \eq{digform}  making the coefficients depend on the matrix entry.
%%%%%%
\subsubsection*{Exact $\Delta H_n$ opertors}

A similar diagrammatic representation can be used to calculate the exact $\Delta \widehat{H}_n$ operator from which one can easily get $\Delta H_n$. The prescription to follow is very similar to the one for the local case, where one starts drawing the same diagrams and putting vertical lines between every pair of vertices. The only difference is that now one extends the external lines to left and right in all possible combinations for each diagram drawn and also assigns a momentum to the external lines. 
For the diagram in (\ref{exdi}) this means
\be
\begin{minipage}[h]{0.16\linewidth}
        \vspace{0pt}
        \includegraphics[width=\linewidth]{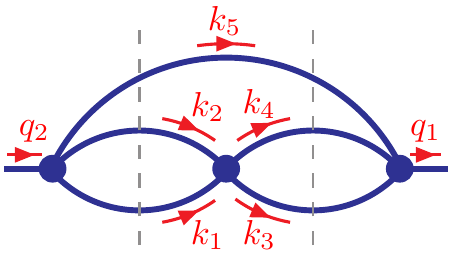} 
   \end{minipage} 
   \quad\quad 
   \begin{minipage}[h]{0.16\linewidth}
        \vspace{0pt}
        \includegraphics[width=\linewidth]{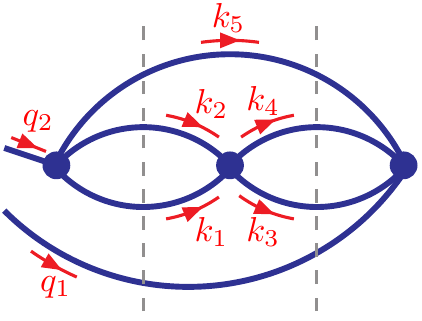} 
   \end{minipage} \,
   \quad\quad 
   \begin{minipage}[h]{0.16\linewidth}
        \vspace{0pt}
        \includegraphics[width=\linewidth]{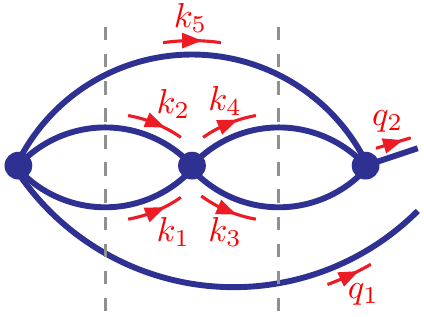} 
   \end{minipage} \,
   \quad\quad 
   \begin{minipage}[h]{0.16\linewidth}
        \vspace{0pt}
        \includegraphics[width=\linewidth]{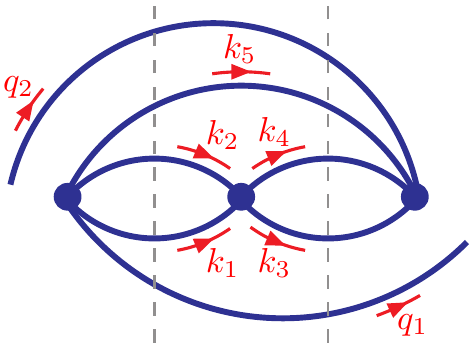} 
   \end{minipage} \, .
   \label{4exactcontr}
\ee 
Now, the operator corresponding to a given diagram with $n$ vertices, $N$ propagators, $A$ external lines starting left and $B$ external lines starting right is
\be
\kappa s g^n \sum_{k's,q's} \frac{1}{\prod_{i=1}^N (2L\om_{k_i})}\prod_{p=1}^{n-1} \frac{\theta\big(\om_{rs}+\sum_{Q_j\in \{s_p\}}\om_{Q_j}-E_L\big)}{\cE-\om_{rs}-\sum_{Q_j \in \{s_p\}}\om_{Q_j}} \,\prod_{\alpha=1}^n L \delta_{\alpha} \prod_{r=A+1}^{A+B}  \frac{a^{\dagger}_{q_r}} {\sqrt{2L\om_{q_r}}}\prod_{l=1}^{A}  \frac{a_{q_l} }{\sqrt{2L\om_{q_l}}} 
\, , \label{digformExact}\ee 
where the sums over $k's,\,q's$ sum over all possible momenta for a given $k_i,\,q_i$. Then, each of the  $n-1$ sets of momenta $\{{s_p}\}$ consists in the momenta of the lines that are cut by each vertical line. For the first diagram from the left in (\ref{digformExact}) these would be $s_1=\{k_1,k_2,k_5\}$ and $s_2=\{k_3,k_4,k_5\}$, and for the second one $s_1=\{q_1, k_1,k_2,k_5\}$ and $s_2=\{q_1, k_3,k_4,k_5\}$. The symbol $\delta_\alpha$ stands for a Kronecker delta that imposes momentum conservation at each vertex $\alpha$.  The symbol $\om_{rs}$ depends on the energy of the states $\bra{E_r}|$, $|\ket{E_s} $ on which $a$ and $a^{\dagger}$ act i.e. it is different for each entry $\big(a^{\dagger}_{-q_r} a_{q_l} \big)_{rs}$,  and is given by $w_{rs}\equiv E_{rs}-\frac{1}{2}\sum_{i=1}^{A+B}\om_{q_i}$ where  $E_{rs}=(E_r+E_s)/2$. As before $s$ is a symmetry factor that counts all the ways that the lines of the vertices can be connected to form the diagram. Lastly $\kappa$ counts all the equivalent ways that the external lines coming out from the same vertex can be ordered left and right, for the diagrams in (\ref{digformExact}) is is always one, since there is only one external line per vertex.  Applying this recipe to the first and second diagrams in (\ref{digformExact})  one has
{\small
\bea
\begin{minipage}[h]{0.16\linewidth}
        \vspace{0pt}
        \includegraphics[width=\linewidth]{./figs/home_gordoNL2.pdf} 
   \end{minipage} &=& s^{221}_4\,  g^3\sum_{k_1, \cdots, k_5} \sum_{q_1,q_2}  \frac{L^2\delta_{ k_1+k_2+k_5,q_1}^{k_3+k_4+k_5,q_2} }{ \prod_{i=1}^5 (2L\om_{k_i}) }  \frac{\theta(\om_{rs}+\om_{k_1}+\om_{k_2}+\om_{k_5}-E_L)}{\cE-\om_{rs}-\om_{k_1}-\om_{k_2}-\om_{k_5}}    \nonumber \\
 && \  \frac{\theta(\om_{rs}+\om_{k_3}+\om_{k_4}+\om_{k_5}-E_L)}{\cE-\om_{rs}-\om_{k_3}-\om_{k_4}-\om_{k_5}}\, L\delta_{k_3+k_4,k_1+k_2}  \frac{a^{\dagger}_{q_1} a_{q_2}}{2 L \sqrt{\om_{q_1} \om_{q_2}}}
       \, ,
       \eea
}
{\small
\bea
\begin{minipage}[h]{0.16\linewidth}
        \vspace{0pt}
        \includegraphics[width=\linewidth]{./figs/home_gordoNL1.pdf} 
   \end{minipage} &=& s^{221}_4\,  g^3\sum_{k_1, \cdots, k_5} \sum_{q_1,q_2}  \frac{L^2\delta_{k_1+k_2+k_5+q_1,0}^{k_3+k_4+k_5+q_1,0} }{ \prod_{i=1}^5 (2L\om_{k_i}) }  \frac{\theta(\om_{rs}+\om_{k_1}+\om_{k_2}+\om_{k_5}+\om_{q_1}-E_L)}{\cE-\om_{rs}-\om_{k_1}-\om_{k_2}-\om_{k_5}-\om_{q_1}}    \nonumber \\
 && \  \frac{\theta(\om_{rs}+\om_{k_3}+\om_{k_4}+\om_{k_5}+\om_{q_1}-E_L)}{\cE-\om_{rs}-\om_{k_3}-\om_{k_4}-\om_{k_5}-\om_{q_1}}\,  L \delta_{q_1+q_2,0} \frac{a^{\dagger}_{-q_1} a_{q_2}}{2 L \sqrt{\om_{q_1} \om_{q_2}}}
       \, ,
 \eea
}
where $\om_{rs} = E_{rs}-(\om_{q_1}+\om_{q_2})/2$ and the symmetry factor $s_p^{mnv}$ is given in \eq{symfact}.

With this set of rules one can easily get the expression for $\Delta \widehat{H}_2$ and $\Delta \widehat{H}_3$ for the $\phi^2$ and $\phi^4$ theories. Then one finds $\Delta H_2$ and $\Delta H_3$  by keeping only the contributions with all poles $\cE > E_T$.

%%%%%%%%%%%%%%%%%%%%
%%%%%%%%%%%%%%%%%%%%%%%%%%%%%%%%%%%%%%%%%%%%%
\section{$\Delta H$ for the $\phi^2$ perturbation}
\label{app1}
\subsection{Two-point correction}

In this section we give the full expressions of the $\Delta \widehat H_2$ corrections for the scalar theory with potential $V=g_2\int dx \phi^2$. Recall that the symmetry factor is given by $s_p=\binom{2}{p}^2 p!$. We will use the prescription $E_{rs}=(E_r+E_s)/2$ where $E_r$ and $E_s$ are $H_0$ eigenvalues.
\bea
\Delta \widehat H_2^\mathds{1} (\cE)_{rs} &=& g_2^2 s_2 \frac{1}{2^2}\sum_k \frac{1}{\om_k^2}\frac{1}{\cE -E_{rs} -2\om_k} \delta_{rs} \\[.2cm]
%%%%%%%%%%%%%%%
\Delta \widehat H_2^{\phi^2} (\cE)_{rs} &=& g_2^2 s_1 \frac{1}{2^2}\sum_q \frac{1}{\om_q^2}\bigg[ \left( a_qa_{-q} \frac{1}{\cE-E_{rs}-\om_q} + \text{h.c.}  \right) \,  \nonumber \\
&+&\, \ad_q a_q \left( \frac{1}{\cE-E_{rs}-2\om_q}+\frac{1}{\cE-E_{rs}}  \right)   \bigg] \\
%%%%%%%%%%%%%%%
\Delta \widehat H_2^{\phi^4} (\cE)_{rs} &=& g_2^2 s_0 \frac{1}{2^2} \sum_{q_1,q_2}\frac{1}{\om_{q_1} \om_{q_2}}\bigg[ \,\,a_{q_1}a_{q_2}a_{-q_1}a_{-q_2} \frac{1}{\cE-E_{rs} -\om_{q_1}+\om_{q_2}} +\text{h.c.} \nonumber\\
&&\phantom{g^2 s_0 \frac{1}{2^2}\sum_{q_1,q_2}} +2\,  \ad_{q_1}a_{q_1} a_{q_2}a_{-q_2} \left( \frac{1}{\cE-E_{rs}+\om_{q_2}} + \frac{1}{\cE-E_{rs}-\om_{q_2}} \right) +\text{h.c.}\nonumber\\
&&\phantom{g^2 s_0 \frac{1}{2^2}\sum_{q_1,q_2}} + \ad_{q_1}\ad_{-q_1}a_{q_2}a_{-q_2} \left( \frac{1}{\cE-E_{rs} +\om_{q_1}+\om_{q_2}} +\frac{1}{\cE-E_{rs} -\om_{q_1}-\om_{q_2}}\right) \nonumber\\
&&\phantom{g^2 s_0 \frac{1}{2^2}\sum_{q_1,q_2}} +4\, \ad_{q_1}\ad_{q_2}a_{q_1}a_{q_2} \frac{1}{\cE-E_{rs}} \,  \bigg]  \, . 
\eea

\subsection{Three-point correction}

In this section we give the full expressions of the $\Delta H_3$ corrections for the scalar theory with potential $V=g_2\int dx \phi^2$. Recall that the symmetry factor is given by \be
s^{mnv}_p=\frac{p!^3}{(p-m-n)!(p-m-v)!(p-n-v)! m!n!v!}\, .
\ee
We use the notation $\Delta H_3= \Delta H_3^{\mathds{1}}+\Delta H_3^{\phi^2}+\Delta H_3^{\phi^4}+\Delta H_3^{\phi^6}$,  where 
\bea
\Delta  H_3^{\mathds{1}}(\cE)_{rs} &=& \frac{g_2^3 s_2^{111}}{2^3} \sum_k \frac{1}{\om_{k_1}\om_{k_2}\om_{k_3}} G_0(k_1,k_2,k_3,E_T) \, ,\label{app31}  \\[.2cm]
%%%%%%%%%%%%%%%%%%%%%%%%%%%%%%%%%%%%%%%%%%%%%%
\Delta  H_3^{\phi^2}(\cE)_{rs} &=&  \frac{g_2^3}{2^3} \sum_{k,q} \frac{1}{\om_{k_1}\om_{k_2}}\frac{1}{\sqrt{\om_{q_1}\om_{q_2}}}\big[ s_2^{200} G_{2,1} (k_1,k_2,q_1,q_2,E_T)  \nonumber\\&& + s_2^{110} G_{2,2} (k_1,k_2,q_1,q_2,E_T)\big] \, ,   \\[.2cm]
%%%%%%%%%%%%%%%%%%%%%%%%%%%%%%%%%%%%%%%%%%%%%%
\Delta  H_3^{\phi^4}(\cE)_{rs} &=& \frac{g_2^3 s_2^{100}}{2^3}  \sum_{k,q} \frac{1}{\om_k} \frac{1}{\sqrt{\om_{q_1}\cdots\om_{q_4}}} G_4(k,q_1,\dots,q_4)  \, ,   \\[.2cm]
%%%%%%%%%%%%%%%%%%%%%%%%%%%%%%%%%%%%%%%%%
\Delta  H_3^{\phi^6}(\cE)_{rs} &=&  \frac{g_2^3 s_2^{000}}{2^3}\sum_q \frac{1}{\sqrt{\om_{q_1}\cdots\om_{q_6}}} G_6(q_1,\dots,q_6)  \, , \label{app33}\eea
where
\bea
G_0 &=&    \delta_{k_1+k_2,0}\delta_{k_1+k_3,0} \, [f_0]^{12}[f_0]^{13} \, ,   \\[.2cm]
%%%%%%%%%%%%%%%%%%%%%
G_{2,1} &=& a_{q_1}a_{q_2}\,\delta_0 \delta_{k_1+k_2,0}
\, [f_2]^{12}_{12}[f_2]^{12} \, +\text{h.c.}+ 2\ad_{q_1}a_{q_2}\,\delta_1 \delta_{k_1+k_2,0}\, [f_2]^{12}_{1}[f_2]^{12}_{2} \, ,  \\[.2cm]
G_{2,2} &=& \ad_{q_1}a_{q_2} \,\delta_1 \delta_{k_1+q_1,0}\delta_{k_1+k_2,0}\, \left( [f_2]^1_{12}[f_2]^{12}_1 + [f_2]^1_{12}[f_2]^{12}_2 + [f_2]^1_{12}[f_2]^2_{12} \right) \, ,    \\[.2cm]
%%%%%%%%%%%%%%%
G_{4} &=& \ad_{q_1}a_{q_2}a_{q_3}a_{q_4} \,\delta_1 \delta_{k+q_1,0} \delta_{q_3+q_4,0} [f_4]^{1}_{1234} [f_4]^1_{12}\, +\text{h.c.}  \,  \nonumber   \\
&+& 2\ad_{q_1}\ad_{q_2}a_{q_3}a_{q_4} \,\delta_2  \delta_{q_1-q_3,0} \delta_{k+q_2,0} [f_4]^{1}_{124} [f_4]^1_{234}\nonumber \\
&+& \ad_{q_1} \ad_{q_2}a_{q_3}a_{q_4} \, \delta_2 \left( \delta_{q_1+q_2,0}\delta_{k+q_4,0}[f_4]_{1234}[f_4]^1_{124} +\delta_{q_3+q_4,0} \delta_{k+q_1,0}[f_4]_{1234}[f_4]^1_{124} \right)\,  ,  \quad \quad \\[.2cm]
G_6 &=& \ad_{q_1}\ad_{q_2}a_{q_3}a_{q_4}a_{q_5}a_{q_6} \,\delta_2 \delta_{q_1+q_2,0} \delta_{q_3+q_4,0} [f_6]_{123456} [f_6]_{1234}\, +\text{h.c.}  \,     \nonumber\\
&+& 2\ad_{q_1}\ad_{q_2}\ad_{q_3}a_{q_4}a_{q_5}a_{q_6} \, \delta_3 \delta_{q_1+q_2,0} \delta_{q_5+q_6,0} [f_6]_{12356} [f_6]_{12456}\,  \, .
\eea
We have defined   $w_{rs}^p\equiv E_{rs}-\frac{1}{2}\sum_{i=1}^p\om_{q_i}$,   $\delta_{d}\equiv\delta_{\Sigma_{i=1}^d q_i,\Sigma_{j=d+1}^p q_j}$ (the Kronecker delta that  imposes momentum conservation to the creation/annihilation operators) and
\be
[f_p]_{ Q }^{K }=\frac{\theta(\om_{rs}^p+\Sigma_{i\in \{Q\}}\om_{q_i}+\Sigma_{i\in \{K\}}\om_{k_i}-E_T)}{\cE-\om_{rs}^p-\sum_{i\in\{Q \}}\om_{q_i}-\sum_{i\in \{K\} }\om_{k_i}} \, . 
\ee

\section{$\Delta H$ for the $\phi^4$ theory}
\label{deltahforphi4app}
In this appendix we give the exact two-point correction and the first terms in the local expansion of the three-point correction. Getting the exact three-point correction would be straightforward. 

\subsection{Two-point correction}
In this appendix we give the full expressions of the $\Delta H_2$ for the $\phi^4$ theory. Using the notation $\Delta H_2=\sum_{n=0}^{8} \Delta H_2^{\phi^n}$ we have
\bea
 \Delta  H_2^{\mathds{1}}(\cE,E_T) &=&\frac{s_4g^2}{2^4L^2}\sum_{k_1k_2k_3k_4} \frac{1}{\omega_{k_1}\omega_{k_2}\om_{k_3}\om_{k_4}} F_{0}(k_1,k_2,k_3,k_4,E_T)  \, , \label{F1}  \\[.2cm]
\Delta  H_2^{\phi^2} (\cE,E_T)&=&\frac{s_{3}g^2}{2^4L^2} \sum_{k_1,k_2,k_3 }\sum_{q_1,q_2}\,  \frac{1}{\omega_{k_1}\omega_{k_2}\om_{k_3}}\frac{1}{\sqrt{\om_{q_1}\om_{q_2}}} F_{2}(k_1,k_2,k_3,q_1,q_2,E_T)   \label{phi2phi4} \, , \\[.2cm]
\Delta  H_2^{\phi^4} (\cE,E_T)&=&\frac{s_{2}g^2}{2^4L^2} \sum_{k_1,k_2  } \sum_{q_1, q_2, q_3,q_4}\, \frac{1}{\om_{k_1}\om_{k_2}   }\frac{1}{\sqrt{\om_{q_1}\cdots\om_{q_4}}} F_4(k_1,k_2,q_1,\dots,q_4,E_T)   \, ,\\[.2cm]
\Delta  H_2^{\phi^6} (\cE,E_T)&=&\frac{s_{1}g^2}{2^4L^2}  \sum_k \sum_{q_1,\dots,q_6}\,\frac{1}{\om_k}\frac{1}{\sqrt{\om_{q_1}\dots\om_{q_6}}}  F_6(k,q_1,q_2,\dots ,q_6,E_T) \label{F6}  \, , \\[.2cm]
\Delta  H_2^{\phi^8} (\cE,E_T)&=&\frac{s_{0}g^2}{2^4L^2}  \sum_{q_1,\dots,q_8}\, \frac{1}{\sqrt{\om_{q_1}\cdots\om_{q_8}}} F_8(q_1,q_2,\dots ,q_8,E_T)  \label{F8}
\eea
The $F_i$ functions are given by 
\bea
F_0&=& \delta_{k_1+k_2+k_3+k_4,0}\, [f_0]^{1234} \\[.3cm]
%%%%% phi2 %%%%%%
F_2&=& 
%% aa %%
 \ad_{q_1} a_{q_2}\,\delta_1\, \delta_{k_1+k_2+k_3,q_1} \left(\,  [f_2]^{123}+[f_2]^{123}_{12}\,  \right)+a_{q_1}a_{q_2}\,\delta_0 \, \delta_{k_1+k_2+k_3,q_1}\, [f_2]^{123}_{2} \, +\, \text{h.c.}   \,  \\[.3cm]
%%%%%% phi4 %%%%%
F_4&=& 
%% aaaa %%
a_{q_1}a_{q_2}a_{q_3}a_{q_4}\,\delta_0\, \delta_{k_1+k_2,q_1+q_2} [f_4]^{12}_{34}+\text{h.c.}\nonumber\\[.2cm]
%% ad aaa %%%
&+& 2 \ad_{q_1}a_{q_2}a_{q_3}a_{q_4}\,\delta_1\, \left( \, \delta_{k_1+k_2,q_1-q_2}  [f_4]^{12}_{2} + \delta_{k_1+k_2,-q_1+q_2}  [f_4]^{12}_{134}\, \right)+\text{h.c.}\nonumber\\[.2cm]
%% adad aa %%
&+& \ad_{q_1}\ad_{q_2}a_{q_3}a_{q_4} \,\delta_2\, \big(\, \delta_{k_1+k_2,q_1+q_2} \, [f_4]^{12}+  \delta_{k_1+k_2,-q_1-q_2}  [f_4]^{12}_{1234} \,  \, +\, 4\, \delta_{k_1+k_2,q_1-q_3} [f_4]^{12}_{14} \big) \\
%%%%%%%%%% phi 6 %%%%%%%%%%%
F_6&=&
%%%ad aaaaa %%%
\ad_{q_1}a_{q_2}a_{q_3}a_{q_4}a_{q_5}a_{q_6}\,\delta_1 \, \delta_{k,q_2+q_3-q_1}\, 3\, [f_6]^{1}_{1456}\,+\text{h.c.}\nonumber\\[.2cm]
%%% adad aaaa %%%
&+& \ad_{q_1}\ad_{q_2}a_{q_3}a_{q_4}a_{q_5}a_{q_6}\,\delta_2 \left( \, 9\, \delta_{k,q_3+q_4-q_1} [f_6]^1_{156}+3\, \delta_{k,q_3-q_1-q_2} [f_6]^1_{12456}\,\right)\, +\text{h.c.}  \\[.2cm]
%% adadad aaa %%%
&+& \ad_{q_1}\ad_{q_2}\ad_{q_3}a_{q_4}a_{q_5}a_{q_6}\,\delta_3\, \big(\, 9\, \delta_{k,q_4+q_5-q_1} [f_6]^1_{16} +9  \delta_{k,-q_4-q_5+q_1}[f_6]^1_{2345}  \,+ \delta_{k+q_1+q_2+q_3,0} [f_6]^1_{123456} \, \big)  \nonumber\\[.3cm]
%%%%%%%%%% phi 8 %%%%%%%%%%%%
F_8&=& 
%% adad aaaaaa %%
\ad_{q_1}\ad_{q_2}a_{q_3}a_{q_4}a_{q_5}a_{q_6}a_{q_7}a_{q_8}\,\delta_2 \, 6 \delta_{q_1+q_2-q_3-q_4,0} [f_8]_{125678}\, +\, \text{h.c.}\nonumber\\[.2cm]
%% adadad aaaaa %%
&+& \ad_{q_1}\ad_{q_2}\ad_{q_3}a_{q_4}a_{q_5}a_{q_6}a_{q_7}a_{q_8}\,\delta_3 \big( 24 \delta_{q_1+q_2-q_4-q_5} [f_8]_{12678}\, +\,4 \delta_{q_1+q_2+q_3-q_4} [f_8]_{1235678}  \,\big) +\text{h.c.}\nonumber\\[.2cm]
%% adadadad aaaa %%
&+& \ad_{q_1}\ad_{q_2}\ad_{q_3}\ad_{q_4}a_{q_5}a_{q_6}a_{q_7}a_{q_8}\,\delta_4 \big(  \,16 \delta_{q_1-q_5-q_6-q_7,0}\left( [f_8]_{18} + [f_8]_{234567} \right) \nonumber\\[.2cm]
&+& 36 \delta_{q_1+q_2-q_5-q_6,0}[f_8]_{1278} + \delta_{q_1+q_2+q_3+q_4,0} [f_8]_{12345678} \,\big)  \, . 
\eea
We have defined   $w_{rs}^p\equiv E_{rs}-\frac{1}{2}\sum_{i=1}^p\om_{q_i}$,   $\delta_{d}\equiv\delta_{\Sigma_{i=1}^d q_i,\Sigma_{j=d+1}^p q_j}$ (the Kronecker delta that  imposes momentum conservation to the creation/annihilation operators) and
\be
[f_p]_{ Q }^{K }=\frac{\theta(\om_{rs}^p+\Sigma_{i\in \{Q\}}\om_{q_i}+\Sigma_{i\in \{K\}}\om_{k_i}-E_T)}{\cE-\om_{rs}^p-\sum_{i\in\{Q \}}\om_{q_i}-\sum_{i\in \{K\} }\om_{k_i}} \, . 
\ee

%%%%%%%%%%%%%%%%%%%%%%%%%%%%%%%%%%%%%%%%%%%%%%%%%%%%%%%%%%%%%%%%%%%%%%%%%%%%%%%%%%%%%%%%%%%%%%%%%%%%%%%%%%%%%%%%%%%%%%%%%%%%%%%%%%%%%%%%%%%%%%%%%%%%%%%%%%%%%%%%%%%%%%%%%%%%%%%%%%%%%%%%%%%%%%%%%%%%%%%%%%%%%%%%%%%%%%%%%%%%%%%%%%%%%%%%%%%%%%%%%%%%%%%%%%%%%%%%%%%%%%

\end{document}